\documentclass[nofootinbib,prd,aps,superscriptaddress,preprintnumbers,twocolumn,showpacs]{revtex4-1}
\usepackage[dvipdfmx]{graphicx}
\usepackage{amsmath,amssymb,amsthm,bm,bigdelim,multirow}
\usepackage{color}

\newcommand{\kslash}{k\kern-1ex /}
\newcommand{\pslash}{p\kern-1ex /}
\newcommand{\qslash}{q\kern-1ex /}
\newcommand{\lslash}{l\kern-1ex /}
\newcommand{\sslash}{s\kern-1ex /}
\newcommand{\Dslash}{D\kern-1.2ex /}

\newcommand{\tr}{{\rm tr}}

\newcommand{\beqa}{\begin{eqnarray}}
\newcommand{\eeqa}{\end{eqnarray}}

\newcommand{\bd}{\begin{description}}
\newcommand{\ed}{\end{description}}
\newcommand{\la}{\langle}
\newcommand{\ra}{\rangle}
\newcommand{\ben}{\begin{eqnarray}}
\newcommand{\een}{\end{eqnarray}}
\newcommand{\nn}{\nonumber}
\def\lsim{\raise0.3ex\hbox{$<$\kern-0.75em\raise-1.1ex\hbox{$\sim$}}}
\def\gsim{\raise0.3ex\hbox{$>$\kern-0.75em\raise-1.1ex\hbox{$\sim$}}}
\def\simgt{\rlap{\lower 3.5 pt\hbox{$\mathchar \sim$}}\raise 2.0pt \hbox {$>$}}
\def\simlt{\rlap{\lower 3.5 pt\hbox{$\mathchar \sim$}}\raise 2.0pt \hbox {$<$}}

\newcommand{\csw}{{c_{\rm SW}}}

\newcommand{\crd}{\color{black}}

\begin{document}
\title{2+1 flavor light hadron spectrum with PACS10 configurations}

\author{K.-I.~Ishikawa}
\affiliation{Graduate School of Science, Hiroshima University, Higashi-Hiroshima, Hiroshima 739-8526, Japan}

\author{N.~Ishizuka}
\affiliation{Center for Computational Sciences, University of Tsukuba, Tsukuba, Ibaraki 305-8577, Japan}

\author{Y.~Kuramashi}
\affiliation{Center for Computational Sciences, University of Tsukuba, Tsukuba, Ibaraki 305-8577, Japan}

\author{Y.~Nakamura}
\affiliation{RIKEN Center for Computational Science, Kobe, Hyogo 650-0047, Japan}

\author{Y.~Namekawa}
\affiliation{Department of Computer Sciences, Fukuyama University, Fukuyama, Hiroshima 729-0292, Japan}

\author{K.~Satoh}
\affiliation{Center for Computational Sciences, University of Tsukuba, Tsukuba, Ibaraki 305-8577, Japan}

\author{E.~Shintani}
\affiliation{Graduate School of Engineering, University of Tokyo, Hongo 7-3-1, Bunkyo, Tokyo, Japan}

\author{Y.~Taniguchi}
\affiliation{Center for Computational Sciences, University of Tsukuba, Tsukuba, Ibaraki 305-8577, Japan}

\author{N.~Ukita}
\affiliation{National Institute of Technology, ICHINOSEKI College, Ichinoseki, Iwate 021-8511, Japan}

\author{H.~Watanabe}
\affiliation{Keio University Sustainable Quantum Artificial Intelligence Center (KSQAIC), Keio University, Minato, Tokyo 108-8345, Japan}

\author{T.~Yamazaki}
\affiliation{Institute of Pure and Applied Sciences, University of Tsukuba, Tsukuba, Ibaraki 305-8571, Japan}
\affiliation{Center for Computational Sciences, University of Tsukuba, Tsukuba, Ibaraki 305-8577, Japan}

\author{T.~Yoshi\'e}
\affiliation{Center for Computational Sciences, University of Tsukuba, Tsukuba, Ibaraki 305-8577, Japan}

\collaboration{PACS Collaboration}
\noaffiliation

\begin{abstract}
We present the results for the hadron spectrum, the pseudoscalar meson decay constants, and the quark mass ratio in the continuum limit using the ``PACS10'' gauge configurations generated with the $O(a)$-improved Wilson quark action and the Iwasaki gauge action. Numerical simulations are carried out at three lattice spacings keeping the space-time volumes over (10 fm$)^4$ in 2+1 flavor QCD at the physical point. We calculate  physical quantities with two choices of conventions to define pseudoexperimental values expected in the isosymmetric QCD. 
In order to investigate the finite size effects on the physical quantities we have additionally generated the (5.4 fm$)^4$ lattice at the coarsest lattice spacing of $a=0.08487(14)$~fm, which is a hypercube with half the length in each direction compared to the PACS10 configurations. 
While the finite size effect is hardly observed in comparison of the hadron effective masses on the (10.9 fm$)^4$ and (5.4 fm$)^4$ lattices, the results for the pseudoscalar meson decay constants show discrepancies beyond the error bars. 
The (5.4 fm$)^4$ lattice also serves for determining the expansion parameters of the pseudoscalar meson masses and decay constants, and the octet baryon masses in terms of the quark masses around the physical point with the aid of the mass reweighting technique. 
The expansion parameters are used for the fine-tuning of the physical point employing the pseudoexperimental masses of the $\pi$ meson, $K$ meson, and the $\Xi$ baryon as physical inputs in each convention. 
 We compare the results for the octet baryon masses and the pseudoscalar decay constants in the continuum limit with the corresponding pseudoexperimental values in both conventions. 
 We also discuss the deviation between two results with different conventions.
 As for the quark mass ratio the result is compared  with the recent FLAG summary.

\end{abstract}
\date{\today}

\preprint{UTHEP-827, UTCCS-P-183, HUPD-2609}

\maketitle

\newpage
\section{Introduction}

The PACS Collaboration has generated the ``PACS10'' configurations with the Wilson-type quarks in 2+1 flavor QCD keeping the physical lattice size over  (10 fm$)^4$ at the physical point with three choices of lattice spacings. The PACS10 configurations have three characteristic features. Firstly, the configurations are generated only at the physical point. This allows us to avoid the systematic errors associated with the chiral extrapolation. Secondly, the extremely large lattice volume $V=L^4$ over (10 fm$)^4$, which reduces the statistical errors in proportion to $1/\sqrt{V}$ and makes the momentum resolution finer  in proportion to $1/L$. Thirdly, three sets of configurations at different lattice spacings allow us to take the continuum limit of physical quantities.

With the use of  partial sets of the PACS10 configurations we have investigated the finite size effect on the hadron masses~\cite{Ishikawa:2018jee,PACS:2019ofv} and calculated various physical quantities~\cite{Shintani:2018ozy,Shintani:2019wai,PACS:2019hxd,Ishikawa:2022ulx,Tsuji:2022ric,Tsuji:2023llh,Aoki:2025taf}. 
In the previous finite size study~\cite{Ishikawa:2018jee,PACS:2019ofv} we performed a comparison between the results on the (10.9 fm$)^4$ and (5.4 fm$)^4$ lattices at the coarsest lattice spacing of $a=0.08487(14)$~fm, the latter of which was additionally generated. The finite size effect on the vector meson effective masses and the baryon effective masses is hardly observed. The situation was somewhat confusing in the pseudoscalar meson sector. The direct comparison with the same hopping parameters on both lattices yields the about 2\% and 5\% deviation for the pion mass and the axial ward identity (AWI) ud quark mass, respectively. After making the AWI ud quark mass on the (5.4 fm$)^4$ lattice equal to that on the (10.9 fm$)^4$ lattice with the mass reweighting technique, the results for the pion mass show consistency between two lattices while the finite size effect on the pseudoscalar meson decay constants is roughly three times larger than the chiral perturbation (ChPT) prediction. 

In this paper we revisit the finite size effect on the hadron masses and the pseudoscalar meson decay constants increasing the statistics of the  (5.4 fm$)^4$ lattice fourfold. For the measurement of physical quantities we employ two conventions to define the pseudoexperimental values in the isosymmetric QCD. In order to investigate the quark mass dependence of the physical quantities around the physical point we apply the mass reweighting technique to the (5.4 fm$)^4$ lattice. The coefficients of the Taylor expansion in terms of the AWI quark masses are used to tune the physical quantities to the physical point determined with $m_\pi$, $m_K$ and $m_\Xi$ as the physical inputs in each convention. Finally, we extrapolate the physical quantities with each convention at three lattice spacings to the continuum limit and compare our results with the pseudoexperimental values or the FLAG summary.

This paper is organized as follows.  Sec.~\ref{sec:simulation} explains the simulation details including the numerical technique and two conventions to determine the pseudoexperimental values in the isosymmetric QCD. In Sec.~\ref{sec:results_b182} we first investigate the finite size effect on the hadron masses and the pseudoscalar meson decay constants. After that we determine the quark mass dependence of the physical quantities around the physical point using the results on the coarsest lattice with the convention A. The results at other lattice spacings are presented in Sec.~\ref{sec:results_b200b220}. Possible finite size effects due to the topological sectors with less frequent transitions are discussed in Sec.~\ref{sec:topologicalcharge}. 
We take the continuum limit of the physical quantities with the convention A in Sec.~\ref{sec:continuum}. In Sec.~\ref{sec:isoqcd_ed} we repeat calculation of the physical quantities with the convention B and discuss the difference between the results with two conventions. Our summary and outlook are given in Sec.~\ref{sec:summary}.

\section{Simulation details}
\label{sec:simulation}

\subsection{Actions}
\label{subsec:action}

We employ the Iwasaki gauge action~\cite{Iwasaki:1983iya} and 
the stout-smeared $O(a)$-improved Wilson quark action.
The former is composed of a plaquette and a $1\times 2$ rectangle loop with the original link variables $U$:
\ben
S_{\rm g}=\frac{1}{g^2}\left\{ c_0\sum_{\rm plaquette}\tr U_{pl}(U)
+c_1\sum_{\rm rectangle}\tr U_{rtg}(U) \right\}
\label{eq:action_g}
\een
with $c_1=-0.331$ and $c_0=1-8c_1=3.648$.
The latter is expressed as
\begin{widetext}
\ben
S_{\rm quark}&=&\sum_{q={\rm  u,d,s}}\left[
\sum_n {\bar q}_n q_n 
    -\kappa_q \csw \sum_n \sum_{\mu,\nu}\frac{i}{2}
              {\bar q}_n\sigma_{\mu\nu}{\tilde F}_{\mu\nu}(n)q_n
\right.\nn\\
&&\left.
-\kappa_q\sum_n
\sum_\mu \left\{{\bar q}_n(1-\gamma_\mu){\tilde U}_{n,\mu} q_{n+{\hat \mu}}
               +{\bar q}_n(1+\gamma_\mu){\tilde U}^{\dag}_{n-{\hat \mu},\mu} q_{n-{\hat \mu}}\right\}
 \right],
\label{eq:action_q}
\een
\end{widetext}
where ${\tilde U}$ denotes the stout-smeared links with six smearing steps choosing the smearing parameter $\rho = 0.1$~\cite{Morningstar:2003gk}.
We consider the case of a degenerate up and down quark mass
$\kappa_{\rm u}=\kappa_{\rm d}=\kappa_{\rm ud}$.
The Euclidean gamma matrices are defined in terms of
the Minkowski matrices in the Bjorken-Drell convention:
$\gamma_j=-i\gamma_{BD}^j$ $(j=1,2,3)$, 
$\gamma_4=\gamma_{BD}^0$,
$\gamma_5=\gamma_{BD}^5$ and 
$\sigma_{\mu\nu}=\frac{1}{2}[\gamma_\mu,\gamma_\nu]$.
The field strength ${\tilde F}_{\mu\nu}$ in the clover term
is given by
\ben 
{\tilde F}_{\mu\nu}(n)&=&\frac{1}{4}\sum_{i=1}^{4}\frac{1}{2i}
\left({\tilde U}_i(n)-{\tilde U}_i^\dagger(n)\right), \\
{\tilde U}_1(n)&=&{\tilde U}_{n,\mu}{\tilde U}_{n+{\hat \mu},\nu}
         {\tilde U}^\dagger_{n+{\hat \nu},\mu}{\tilde U}^\dagger_{n,\nu}, \\
{\tilde U}_2(n)&=&{\tilde U}_{n,\nu}{\tilde U}^\dagger_{n-{\hat \mu}+{\hat \nu},\mu}
         {\tilde U}^\dagger_{n-{\hat \mu},\nu}{\tilde U}_{n-{\hat \mu},\mu}, \\
{\tilde U}_3(n)&=&{\tilde U}^\dagger_{n-{\hat \mu},\mu}{\tilde U}^\dagger_{n-{\hat \mu}-{\hat \nu},\nu}
         {\tilde U}_{n-{\hat \mu}-{\hat \nu},\mu}{\tilde U}_{n-{\hat \nu},\nu}, \\
{\tilde U}_4(n)&=&{\tilde U}^\dagger_{n-{\hat \nu},\nu}{\tilde U}_{n-{\hat \nu},\mu}
         {\tilde U}_{n+{\hat \mu}-{\hat \nu},\nu}{\tilde U}^\dagger_{n,\mu}.
\een
The improvement coefficient $\csw$ for $O(a)$ improvement was 
determined nonperturbatively in Ref.~\cite{Taniguchi:2012gew}.

\subsection{Configuration generation}
\label{subsec:lattice_param}

We have generated 2+1 flavor QCD gauge configurations at three different lattice spacings.
Their lattice sizes $V=L^3\times T$, $\beta$ values and lattice spacings $a$ are summarized in Table~\ref{tab:para_sim} together with other simulation parameters. Hereafter, we refer to four sets of configurations by the labels of L128, L64, L160, and L256. 
The hopping parameters for the light (degenerate up-down) and strange quarks $(\kappa_{\rm ud},\kappa_{\rm s})$ are carefully adjusted to yield the physical pion and kaon masses with the use of the cutoff $a^{-1}$ determined from the $\Xi$ mass.

  \begin{table*}
    \begin{center}
      \caption{Summary of simulation parameters. Lattice spacing $a$ is determined with convention A.\label{tab:para_sim}}
      \begin{tabular}{c|ccc}
        \hline\hline
        data set & L128/L64 & L160 & L256\\\hline
    $V=L^3\times T$ & $128^4/64^4$ & $160^4$ & $256^4$\\
        $\beta$  & 1.82 & 2.00 & 2.20\\
    $a$ [fm] & 0.08487(14) & 0.06314(09) & 0.04120(18)\\
    $c_{\rm SW}$ & 1.11 & 1.02 & 1.00 \\
    $\kappa_{\rm ud}$ & 0.126117 & 0.125814 & 0.1254872\\
    $\kappa_{\rm s}$ & 0.124902 & 0.124925 & 0.1249349\\
    $\tau$ & 1.0 & 1.0 & 3.0\\
    $\rho_1$ & 0.9997 & 0.9997 & 0.9997\\
    $\rho_2$ & 0.9940 & 0.9940 & 0.9940\\
    block size & (16,16,8,64)/(8,8,16,32) & (20,20,10,80) & (8,8,4,128) \\
    $(N_0,N_1,N_2,N_3,N_4)$ & (8,2,2,2,22)/(8,2,2,2,12)&(8,2,2,2,20) & (12,2,2,2,100) \\
    $N_{\rm RHMC}$ & 8/$-$ & 10 & 14 \\
    $[{\rm min},{\rm max}]$ & [0.000250,1.85]/$-$ & [0.000190,1.90] & [0.000083,1.57]\\
    $N_{\rm poly}$ & $-$/350 & $-$ & $-$\\
    $Z_A$ & 0.9650(68)(95) & 0.9783(21)(81) & 0.98497(156)(255) \\
    \hline\hline    
      \end{tabular}
    \end{center}
  \end{table*}

The degenerated up-down quarks are simulated with the domain-decomposed HMC (DDHMC) algorithm~\cite{Luscher:2003vf,Luscher:2004pav}. The ud quark determinant is separated into the UV and IR parts after the even-odd preconditioning. We also apply the twofold mass preconditioning~\cite{Hasenbusch:2001ne, Hasenbusch:2002ai} to the IR part by splitting it into $\tilde F_{\rm IR}$, $F_{\rm IR}^{\prime}$ and $F_{\rm IR}^{\prime\prime}$. This decomposition is controlled by two additional hopping parameters:
$\kappa^\prime_{\rm ud}=\rho_1\kappa_{\rm ud}$ and 
$\kappa^{\prime\prime}_{\rm ud}=\rho_1\rho_2\kappa_{\rm ud}$. ${\tilde F}_{\rm IR}$ is derived from the action preconditioned with
$\kappa^\prime_{\rm ud}$. The ratio of two preconditioners with 
$\kappa^\prime_{\rm ud}$ and $\kappa^{\prime\prime}_{\rm ud}$ gives
$F^{\prime}_{\rm IR}$.  
$F^{\prime\prime}_{\rm IR}$ is from the heaviest preconditioner 
with $\kappa^{\prime\prime}_{\rm ud}$. 
In the end the force terms  consist of  the gauge force $F_{\rm g}$, the UV force $F_{\rm UV}$ and the three IR forces $F_{\rm IR}^{\prime\prime}$, $F_{\rm IR}^{\prime}$ and $\tilde F_{\rm IR}$. The IR forces are obtained with the mixed precision nested BiCGStab method for the quark solver.
We adopt the multiple time scale integration scheme~\cite{Sexton:1992nu}  in the molecular dynamics (MD) steps. 
The associated step sizes are controlled by a set of integers $(N_0,N_1,N_2,N_3,N_4)$: $\delta\tau_{\rm g}=\tau/(N_0 N_1 N_2 N_3 N_4),$ 
$\delta\tau_{\rm UV}=\tau/(N_1 N_2 N_3 N_4),$ 
$\delta\tau^{\prime\prime}_{\rm IR}=\tau/(N_2 N_3 N_4),$ 
$\delta\tau^{\prime}_{\rm IR}=\tau/(N_3 N_4),$ 
$\delta{\tilde \tau}_{\rm IR}=\tau/N_4$ with $\tau$ the trajectory length.

The strange quark on the L64 lattice at $\beta=1.82$ is simulated using the UVPHMC algorithm with $N_{\rm poly}=350$~\cite{deForcrand:1996ck,Frezzotti:1997ym,Frezzotti:1998eu,JLQCD:2001ucs,PACS-CS:2006wru} where the action is UV-filtered~\cite{Alexandrou:1999ii,deForcrand:1998sv} after the even-odd preconditioning without domain decomposition. We set the step size as
$\delta\tau_{\rm s}=\delta\tau^{\prime\prime}_{\rm IR}$
according to our observation  $||F_{\rm s}||\approx ||F_{\rm IR}^{\prime\prime}||$. 
This algorithm is made exact
by correcting the polynomial approximation with the 
global Metropolis test~\cite{Borici:1995np,Borrelli:1996re,deForcrand:1996dq,Galli:1996mx} at the end of each trajectory.
For other lattice sizes we employ the RHMC algorithm~\cite{Clark:2006fx} to incorporate the strange quark.
The rational approximation order $N_{\rm RHMC}$, the approximation range [min,max], and the approximation tolerance are chosen at fixed values for the force term,  while they are automatically chosen to have the best double precision approximation after computing the min and max eigenvalues in the HMC Hamiltonian computation. The RHMC force time step $\delta \tau_{\rm s}$ is assigned at $\delta \tau^{\prime\prime}_{\rm IR}$.

\subsection{Mass reweighting}
\label{subsec:reweighting}

We investigate the quark mass dependence of the light hadron masses employing the mass reweighting technique.
The reweighting factor for the ud quark is evaluated with a stochastic method introducing 
 a set of independent Gaussian random noises $\eta_i$ $(i=1,\dots,N_\eta)$:
\begin{eqnarray}
 {\rm det}[W_{\rm ud}^2] &=&
\left[\lim_{N_\eta\rightarrow \infty}
 \frac{1}{N_\eta} \sum_{i=1}^{N_\eta} 
  {\rm e}^{-\vert W_{\rm ud}^{-1}\eta_i\vert^2 +\vert \eta_i\vert^2}\right],
 \label{eq:rwfactor_ud}\\
 W_{\rm ud} &=& \frac{D(\kappa_{\rm ud}^*)}{D(\kappa_{\rm ud})},
\label{eq:w_ud}
\end{eqnarray}
where $D(\kappa_{\rm ud}^*)$ is the Wilson-Dirac matrix with a target hopping parameter $\kappa_{\rm ud}^*$. For the strange quark we employ the square root trick, 
\begin{eqnarray}
 {\rm det}[W_{\rm s}] &=&
\left[\lim_{N_\eta\rightarrow \infty}
 \frac{1}{N_\eta} \sum_{i=1}^{N_\eta} 
  {\rm e}^{-\vert W_{\rm s}^{-1}\eta_i\vert^2 +\vert \eta_i\vert^2}\right]^{\frac{1}{2}},
 \label{eq:rwfactor_s}\\
 W_{\rm s} &=& \frac{D(\kappa_{\rm s}^*)}{D(\kappa_{\rm s})}.
\label{eq:w_s}
\end{eqnarray}
To reduce the fluctuation in the stochastic evaluation (\ref{eq:rwfactor_ud}) and (\ref{eq:rwfactor_s}), we employ the determinant breakup technique~\cite{PACS-CS:2009sof, Hasenfratz:2008fg, RBC:2010qam}, in which 
the interval between $\kappa_{q}$ and $\kappa_{q}^*$ is divided into small sub-intervals with $\Delta \kappa_{\rm ud}=0.0000005$ and $\Delta \kappa_{\rm s}=0.0000020$. We choose $N_\eta=12$ following Ref.~\cite{Ishikawa:2018jee}.

\subsection{Measurement of hadronic observables}
\label{subsec:measurement}
\subsubsection{Hadron operators}
\label{subsubsec:operators}

We measure the meson and baryon correlators employing appropriate operators. The PS (pseudoscalar) and vector meson operators are expressed as 
\ben
M_\Gamma^{fg}(x)={\bar q}_f(x)\Gamma q_g(x),
\een
where $f$ and $g$ ($f,g={\rm ud},{\rm s})$ label valence quark flavors. We take $\Gamma=\gamma_5$ for the PS meson and $\Gamma=\gamma_\mu$ $(\mu=1,2,3)$ for the vector meson.

The octet baryon operators are given by 
\ben
{\cal O}^{fgh}_\alpha(x)=\epsilon^{abc}((q_f^a(x))^T C\gamma_5 q_g^b(x))
q_{h\alpha}^c(x),
\een 
where $a,b,c$ are color indices, $C=\gamma_4\gamma_2$ is the charge conjugation matrix and $\alpha=1,2$ label the $z$-component of the spin 1/2 (in Dirac representation $\alpha$ consists of four spinor components). Referring to Ref.~\cite{CP-PACS:2002unz}, the $\Sigma$- and $\Lambda$-like octet baryons are distinguished by the flavor structures:
\ben
\Sigma{\rm -like}\;\; &:& \;\; -\frac{{\cal O}^{[fh]g}+{\cal O}^{[gh]f}}{\sqrt{2}},\\
\Lambda{\rm -like}\;\; &:& \;\; \frac{{\cal O}^{[fh]g}-{\cal O}^{[gh]f}-2{\cal O}^{[fg]h}}
{\sqrt{6}},
\een
where ${\cal O}^{[fg]h}={\cal O}^{fgh}-{\cal O}^{gfh}$. The decuplet baryon operators for the four $z$-components of the spin 3/2 is defined as
\ben
D^{fgh}_{3/2}(x)&=&-\epsilon^{abc}((q_f^a(x))^T C\Gamma_+
q_g^b(x))q_{h1}^c(x),\label{eq:dec1}\\
D^{fgh}_{1/2}(x)&=&
\epsilon^{abc}[((q_f^a(x))^T C\Gamma_0q_g^b(x))q_{h1}^c(x)\nn\\
&&-((q_f^a(x))^T C\Gamma_+q_g^b(x))q_{h2}^c(x)]/\sqrt{3},\label{eq:dec2}\\
D^{fgh}_{-1/2}(x)&=&
\epsilon^{abc}[((q_f^a(x))^T C\Gamma_0q_g^b(x))q_{h2}^c(x)\nn\\
&&+((q_f^a(x))^T C\Gamma_-q_g^b(x))q_{h1}^c(x)]/\sqrt{3},\label{eq:dec3}\\
D^{fgh}_{-3/2}(x)&=&\epsilon^{abc}((q_f^a(x))^T C\Gamma_-
q_g^b(x))q_{h2}^c(x),\label{eq:dec4}
\een
where $\Gamma_{\pm}=(\gamma_1\mp i\gamma_2)/2$, $\Gamma_0=\gamma_3$ and the flavor structures should be symmetrized. 
The correlator for the decuplet baryon is obtained by the summation over four $z$-components as, 
\begin{eqnarray}
 && \sum_{\vec x}
 \sum_{S=\pm3/2,\pm1/2}
 \left[
 \langle D_{S}^{fgh}(x) \overline{D}_{S}^{fgh}(0)\rangle
 \right]
\end{eqnarray}
with operators in Eq.~(\ref{eq:dec1})--(\ref{eq:dec4}). $\sum_{\vec x}$ represents the summation over spatial lattice site ${\vec x}=(x_1,x_2,x_3)$.

{\crd It can be easily found that the correlators based on the decuplet baryon operators $D_{\pm\frac{3}{2},\pm\frac{1}{2}}^{fgh}(x)$ are equivalent to those constructed using the vector-spinor interpolating operators $B_{\mu\beta}(x)\ (\beta=1,2,3,4)$ projected to the spin-parity of $j^P=\frac{3}{2}^+$. For example, 
the equivalence of the $\Omega$ baryon correlator is written as
\begin{eqnarray}
 && \sum_{\vec x}
 \sum_{S=\pm3/2,\pm1/2}
 \left[
 \langle D_{S}^{sss}(x) \overline{D}_{S}^{sss}(0)\rangle
 \right]\nonumber\\
&& =
\sum_{\vec x}
\sum_{\mu,\nu=1,2,3,4}
{\rm tr} \left[P^{\frac{3}{2}^+}_{\mu\nu}
\langle
 B_{\nu}^{\Omega}(x) \overline B_{\mu}^{\Omega}(0)
\rangle\right]
\label{equiv_corr}
\end{eqnarray}
with 
\begin{eqnarray}
 B_{\mu\beta}^{\Omega}(x) &=& \epsilon^{abc}((s^a(x))^T C\gamma_\mu
s^b(x))s_\beta^c(x) / \sqrt{2},
\end{eqnarray}
where 
${\rm tr}$ denotes the trace over spinor indices and the spin-parity projection operator $P_{\mu\nu}^{\frac{3}{2}^+}$ is defined by $P_{\mu\nu}^{\frac{3}{2}^+}=\left(\frac{1+\gamma_4}{2}\right)\left(\delta_{\mu\nu}-\frac{1}{3}\gamma_\mu\gamma_{\nu}\right)$ for $\mu,\nu=1,2,3$ and $P_{44}^{\frac{3}{2}^+}=0$ for the zero spatial momentum state.
The vector-spinor operators for other decuplet baryons $\Delta, \Sigma^*$, $\Xi^*$ are defined as
\begin{eqnarray}
 B_{\mu\beta}^{\Delta}(x) &=&
 \epsilon^{abc}((u^{a}(x))^TC\gamma_{\mu}u^b(x))u_{\beta}^c(x) / \sqrt{2},
 \\
 B_{\mu\beta}^{\Sigma^*}(x) &=&
 \epsilon^{abc}
 [
  2((u^{a}(x))^TC\gamma_{\mu}s^b(x))u_{\beta}^c(x)\nonumber \\&&
  +((u^{a}(x))^TC\gamma_{\mu}u^b(x))s_{\beta}^c(x)
 ] / \sqrt{6},
 \\
 B_{\mu\beta}^{\Xi^*}(x) &=&
 \epsilon^{abc}
 [
  2((u^{a}(x))^TC\gamma_{\mu}s^b(x))s_{\beta}^c(x)\nonumber \\&&
  +((s^{a}(x))^TC\gamma_{\mu}s^b(x))u_{\beta}^c(x)
 ] / \sqrt{6}.
\end{eqnarray}
The equivalence as shown with Eq.~(\ref{equiv_corr}) also holds for $\Delta, \Sigma^*$ and $\Xi^*$. 
}
  
\subsubsection{PS meson decay constants and quark masses}
\label{subsubsec:quarkmass}

We measure the PS meson decay constants and the quark masses using the PS meson operator and axial vector current with zero spatial momentum projection:
\ben
P(t)&=&\sum_{\vec x}{\bar q}_f({\vec x},t)\gamma_5 q_g({\vec x},t), \\
A_{\mu}(t)&=&\sum_{\vec x}{\bar q}_f({\vec x},t)\gamma_\mu\gamma_5 q_g({\vec x},t)
\een 
with $f,g={\rm ud},{s}$. 


The PS meson masses are extracted from the correlation function of $\la P(t)P^\dagger(0)\ra$, whose $t$ dependence is given by
\ben
\la P(t)P^\dagger(0)\ra&=&L^3 C_{PP}W(m_{\rm PS}T) \label{eq:pp}\\
&&\times \left[\exp(-m_{\rm PS}t)+\exp(-m_{\rm PS}(T-t))\right], \nn
\een
where $W(m_{\rm PS}T)$ with $T=64$ and 128 denotes the contribution from the PS meson propagation wrapping around the lattice in the time direction:
\ben
W(m_{\rm PS}T)&=&1+\exp(-m_{\rm PS}T)+\exp(-2m_{\rm PS}T)+\cdots \nn\\
&=&\frac{1}{1-\exp(-m_{\rm PS}T)}.
\een

The PS meson decay constant is defined by
\ben
&Z_A\left(1+b_A\frac{Z_A(m_f^{\rm AWI}+m_g^{\rm AWI})}{2Z_PZ_m}\right)
\sqrt{2\kappa_f}\sqrt{2\kappa_g} \label{eq:zm}
\left\vert\la 0|A_4|{\rm PS}\ra\right\vert \nn\\
&=f_{\rm PS}m_{\rm PS}
\een
with $|{\rm PS}\ra$ the PS meson state at rest and $Z_A$, $Z_P$ and $Z_m$ denote the renormalization factors for the axial vector current, the pseudoscalar density and the quark mass, respectively. $b_A$ is the $O(ma)$ improvement coefficient of the axial vector current. 
We evaluate $f_{\rm PS}$ with the following combination:
\ben
f_{\rm PS}&=&Z_A\left(1+b_A\frac{Z_A(m_f^{\rm AWI}+m_g^{\rm AWI})}{2Z_PZ_m}\right) \nn \\
&&\times \sqrt{2\kappa_f}\sqrt{2\kappa_g}
\frac{\sqrt{2}\left\vert C_{A P}\right\vert}
{\sqrt{ m_{\rm PS}  \left\vert C_{PP}\right\vert}},
\een
where we extract $m_{\rm PS}$, $C_{A P}$, $C_{PP}$ from a simultaneous fit of Eq.~(\ref{eq:pp}) and 
\ben
 \la A_4(t)P^\dagger(0)\ra&=&L^3 C_{AP}W(m_{\rm PS}T) 
 \label{eq:ap}\\
&&\times \left[\exp(-m_{\rm PS}t)-\exp(-m_{\rm PS}(T-t))\right]\nn
\een
with a common fit range.

For the $O(ma)$ improvement on the L128 and L64 lattices we take $b_A=1$ and $Z_m=0.9950(111)(89)$ and $Z_P=1.01317(43)$ at $\beta=1.82$, which are nonperturbatively determined by the Schr{\"o}dinger functional scheme~\cite{Ishikawa:2015fzw}.
Since the values of $Z_m$ and $Z_P$ at $\beta=1.82$ differ from one by at most around 1\%, we set $Z_m=Z_P=1$ on the L160 and L256 lattices. This approximation gives negligible effects on the magnitude of the $O(ma)$ correction because of small values of $m_f$ and $m_g$ in lattice units. 
The values of $Z_A$, which are listed in Table~\ref{tab:para_sim}, are also determined by the Schr{\"o}dinger functional scheme~\cite{Ishikawa:2015fzw}.
Since the value of $c_A$, which is the $O(a)$ improvement factor, is consistent with zero within the statistical error~\cite{Taniguchi:2012gew}, we assume the leading cutoff effect on $f_{\rm PS}$ is $O(a^2)$. 

We define the bare AWI quark mass as
\ben
m^{\rm AWI}_f+ m^{\rm AWI}_g=\frac{\langle 0 |\nabla_4
  A_4|{\rm PS} \rangle}{\langle 0| P | {\rm PS}\rangle},
\label{mAWTI}
\een
which is evaluated with
\ben
m^{\rm AWI}_f+m^{\rm AWI}_g&=&m_{\rm PS}
\left\vert \frac{C_{AP}}{C_{PP}}\right\vert.
\een
We extract $m_{\rm PS}$, $C_{A P}$, $C_{PP}$ from a simultaneous fit of Eqs.~(\ref{eq:pp}) and  (\ref{eq:ap}) with a common fit range.

\subsubsection{Crystal source method}
\label{subsubsec:crystal}

In order to reduce the statistical errors of hadron propagators, we employ the crystal source method~\cite{PACS:2019ofv}. 
First, we partition the three-dimensional source space into cubic cells, each having an edge length of $b_{\rm cell} = L/n_{\rm div} \in \mathbb{Q}$ with $n_{\rm div} \in \mathbb{N}$ and a crystal structure like simple cubic (SC), body-centered cubic (BCC), or face-centered cubic (FCC).
Next, we pack equal-sized balls in a non-overlapping manner, with each ball’s center positioned at each lattice site of the cells. For the SC, BCC, and FCC structures, the radius, and the number of the balls $(r_{\rm ball}, n_{\rm ball})$ are given by $(b_{\rm cell}/2, n^3_{\rm div})$, $(\sqrt{3}b_{\rm cell}/4, 2n^3_{\rm div})$, and $(\sqrt{2}b_{\rm cell}/4, 4n^3_{\rm div})$, respectively. The crystal source $\Psi(\vec{x})$ with the Coulomb gauge fixing is determined by applying  $\mathbb{Z}_3$ noise $\eta(\vec{y})$ to the balls as 
\ben
\Psi(\vec{x}) = \sum_{\vec{y}\in\Lambda_{\rm cell}} \psi(\vec{x},\vec{y}) \eta(\vec{y}).
\een
Here, \(\Lambda _{\rm{cell}}\) denotes the set of lattice sites within the SC, BCC, or FCC cells. Additionally, the smeared wave function $\psi(\vec{x},\vec{y})$ possesses a finite support restricted to the interior, defined as $\psi(\vec{x},\vec{y})=A_q\exp(-B_q|\vec{x}-\vec{y}|)$ $(q={\rm ud, s})$ under the condition \(\vert{}\vec{x} - \vec{y}\vert{} < r_{\rm{ball}}\). Quark contractions for hadron operators are practically limited to each ball as a consequence of $\mathbb{Z}_3$ noise. The number of $\mathbb{Z}_3$ noises are listed in Table~\ref{tab:para_src} together with the number of source points.

The vector meson and baryon propagators are computed using the crystal source method, with the parameters listed in Table~\ref{tab:para_src}. 
The solver algorithm employed is the mixed-precision BiCGStab method, with a tolerance of $|Dx-b|/|b| < 10^{-8}$ for L64, L128, and L160, and $|Dx-b|/|b| < 10^{-10}$ for L256.
In contrast, the PS meson propagators are obtained using the wall source method without gauge fixing~\cite{Kuramashi:1993ka,Fukugita:1994ve}. The solver algorithm and the tolerance of $|Dx-b|/|b|$ are the same as those for vector meson and baryon propagators. 

We calculate the hadronic observables at every 10, 10, 5, and 4 trajectories for L128, L64, L160, and L256 lattices.
The total number of the measurements are 40960, 25600, 19200 and 640 for L128, L64, L160 and L256 using 20, 800, 20 and 20 configurations, respectively. 
The statistical errors are estimated with the jackknife method. After investigating the binsize dependence, we have chosen one configuration for the jackknife binsize on L128, L160 and L256 lattices and five ones on L64 lattice.

  \begin{table*}
    \begin{center}
      \caption{Crystal source parameters. Total measurements is the number of source points multiplied by the number of $\mathbb{Z}_3$ noises. \label{tab:para_src}}
      \begin{tabular}{c|cccc}
        \hline\hline
        data set & L128 & L64 & L160 & L256\\\hline
    crystal structure & BCC & SC & FCC & BCC \\
    $n_{\rm div}$ & 3 & 2 & 2 & 3 \\
    $(b_{\rm cell}, r_{\rm ball}, n_{\rm ball})$ & 
     $(128/3, 32/\sqrt{3}, 54)$ & $(32, 16, 8)$ 
    & $(80, 20\sqrt{2}, 32)$ & $(256/3, 64/\sqrt{3}, 54)$\\ 
    $(A_{\rm ud}, B_{\rm ud})$ &(1.0, 0.03) &(1.0, 0.00) &(1.0, 0.08)&(1.0, 0.00)\\
    $(A_{\rm s}, B_{\rm s})$  &(1.0, 0.30) &(1.0, 0.22)&(1.0, 0.14)&(1.0, 0.08)\\
    \# of source points & 10240 & 25600 & 6400 & 640\\ 
    \# of $\mathbb{Z}_3$ noises  & 4 & 1 & 3 & 1\\ 
    Total measurements  & 40960 & 25600 & 19200 & 640\\ 
    \hline\hline    
      \end{tabular}
    \end{center}
  \end{table*}

\subsubsection{pseudoexperimental values in isosymmetric QCD}
\label{subsubsec:isoqcd}

Since our simulation is performed in the isosymmetric QCD (isoQCD) with the degenerated up-down quarks neglecting the QED effects, there are some ambiguities in the definition of pseudoexperimental values to be compared with the lattice results. In Table~\ref{tab:pexp_isoqcd} we 
present two conventions for pseudoexperimental values in isoQCD, which are obtained with different combinations of experimental values in the real world. In Secs.~\ref{sec:results_b182}, \ref{sec:results_b200b220}, \ref{sec:topologicalcharge} and \ref{sec:continuum} we first analyze the simulation data with the convention A that has been historically employed in PACS Collaboration and its predecessors~\cite{PACS-CS:2009sof,PACS-CS:2008bkb,CP-PACS:2001vqx,CP-PACS:2002unz}. After that  we reanalyzed the data with the convention B, which is based on the FLAG definition of the isospin symmetric point called ``Edinburgh consensus''~\cite{nfzp-p5dn}, and compare both results in Sec.~\ref{sec:isoqcd_ed}.

 \begin{table*}
    \begin{center}
      \caption{Pseudoexperimental values in isosymmetric QCD (isoQCD), which are defined by combinations of experimental values in the real world. We use $m_\pi$, $m_K$ and $m_\Xi$ as the physical inputs. \label{tab:pexp_isoqcd}}
      \begin{tabular}{c|cccc}
        \hline\hline
         & \multicolumn{2}{c}{convention A} & \multicolumn{2}{c}{convention B} \\ 
         & definition & [MeV] & definition & [MeV] \\\hline
  $m_\pi$ & $m_{\pi^+}$ & 135.0 & $m_{\pi^+}$ & 135.0  \\
  $m_K$ & $m_{K^0}$ & 497.6 & $\frac{1}{2}\{m_{K^0}+m_{K^+}\}$  & 494.6  \\
  $m_\rho$  & $m_{\rho^0}$ & 775.3 & $m_{\rho^0}$  & 775.3  \\
  $m_{K^*}$ & $m_{K^{*0}}$ & 895.6 & $\frac{1}{2}\{m_{K^{*0}}+m_{K^{*+}}\}$  & 893.6  \\
  $m_\phi$ & $m_\phi$ & 1019.5 & $m_\phi$  & 1019.5  \\
  $m_N$ & $m_n$ & 939.6 & $\frac{1}{2}\{m_p+m_n\}$  & 938.9  \\
  $m_\Lambda$ & $m_{\Lambda^0}$ & 1115.7 & $m_{\Lambda^0}$  & 1115.7  \\
  $m_\Sigma$ & $m_{\Sigma^0}$ & 1192.6 & $\frac{1}{3}\{m_{\Sigma^+}+m_{\Sigma^0}+m_{\Sigma^-}\}$  & 1193.2  \\
  $m_\Xi$ & $m_{\Xi^0}$ & 1314.8 & $\frac{1}{2}\{m_{\Xi^0}+m_{\Xi^-}\}$  & 1318.3  \\
  $m_\Delta$ & $m_\Delta$ & 1232.0 & $m_\Delta$  & 1232.0  \\
  $m_{\Sigma^*}$ & $m_{\Sigma^{*0}}$ & 1383.7 & $\frac{1}{3}\{m_{\Sigma^{*+}}+m_{\Sigma^{*0}}+m_{\Sigma^{*-}}\}$  & 1384.6  \\
  $m_{\Xi^*}$ & $m_{\Xi^{*0}}$ & 1531.8 & $\frac{1}{2}\{m_{\Xi^{*0}}+m_{\Xi^{*-}}\}$  & 1533.4  \\
  $m_\Omega$ & $m_{\Omega}$ & 1672.5 & $m_{\Omega}$  & 1672.5  \\
  $f_{\pi}$ & $f_{\pi^+}$ & 130.5 & $f_{\pi^+}$  & 130.5  \\
   \hline\hline
      \end{tabular}
    \end{center}
  \end{table*}

\section{Results at $\beta=1.82$}
\label{sec:results_b182}
\subsection{Finite size effect on hadron masses}
\label{subsec:fse_b182}
\subsubsection{PS meson sector}
\label{subsubsec:fse_ps}

\begin{figure}[h]
\begin{center}
\includegraphics[width=75mm,angle=0]{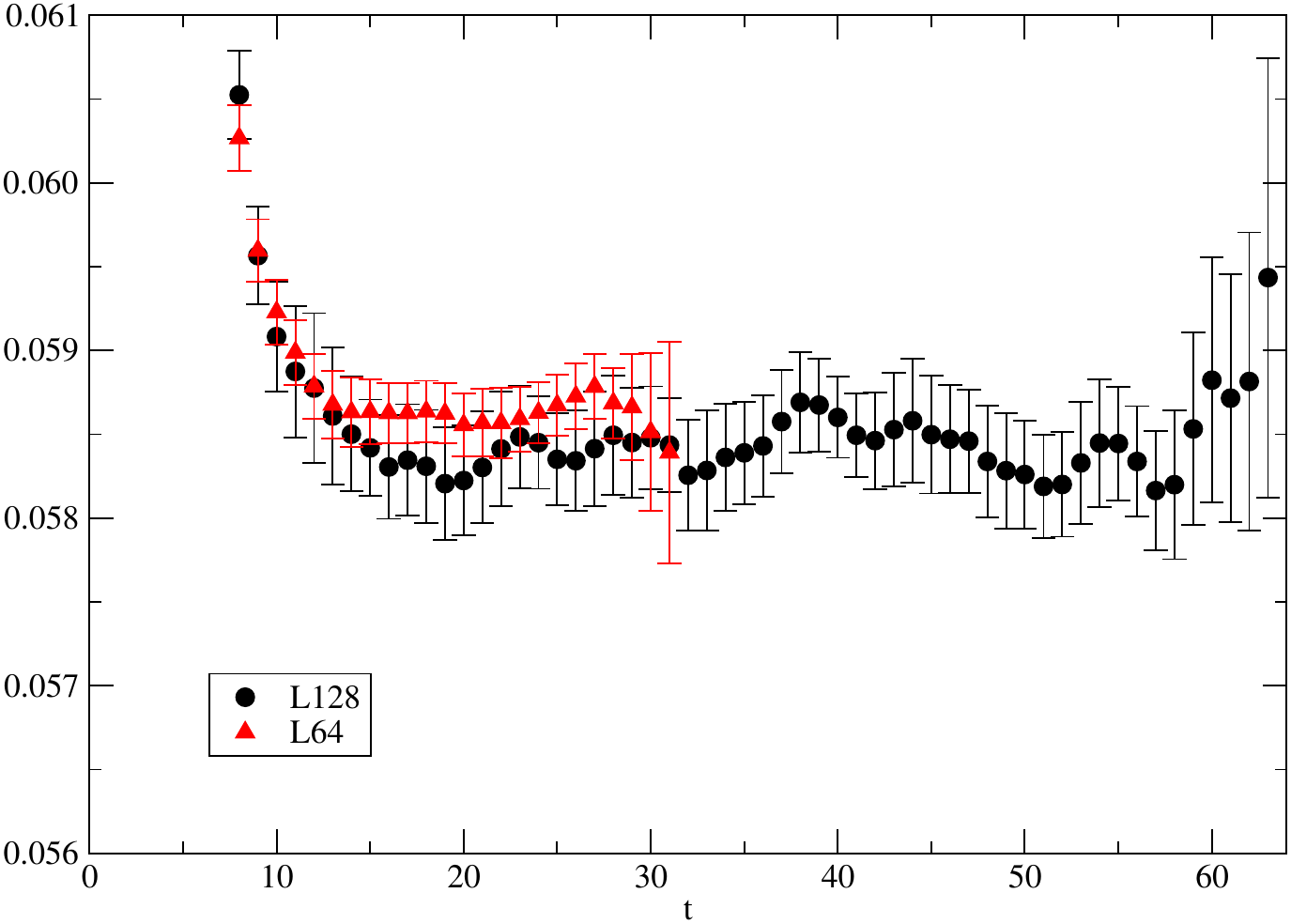}\\
\includegraphics[width=75mm,angle=0]{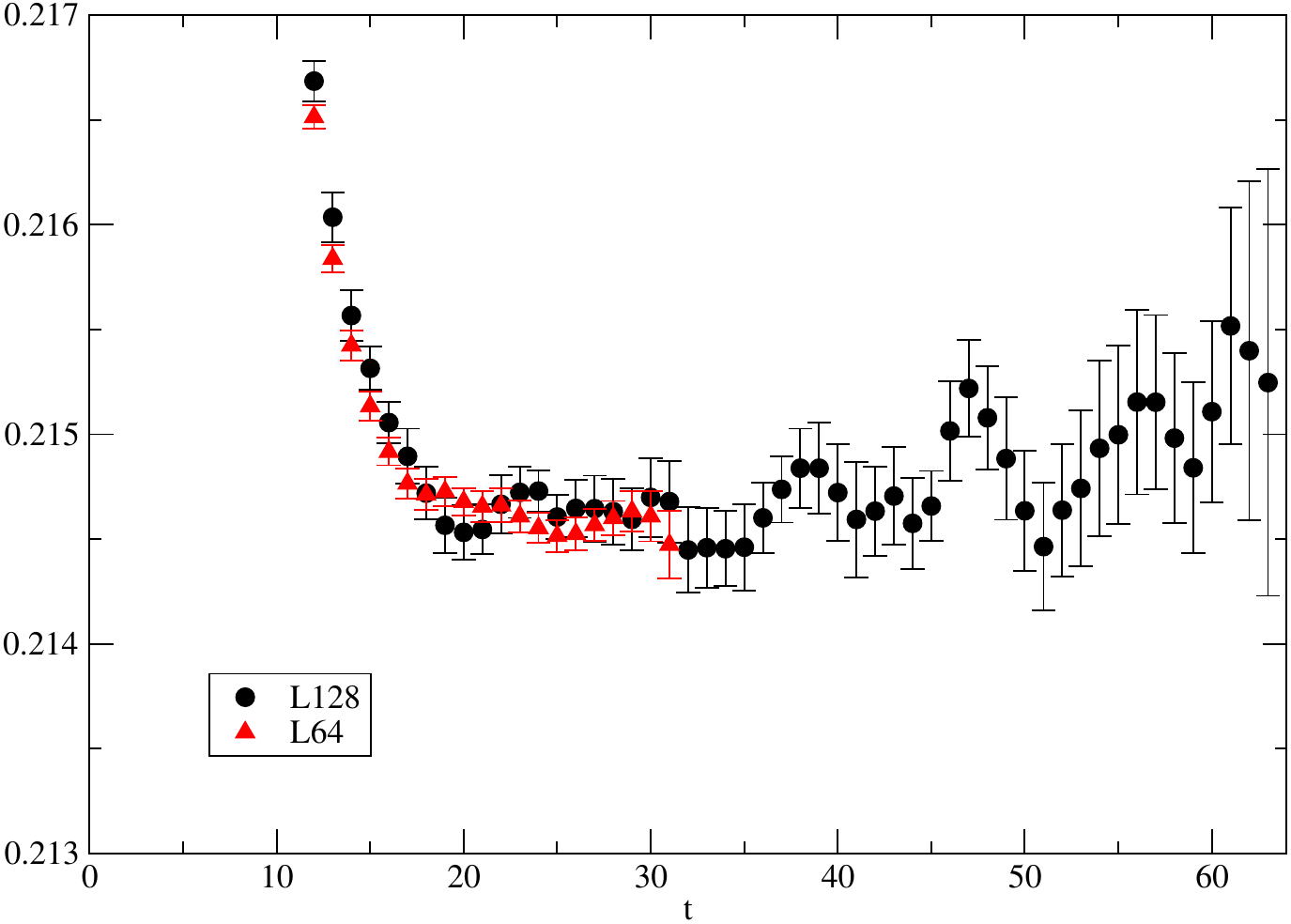}
\end{center}
\vspace{-5mm}
\caption{Comparison of the effective masses for $\pi$ (top) and $K$ (bottom) mesons on the L128 and L64 lattices at $\beta=1.82$ with $(\kappa_{\rm ud},\kappa_{\rm s})=(0.126117,0.124902)$.}
\label{fig:fse_effmass}
\end{figure}

\begin{figure}[h]
\begin{center}
\includegraphics[width=75mm,angle=0]{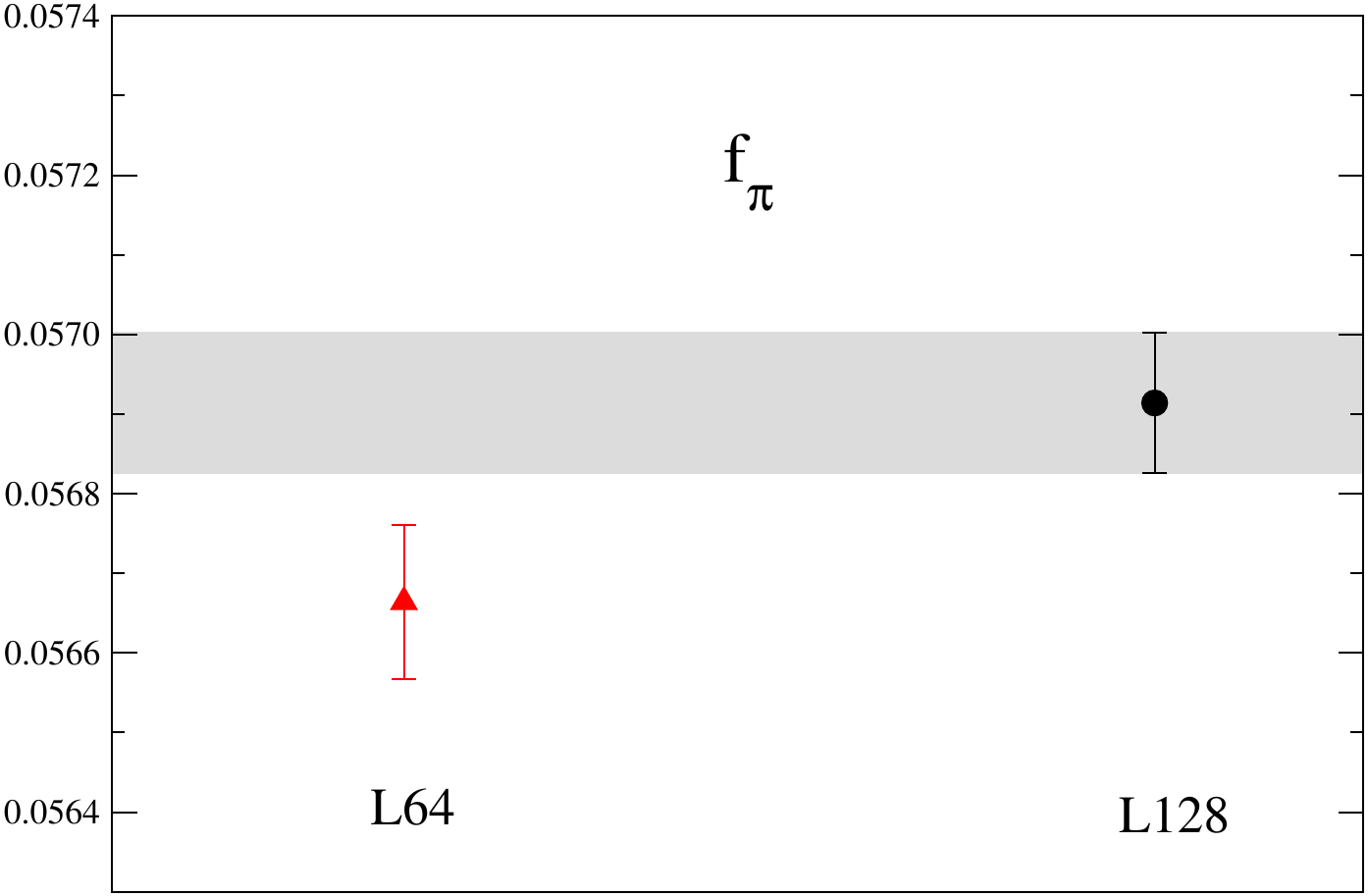}\\
\includegraphics[width=75mm,angle=0]{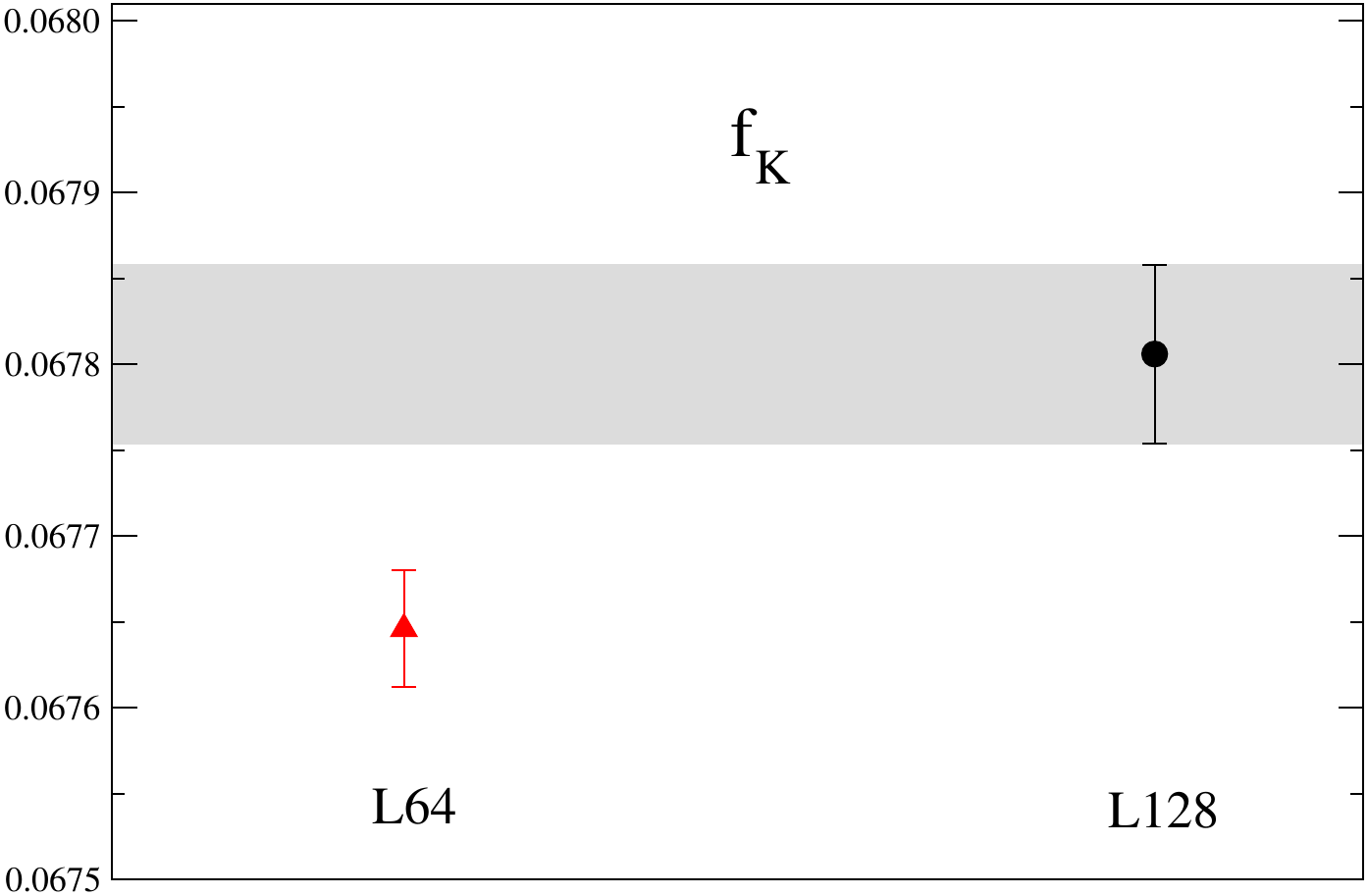}\\
\includegraphics[width=75mm,angle=0]{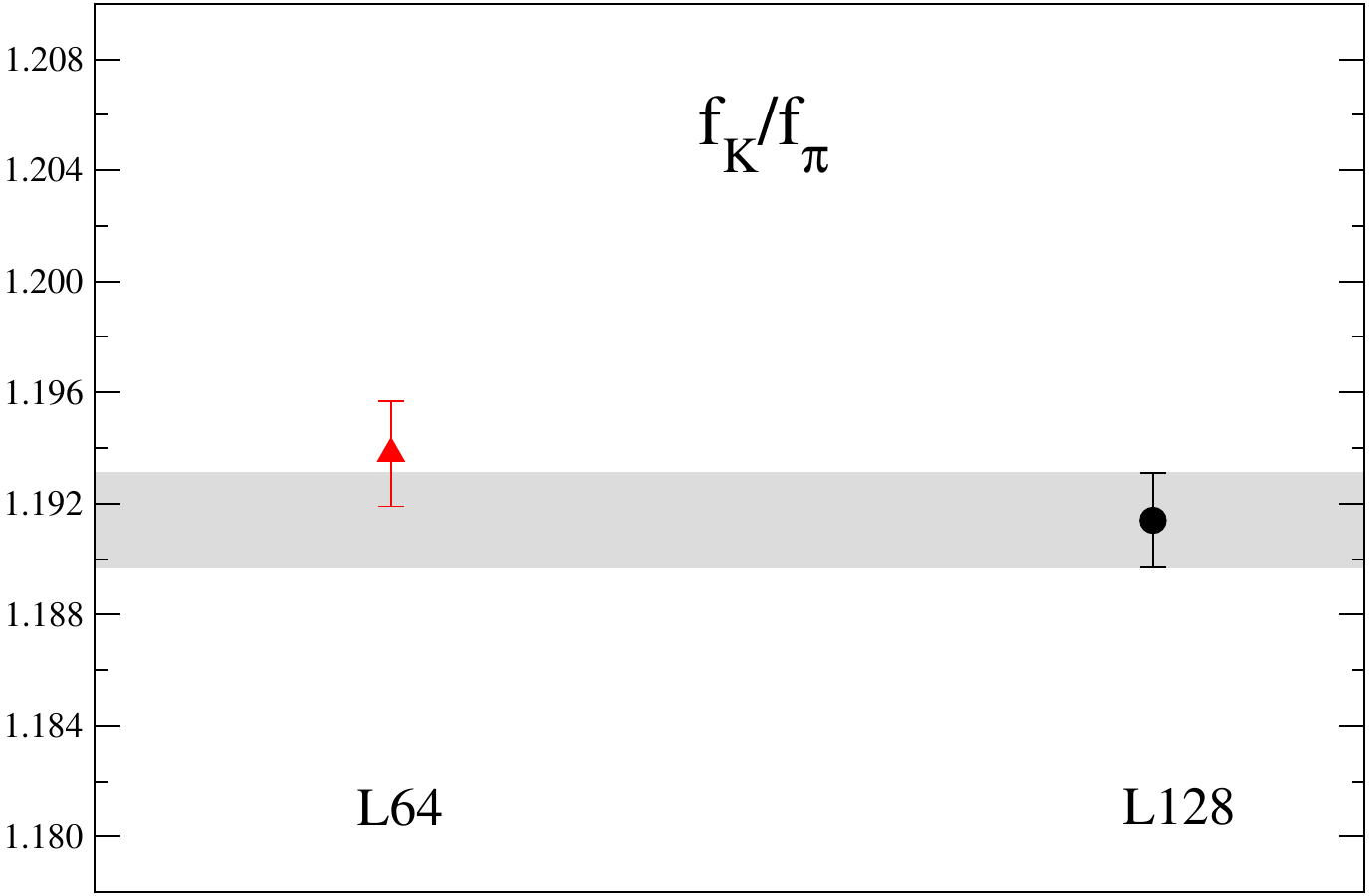}
\end{center}
\caption{Comparison of $f_\pi$ (top), $f_K$ (middle), and $f_K/f_{\pi}$ (bottom) on the L128 and L64 lattices at $\beta=1.82$ with $(\kappa_{\rm ud},\kappa_{\rm s})=(0.126117,0.124902)$. Shaded band denotes the error band of the L128 result.}
\label{fig:fse_decayconst}
\end{figure}

We first investigate the finite size effect on the PS meson sector using L128 and L64 lattices at $(\kappa_{\rm ud},\kappa_{\rm s})=(0.126117,0.124902)$. Figure~\ref{fig:fse_effmass} shows the effective masses for the $\pi$ and $K$ mesons. In both cases the results on the L128 and L64 lattices show consistency within the error bars indicating negligible finite size effects. In Table~\ref{tab:hmass} we summarize the fit results for the PS meson masses together with their fit ranges. 
We also list the AWI quark masses $m_{\rm ud,s}$, which show consistency between the L128 and L64 lattices as expected. For the decay constants, we plot the results on the L128 and L64 lattices in Fig.~\ref{fig:fse_decayconst}. Their numerical values are presented in Table~\ref{tab:hmass}. The small finite size effect is observed in both the $\pi$ and $K$ channels. It should be noted that the smaller  values of $f_{\pi,K}$ on smaller spatial lattice volume is an expected feature from the ChPT prediction. 
The magnitude of deviation is $-$0.44(32)\% for $f_\pi$ and $-$0.24(13)\% for $f_K$.

  \begin{table*}
    \begin{center}
      \caption{Fit results for hadron masses, quark masses, and PS meson decay constants at the simulation point together with the fit range of [$t_{\rm min}$,$t_{\rm max}]$.\label{tab:hmass}}
      \begin{tabular}{c|cccc}
        \hline\hline
        data set & L128 & L64 & L160 & L256\\\hline
  $m_\pi$ & 0.05843(27) & 0.05862(18) & 0.04426(06)  & 0.02965(02)  \\
        & [17,60] & [17,30] & [30,76] & [50,120]\\
  $m_K$ & 0.21468(08) & 0.21459(06) & 0.16200(07)  & 0.10728(04)  \\
       & [20,60] & [20,30] & [35,76] & [50,120]\\
  $m_N$ & 0.4090(24) & 0.4043(73) & 0.3028(11) & 0.2020(40) \\
       & [12,18] & [12,18] & [16,24] & [25,37] \\
  $m_\Lambda$ & 0.4808(05) & 0.4783(25) & 0.3590(05) & 0.2357(10)  \\
       & [12,18] & [12,18] & [16,24] & [25,37] \\
  $m_\Sigma$ & 0.5150(06) & 0.5142(21) & 0.3850(04)  & 0.2528(11)  \\
       & [12,18] & [12,18] & [16,24] & [25,37] \\
  $m_\Xi$ & 0.5664(04) & 0.5651(06) & 0.4246(03)  & 0.2793(05) \\
       & [15,25] & [15,25] & [20,34] & [31,51] \\
  $m_{\rm ud}$ & 0.001366(14) & 0.001376(09) & 0.0010511(38) & 0.0006870(06) \\
        & [17,60] & [17,30] & [30,76] & [50,120]\\
  $m_{\rm s}$ & 0.037935(15) & 0.037923(08) & 0.028757(03) & 0.018262(02) \\
         & [20,60] & [20,30] & [35,76] & [50,120]\\
  $m_{\rm s}/m_{\rm ud}$ & 27.77(29) & 27.56(18) & 27.360(98)  & 26.581(21)  \\
  $f_{\pi}$ & 0.056913(88)(666) & 0.056664(97)(663) & 0.042042(30)(353)  & 0.027250(19)(81)  \\
       & [17,60] & [17,30] & [30,76] & [50,120]\\
  $f_K$ & 0.067806(52)(793) & 0.067646(34)(791) & 0.050311(51)(423) & 0.032809(22)(98)  \\
       & [20,60] & [20,30] & [35,76] & [50,120]\\
  $f_{K}/f_{\pi}$ & 1.1914(17)(02) & 1.1938(19)(02) & 1.1967(12)(01) & 1.20401(085)(003) \\
   \hline\hline
      \end{tabular}
    \end{center}
  \end{table*}

\subsubsection{Comparison with ChPT prediction}
\label{subsubsec:fse_chpt}

\begin{figure}[th]
\vspace{3mm}
\begin{center}
\includegraphics[width=75mm,angle=0]{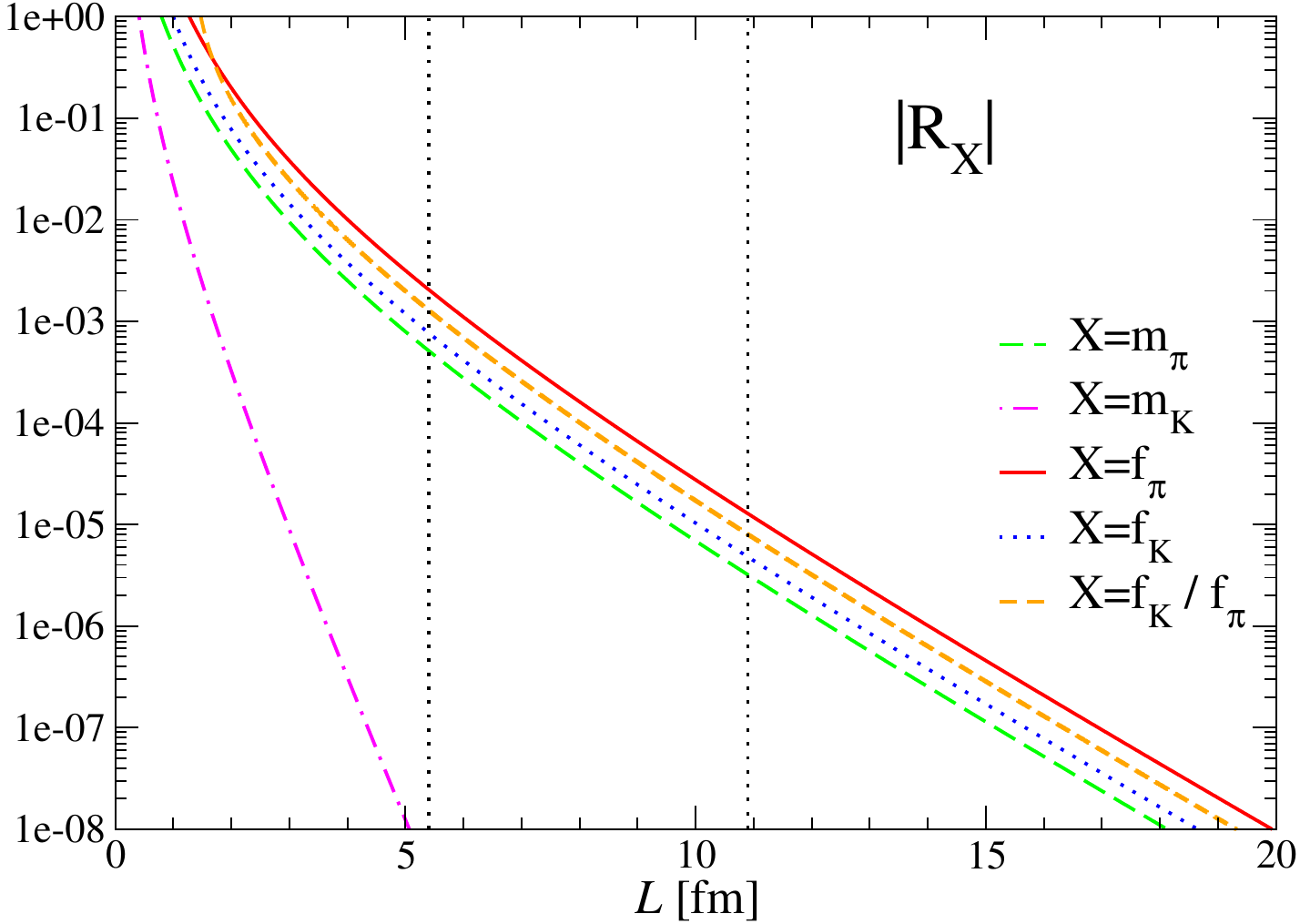}
\end{center}
\vspace{-.5cm}
\caption{$L$ dependence of $\vert R_X\vert $ $(R_{m_{\rm PS}}>0$ and $R_{f_{\rm PS}}<0)$ for $X=m_\pi,m_K,f_\pi,f_K,f_K/f_\pi $ at the physical point. Dotted vertical lines indicate $L=5.4$ fm and 10.9 fm.}
\label{fig:fse_chpt}
\end{figure}

It is instructive to compare our results with the ChPT predictions. The SU(3) ChPT gives the following
one-loop expressions for the finite size effects defined by $R_X=(X(L)-X(\infty))/X(\infty)$ for $X=m_\pi,m_K,f_\pi,f_K$~\cite{Colangelo:2005gd}:
\ben
R_{m_\pi}&=&\frac{1}{4}\xi_\pi{\tilde g}_1(\lambda_\pi)-\frac{1}{12}\xi_\eta{\tilde g}_1(\lambda_\eta),\\
R_{m_K}&=&\frac{1}{6}\xi_\eta{\tilde g}_1(\lambda_\eta),\\
R_{f_\pi}&=&-\xi_\pi{\tilde g}_1(\lambda_\pi)-\frac{1}{2}\xi_K{\tilde g}_1(\lambda_K),\\
R_{f_K}&=&-\frac{3}{8}\xi_\pi{\tilde g}_1(\lambda_\pi)-\frac{3}{4}\xi_K{\tilde g}_1(\lambda_K)-\frac{3}{8}\xi_\eta{\tilde g}_1(\lambda_\eta)
\een
with
\ben
\xi_{\rm PS}&\equiv& \frac{2m_{\rm PS}^2}{(4\pi f_\pi)^2},\\
\lambda_{\rm PS}&\equiv& m_{\rm PS}L,\\
{\tilde g}_1(x)&\equiv&\sum_{n=1}^{\infty}\frac{4m(n)}{\sqrt{n} x}K_1(\sqrt{n} x),
\een
where $K_1$ is the modified Bessel function of the second kind and $m(n)$ denotes the multiplicity of the partition $n=n_x^2+n_y^2+n_z^2$. Figure~\ref{fig:fse_chpt} shows the $L$ dependence of $\vert R_X\vert $ $(X=m_\pi,m_K,f_\pi,f_K,f_K/f_\pi)$ at the physical point defined by the convention A. The values of $R_X $ at $L=10.9$ fm and $5.4$ fm, which are the physical extent of L128 and L64 lattices, respectively, are summarized in Table~\ref{tab:fse_chpt}. 
We find that the finite size effect may be detectable on the L64 lattice, while it should be completely negligible on the L128 lattice. The expected magnitude for the L64 case is 0.20\% for $R_{f_\pi}$, 0.07\% for $R_{f_K}$ 0.12\% for $R_{f_K/f_\pi}$. These numbers should be compared with $-$0.44(32)\%  for $R_{f_\pi}^\prime$, $-$0.24(13)\% for $R_{f_K}^\prime$ and 0.20(30)\% for $R_{f_K/f_\pi}^\prime$ with $R_X^\prime=(X({\rm L64})-X({\rm L128}))/X({\rm L128}))$ in our simulations. The ChPT predictions and the simulation results give the same order of magnitude, though the latter is roughly a factor of two larger than the former.
The discrepancy, though errors are rather large, may be due to the higher-order effects in the SU(3) ChPT, which are shown to enhance the magnitude of the finite size effects for $f_\pi$ and $f_K$~\cite{Colangelo:2005gd}.

\begin{table}[h]
\caption{$R_X$ $(X=m_\pi,m_K,f_\pi,f_K,f_K/f_\pi)$ with $L=5.4$ fm and 10.9 fm at the physical point.}
\begin{center}
\begin{tabular}{crr}
\hline\hline
data set & \multicolumn{1}{c}{L64} & \multicolumn{1}{c}{L128} \\\hline
$m_\pi L$ &   \multicolumn{1}{c}{3.75} & \multicolumn{1}{c}{7.48} \\
$R_{m_{\pi}}$ & $4.971\times 10^{-4}$ & $3.300\times 10^{-6}$ \\
$R_{m_{K}}$ & $3.133\times 10^{-9}$ & $1.709\times 10^{-16}$ \\
$R_{f_{\pi}}$ & $-1.988\times 10^{-3}$ & $-1.320\times 10^{-5}$ \\
$R_{f_{K}}$ & $-7.458\times 10^{-4}$ & $-4.951\times 10^{-6}$ \\
$R_{f_{K}/f_\pi}$ & $ 1.245\times 10^{-3}$ & $8.252\times 10^{-6}$ \\
\hline\hline
\end{tabular}
\end{center}
\label{tab:fse_chpt}
\end{table}%

\subsubsection{Vector meson sector}
\label{subsubsec:fse_v}

\begin{figure}[h]
\begin{center}
\includegraphics[width=70mm,angle=0]{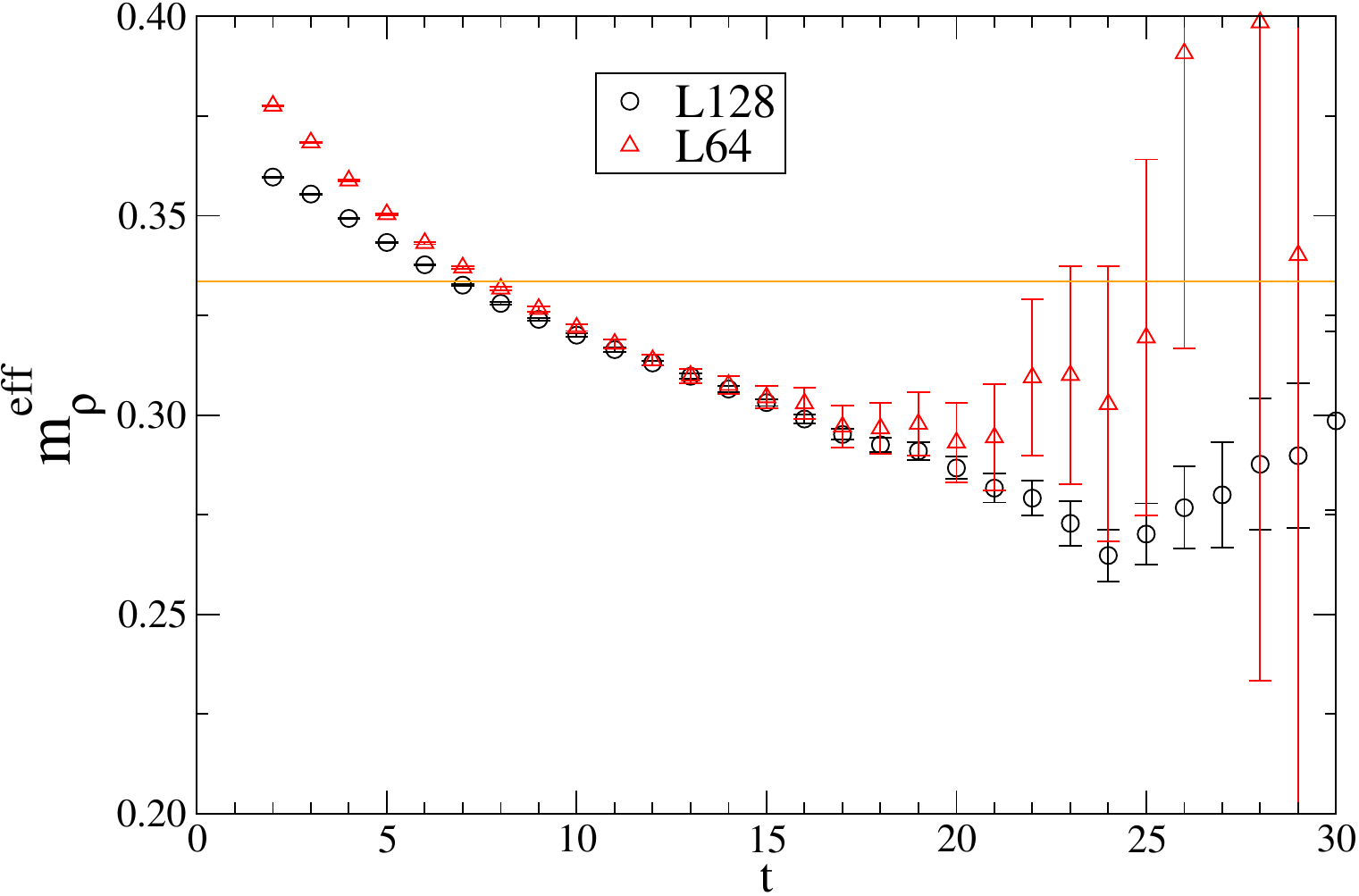}\\
\includegraphics[width=70mm,angle=0]{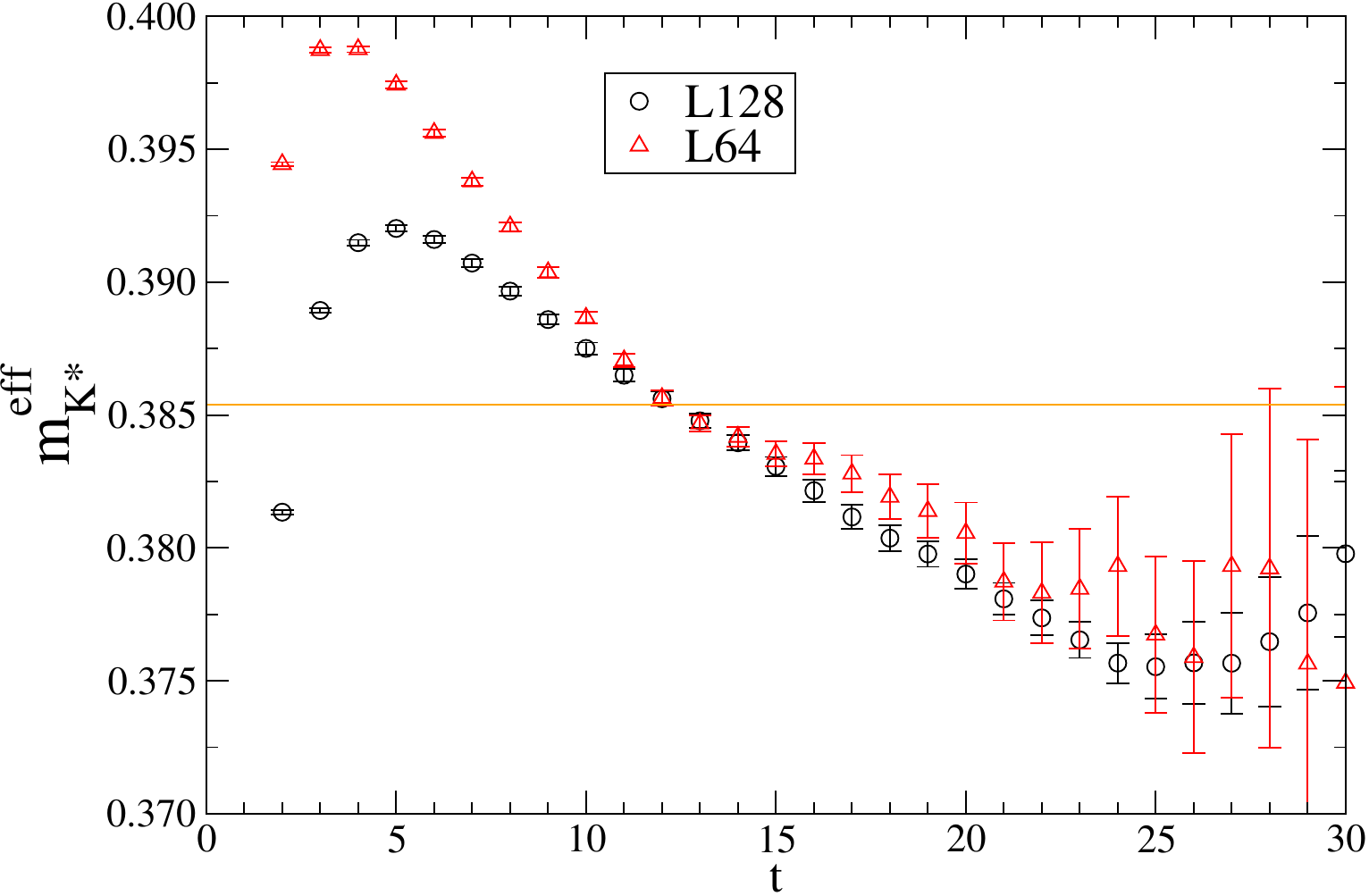}\\
\includegraphics[width=70mm,angle=0]{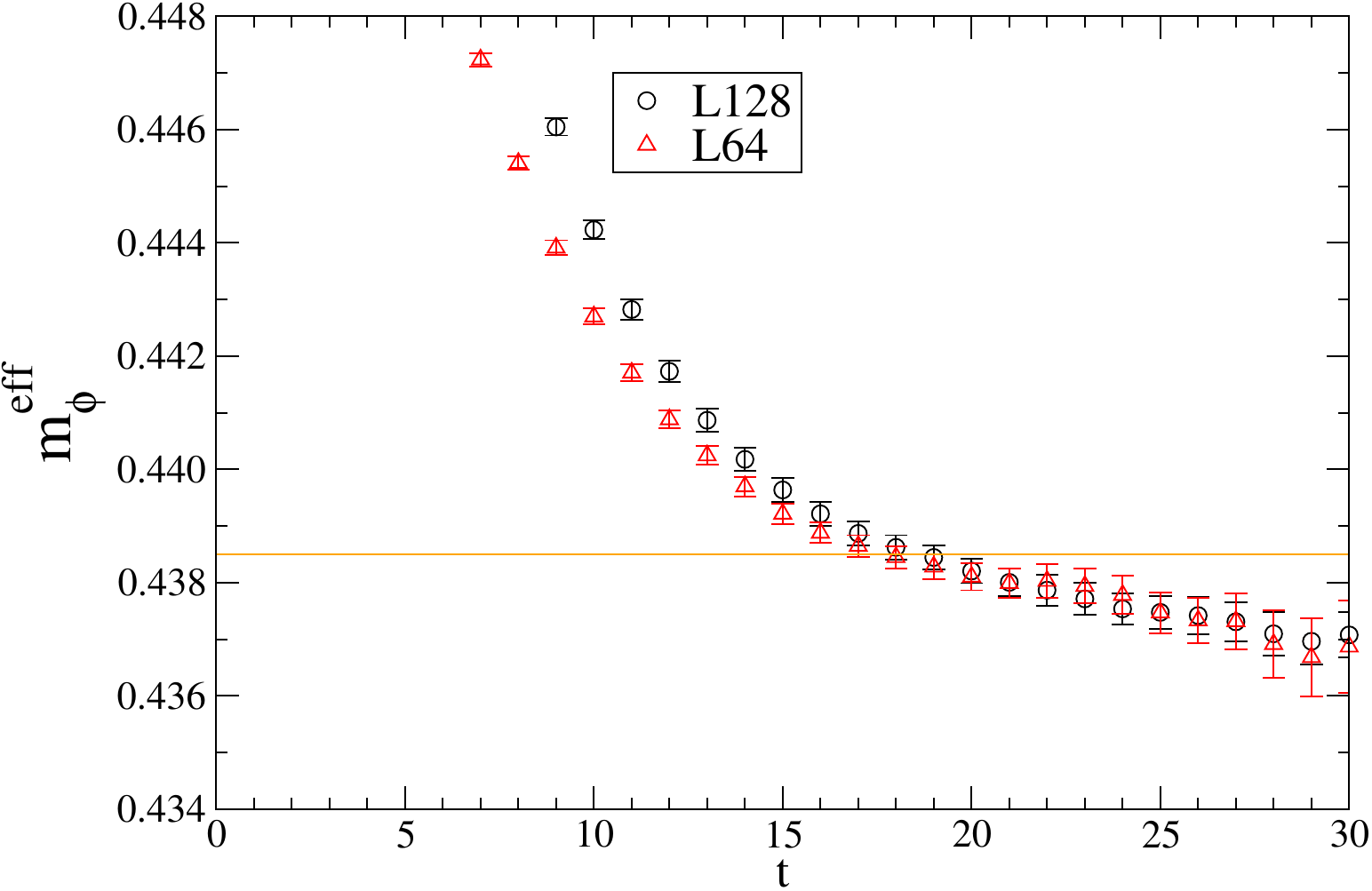}\\
\end{center}
\vspace{-5mm}
\caption{Comparison of effective masses for $\rho$, $K^*$ and $\phi$ (from top to bottom) channels on L128 and L64 lattices at $\beta=1.82$ with $(\kappa_{\rm ud},\kappa_{\rm s})=(0.126117,0.124902)$. The horizontal line denotes the pseudoexperimental resonance level.}
\label{fig:effmass_v}
\end{figure}

In Fig.~\ref{fig:effmass_v} we plot the vector meson effective masses on the L128 and L64 lattices at $(\kappa_{\rm ud},\kappa_{\rm s})=(0.126117,0.124902)$. It is hard to detect the finite size effect in the large time region beyond the statistical errors. In all $\rho$, $K^*$ and $\phi$ channels, the effective masses gradually decrease and go below the pseudoexperimental resonance level denoted by the horizontal line in the large time. This could be due to the contribution from the lower energy level of the $\pi\pi$, $K\pi$ and $KK$ states. We do not try to extract the masses by applying the single exponential fit to the propagators.

\subsubsection{Octet baryon sector}
\label{subsubsec:fse_8b}

\begin{figure}[t!]
\begin{center}
\includegraphics[width=70mm,angle=0]{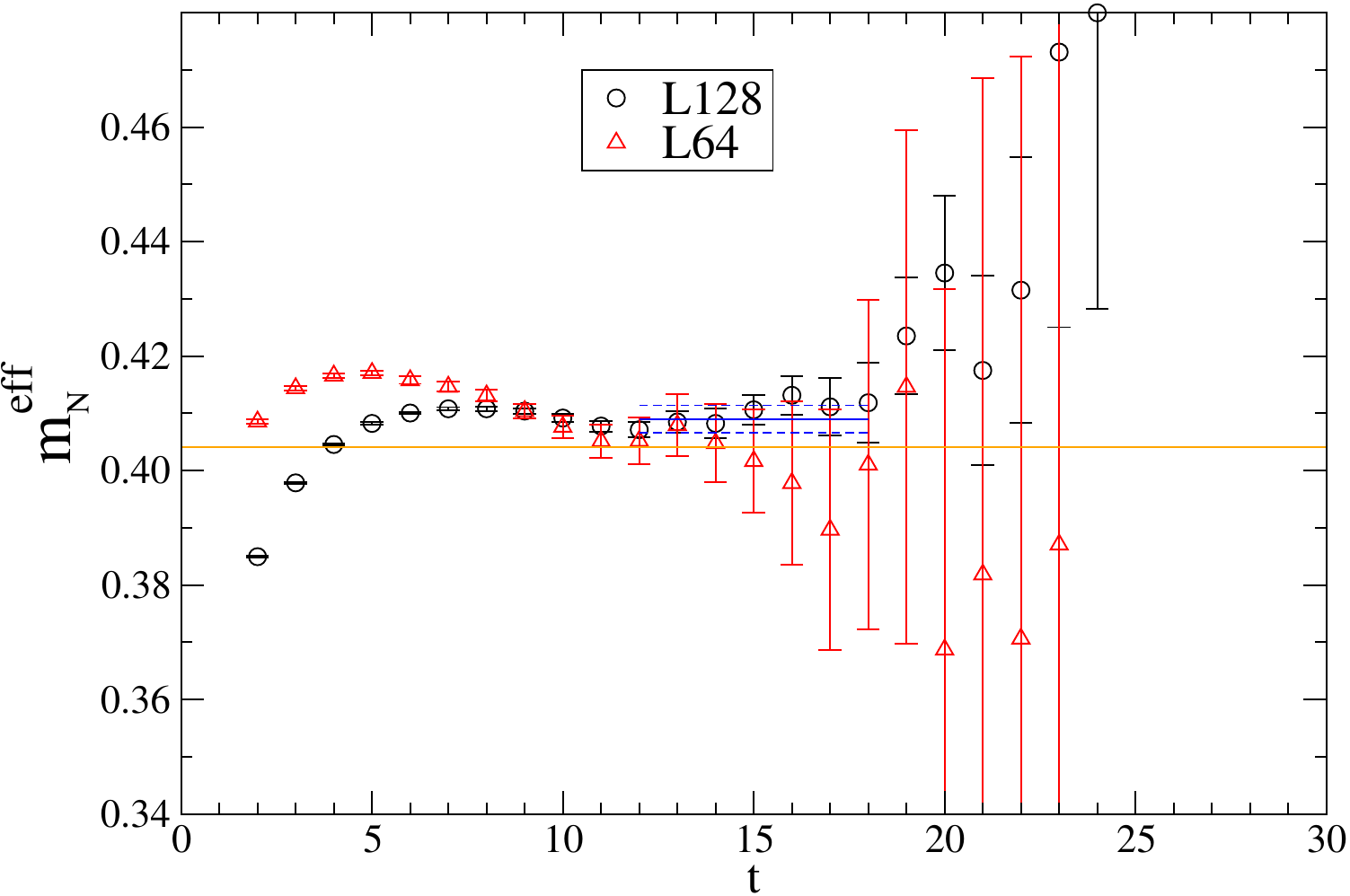}\\
\includegraphics[width=70mm,angle=0]{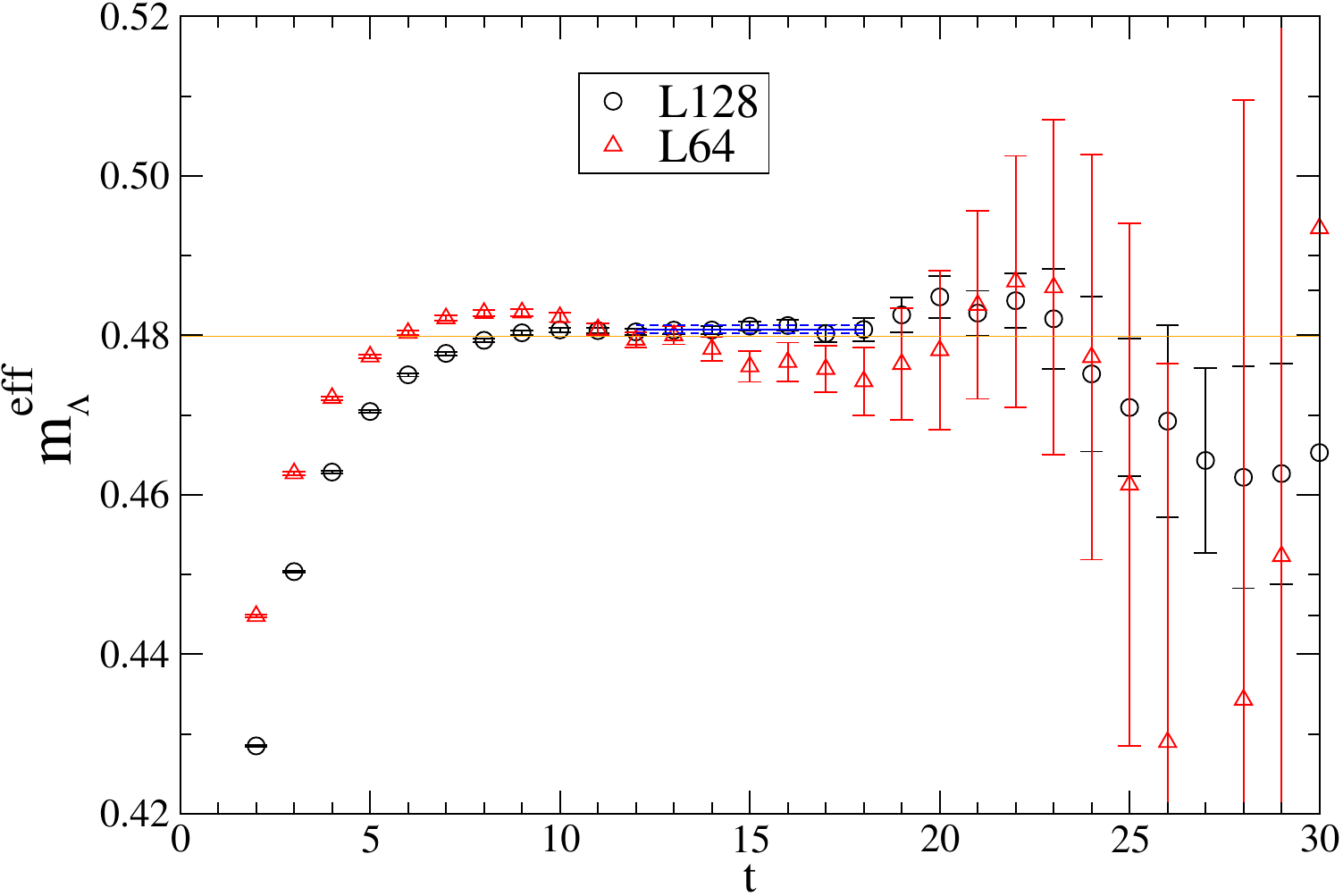}\\
\includegraphics[width=70mm,angle=0]{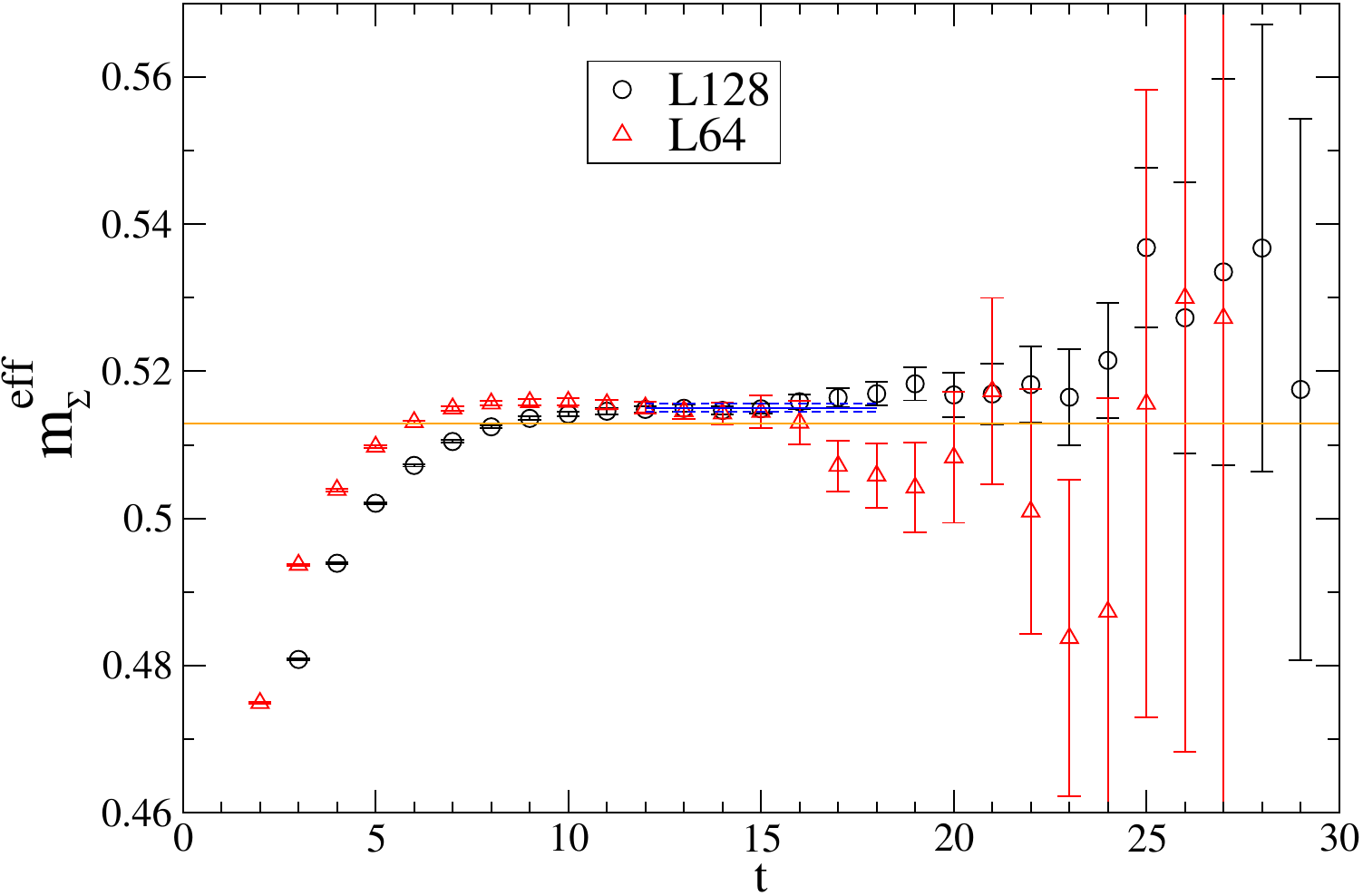}\\
\includegraphics[width=70mm,angle=0]{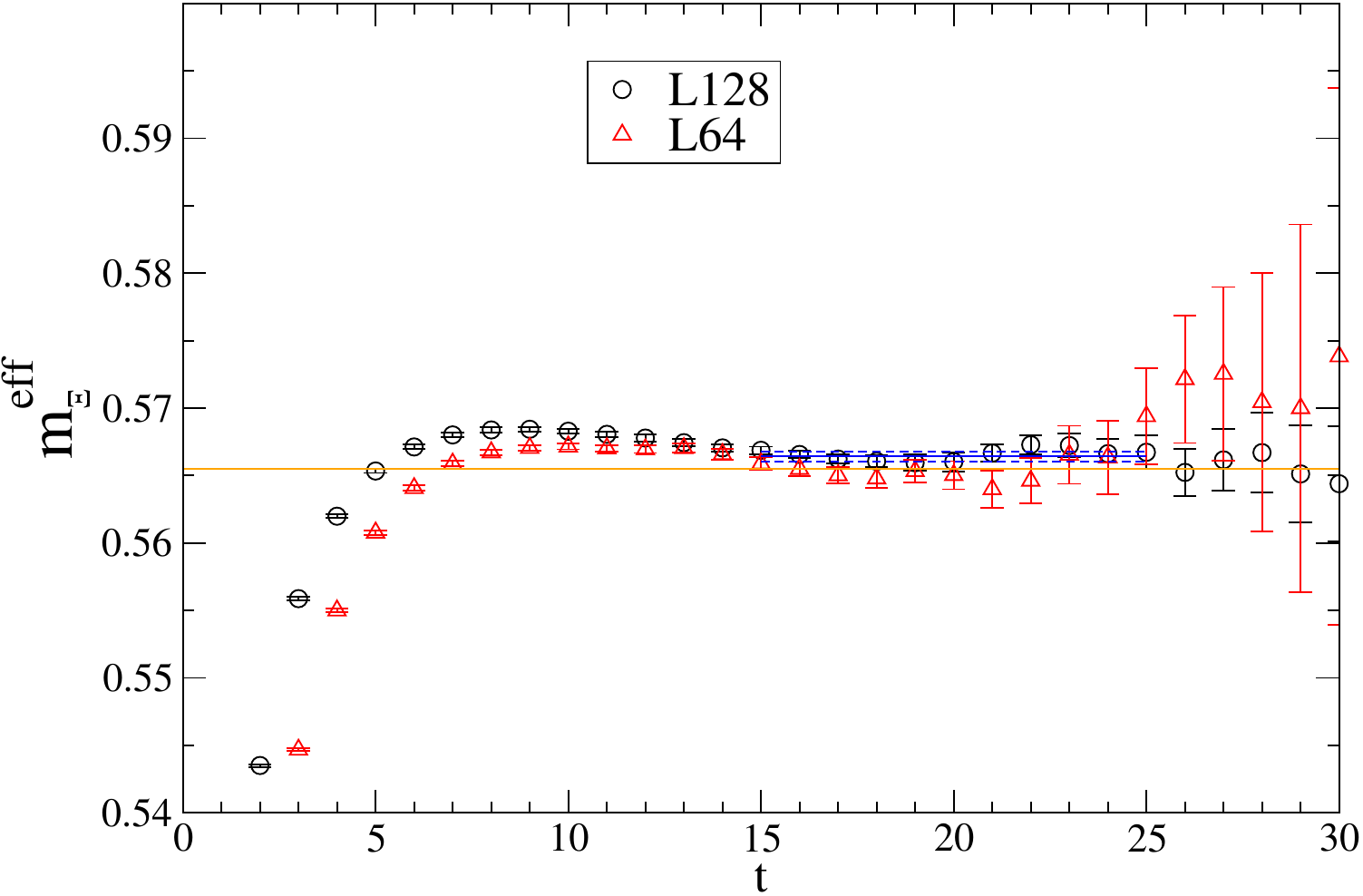}\\
\end{center}
\vspace{-5mm}
\caption{Comparison of effective masses for $N$, $\Lambda$, $\Sigma$ and $\Xi$ (from top to bottom) baryon channels on the L128 and L64 lattices at $\beta=1.82$ with $(\kappa_{\rm ud},\kappa_{\rm s})=(0.126117,0.124902)$. The horizontal line denotes the pseudoexperimental mass.}
\label{fig:effmass_8b}
\end{figure}

Figure~\ref{fig:effmass_8b} shows the octet baryon effective masses on the L128 and L64 lattices at $(\kappa_{\rm ud},\kappa_{\rm s})=(0.126117,0.124902)$. The horizontal line indicates the pseudoexperimental value. The finite size effect is hardly observed in all $N$, $\Lambda$, $\Sigma$, and $\Xi$ channels. The results of the octet baryon masses with a single exponential fit are listed in Table~\ref{tab:hmass}. Since the $\Xi$ effective mass shows a reasonable plateau with the smallest statistical errors among the octet baryons, we use the $\Xi$ mass as a physical input, together with $\pi$ and $K$ meson masses, to determine the physical quark masses and the lattice spacing in Table~\ref{tab:para_sim}.   

\subsubsection{Decuplet baryon sector}
\label{subsec:fse_10b}

\begin{figure}[t!]
\begin{center}
\includegraphics[width=70mm,angle=0]{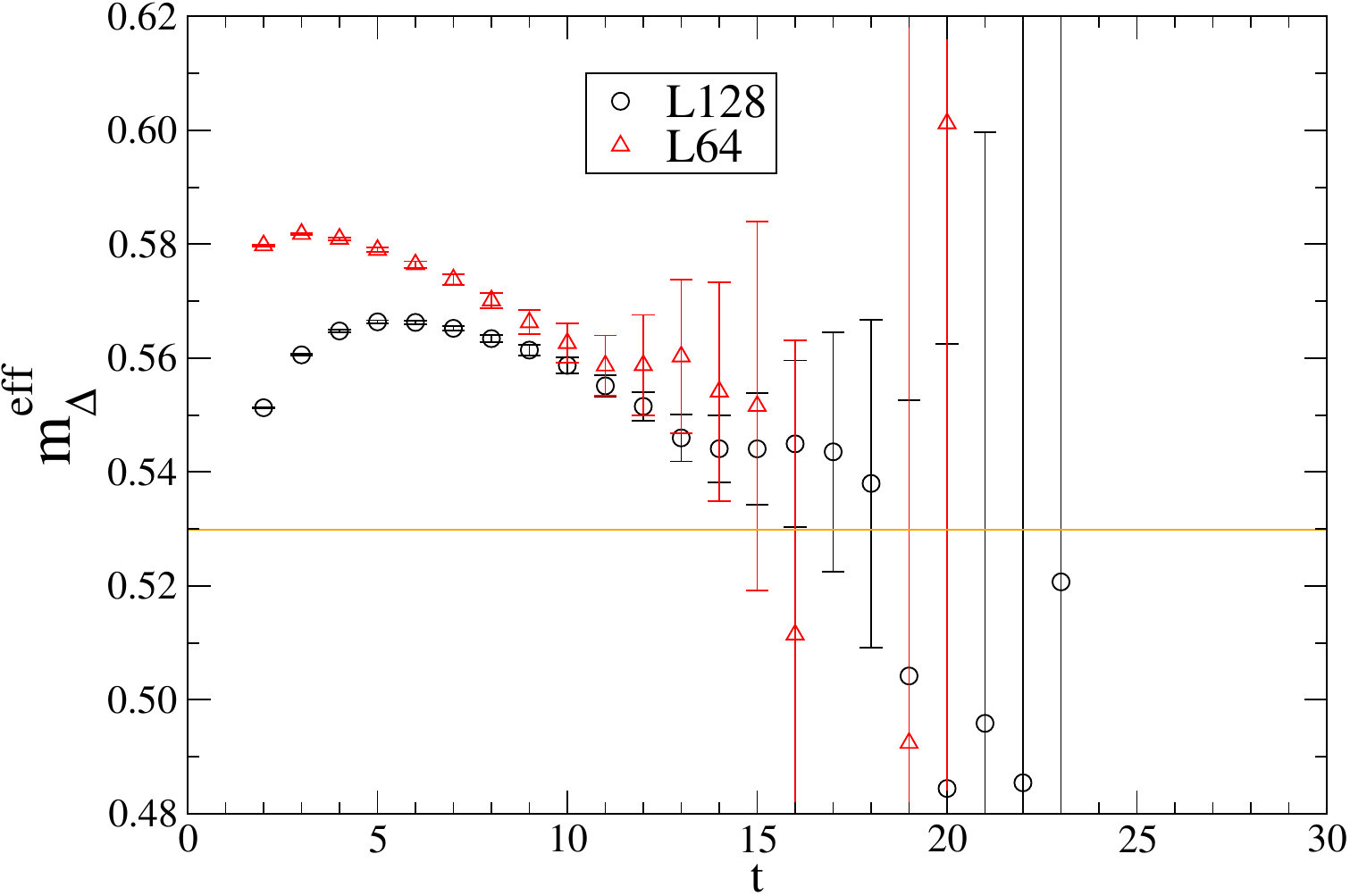}\\
\includegraphics[width=70mm,angle=0]{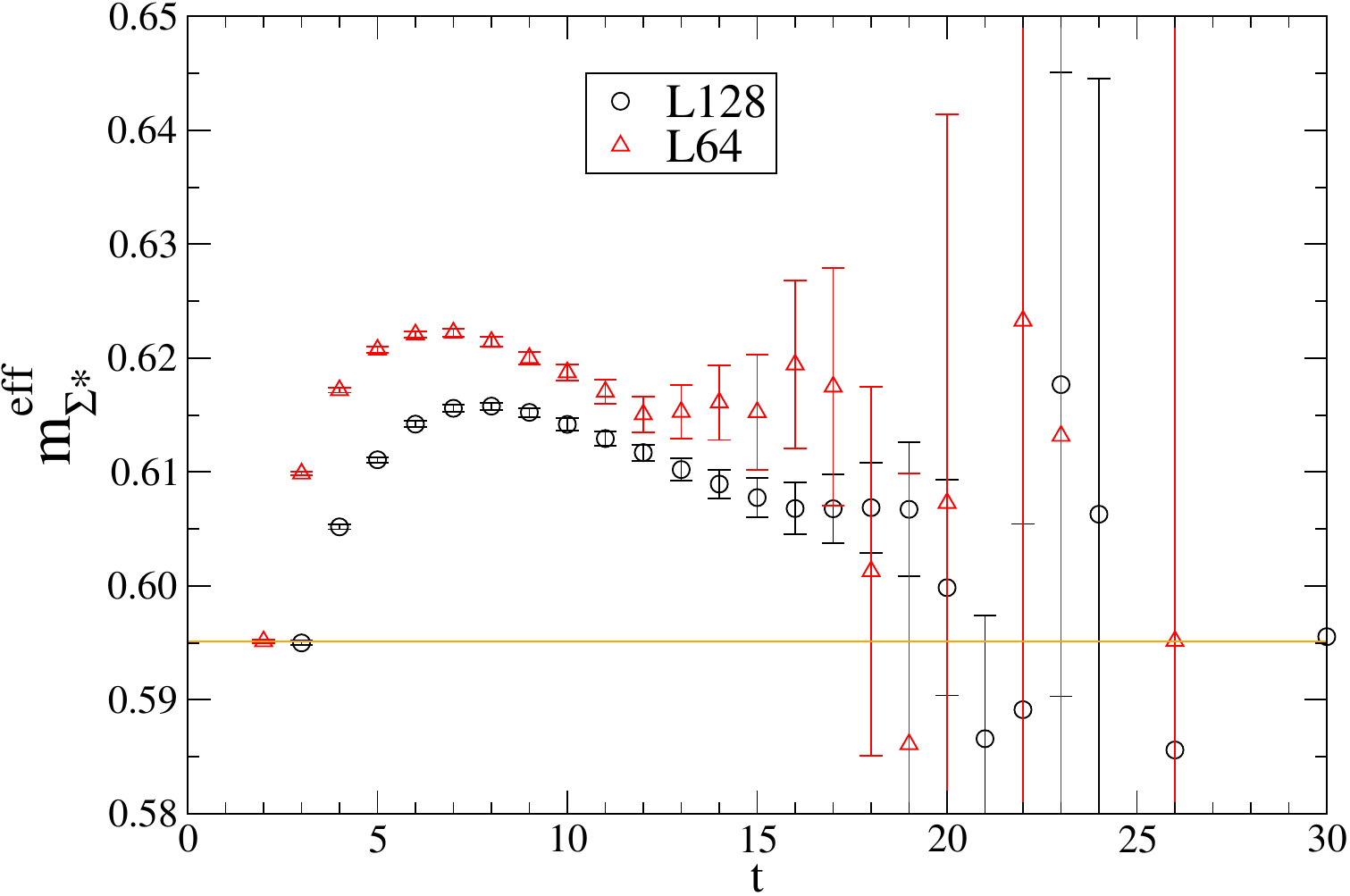}\\
\includegraphics[width=70mm,angle=0]{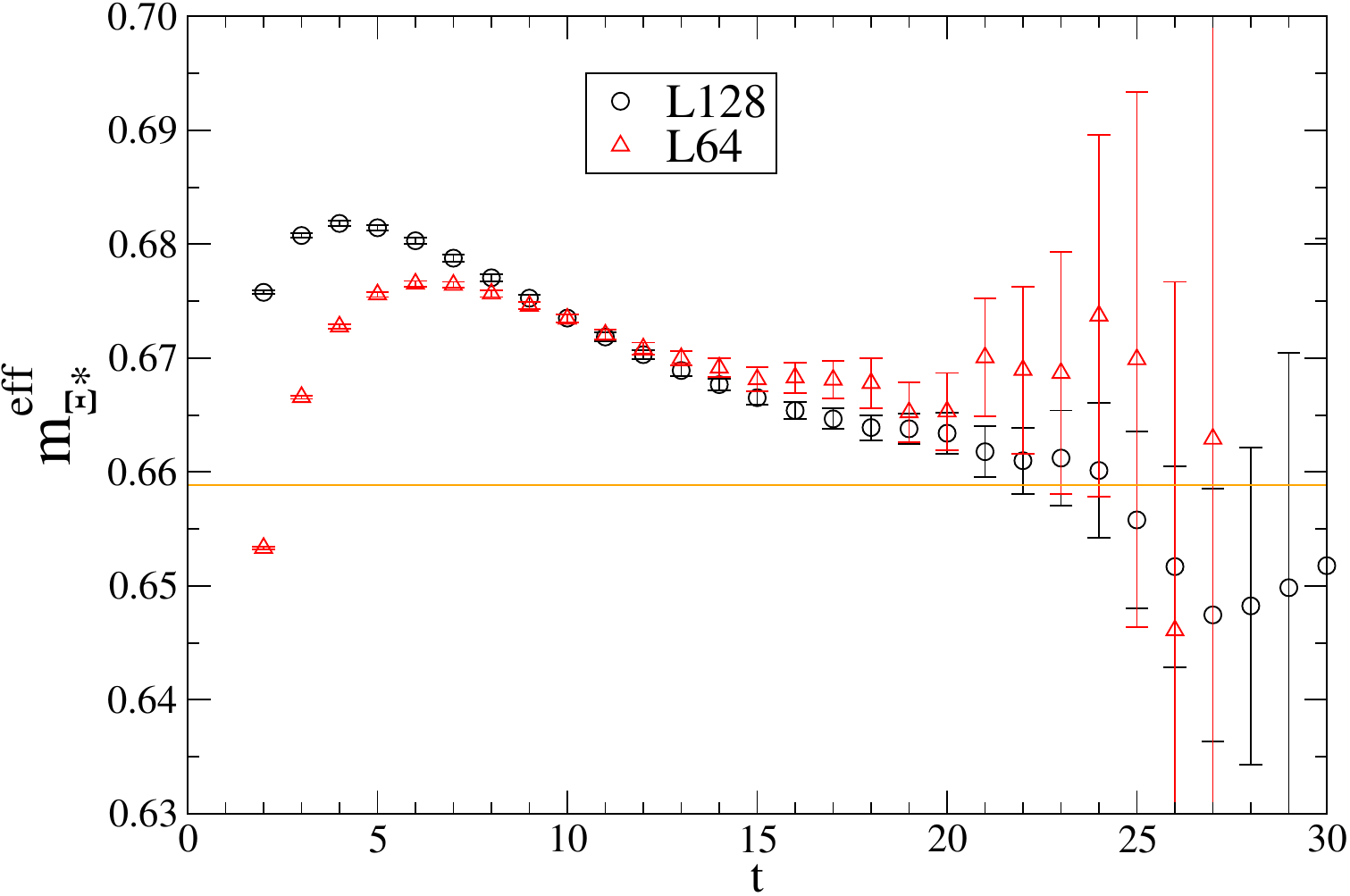}\\
\includegraphics[width=70mm,angle=0]{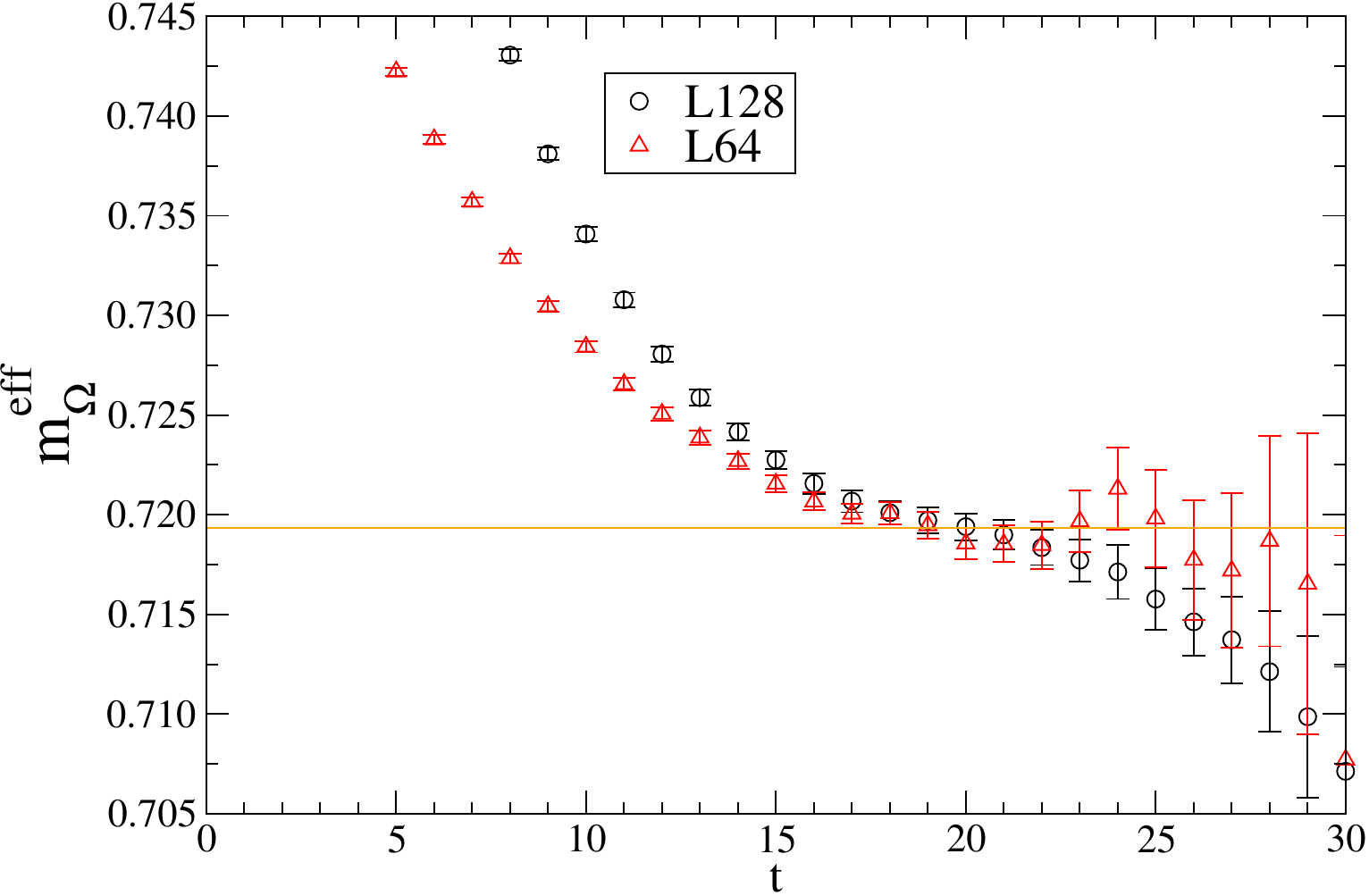}\\
\end{center}
\vspace{-5mm}
\caption{Comparison of effective masses for $\Delta$, $\Sigma^*$, $\Xi^*$ and $\Omega$ (from top to bottom) baryon channels on the L128 and L64 lattices at $\beta=1.82$ with $(\kappa_{\rm ud},\kappa_{\rm s})=(0.126117,0.124902)$. The horizontal line denotes the pseudoexperimental resonance level for $\Delta$, $\Sigma^*$ and $\Xi^*$ and the pseudoexperimental mass for $\Omega$.}
\label{fig:effmass_10b}
\end{figure}

We plot the decuplet baryon effective masses on the L128 and L64 lattices at $(\kappa_{\rm ud},\kappa_{\rm s})=(0.126117,0.124902)$ in Fig.~\ref{fig:effmass_10b}, where the horizontal lines denote the pseudoexperimental resonance levels for the $\Delta$, $\Sigma^*$ and $\Xi^*$ baryons and the pseudoexperimental mass for the $\Omega$ baryon. In this work we are not going into detailed analyses on the resonance states of the $\Delta$, $\Sigma^*$, and $\Xi^*$ baryons. A peculiar finding in Fig.~\ref{fig:effmass_10b} is that the effective mass of the $\Omega$ baryon on the L128 lattice shows monotonic decreases toward the large time region, which is quite similar to the behaviors of the vector meson case. A possibility of the mixing effect with the nearby states was discussed in Ref.~\cite{PACS:2019ofv}. Since there seems no reasonable plateau region, we give up extracting the $\Omega$ baryon mass.

\subsection{Determination of the physical point with convention A}
\label{subsec:physicalpoint}

\begin{figure}[t!]
\begin{center}
\includegraphics[width=70mm,angle=0]{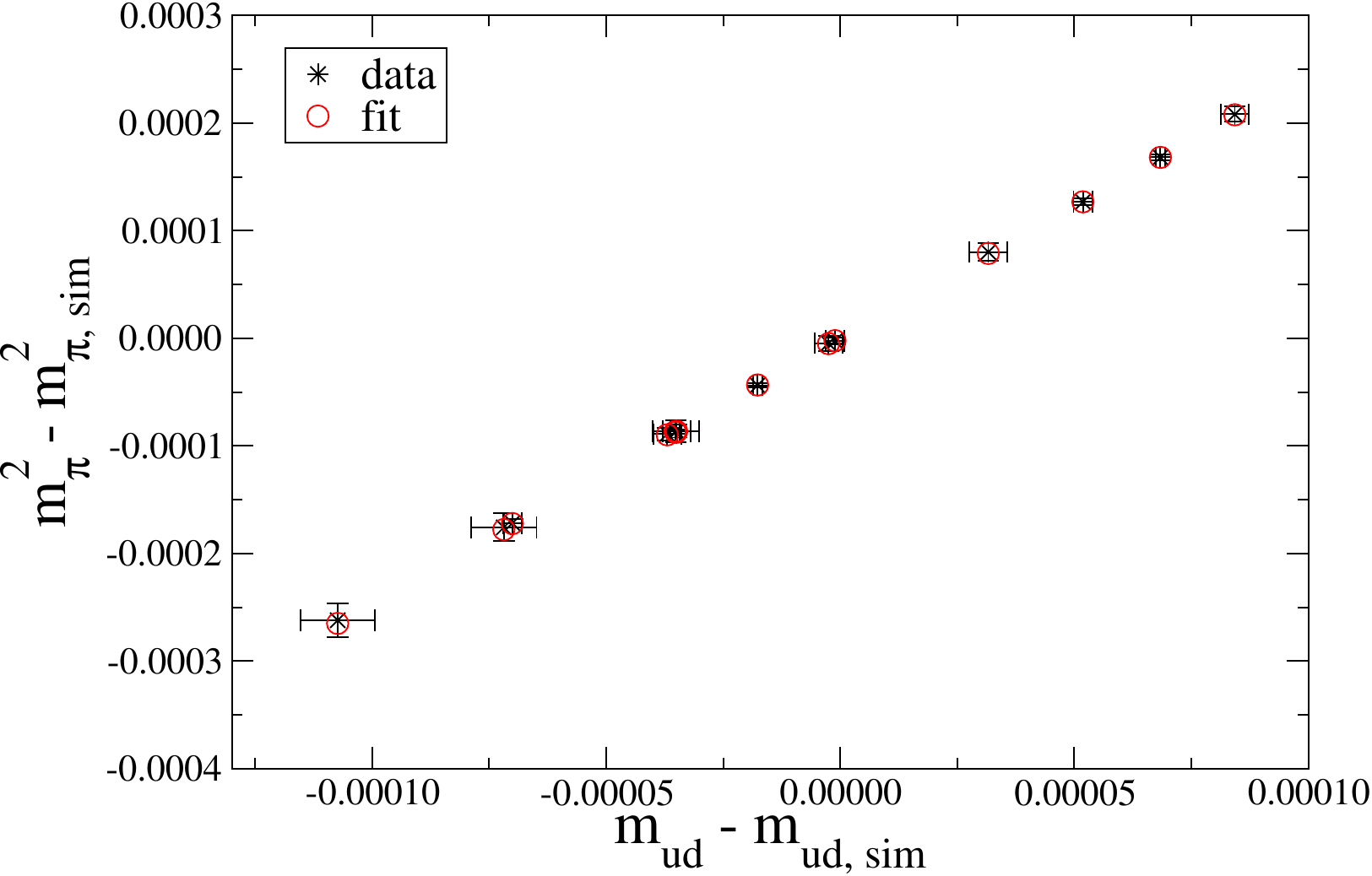}\\
\includegraphics[width=70mm,angle=0]{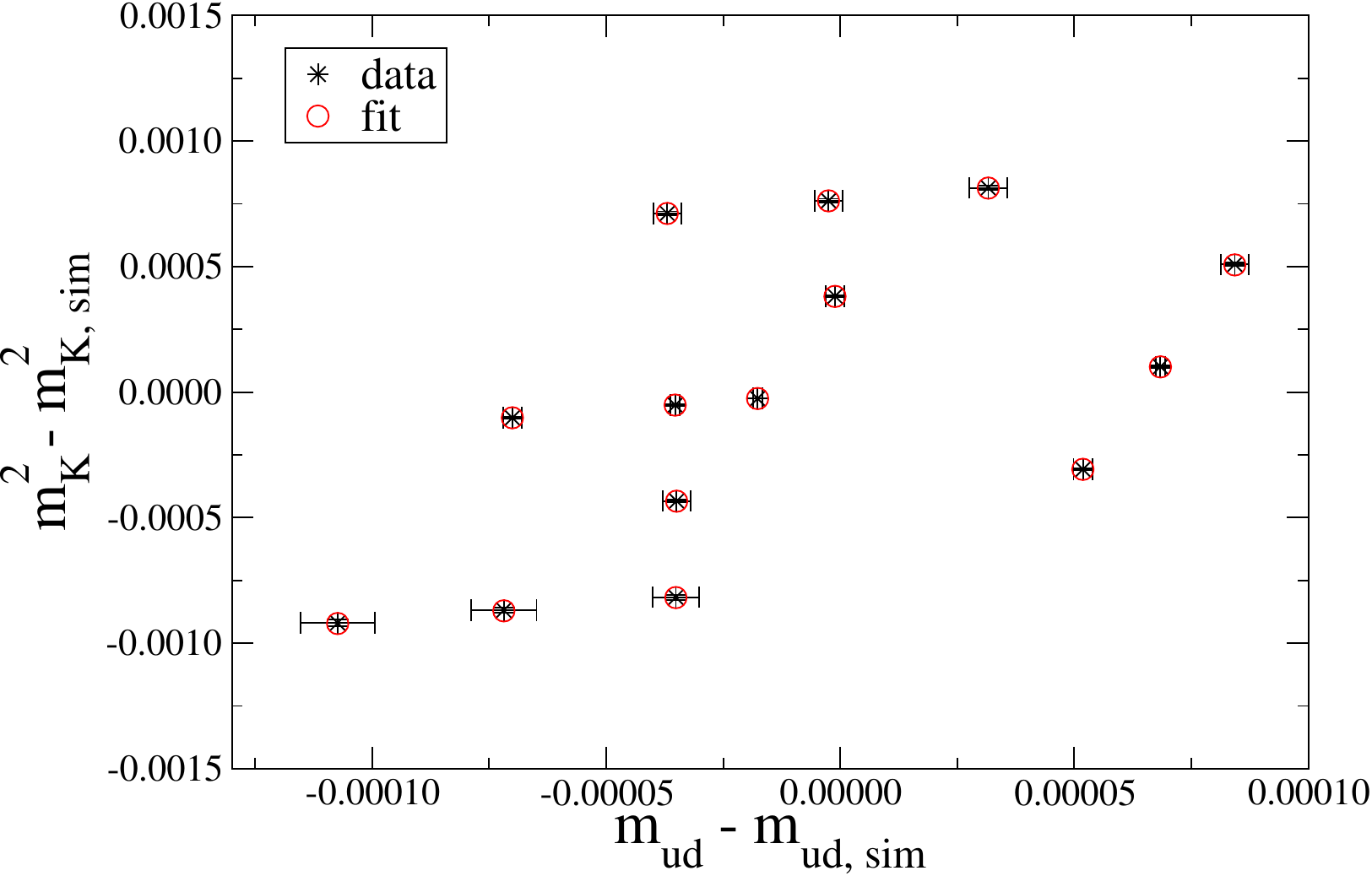}\\
\includegraphics[width=70mm,angle=0]{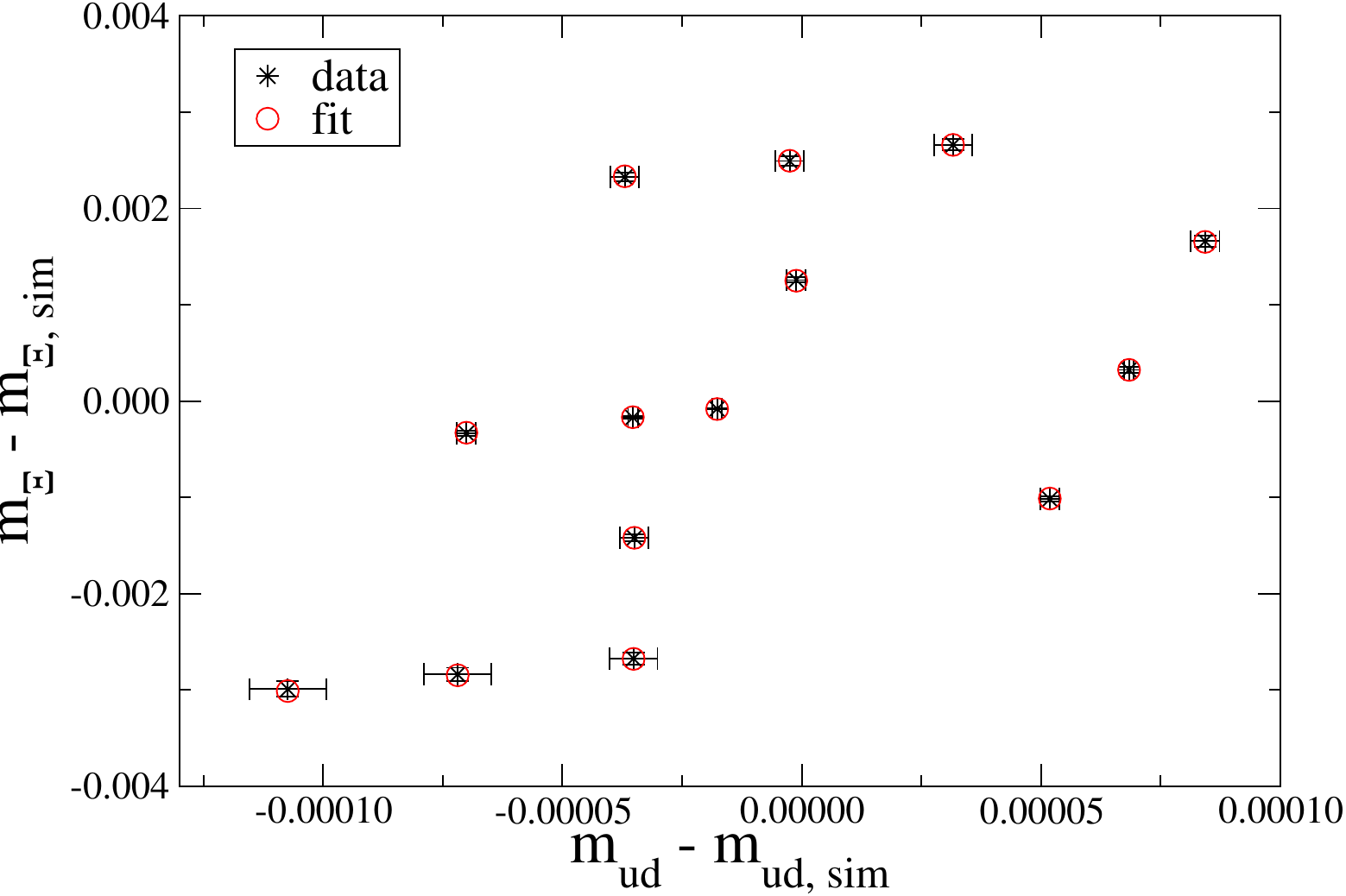}\\
\end{center}
\vspace{-5mm}
\caption{AWI quark mass dependence of $m_\pi^2-m_\pi^2|_{\rm sim}$, $m_K^2-m_K^2|_{\rm sim}$ and $m_\Xi-m_\Xi|_{\rm sim}$ (from top to bottom) on the L64 lattice together with the fit results.}
\label{fig:fit_64}
\end{figure}

\begin{figure}[t!]
\begin{center}
\includegraphics[width=75mm,angle=0]{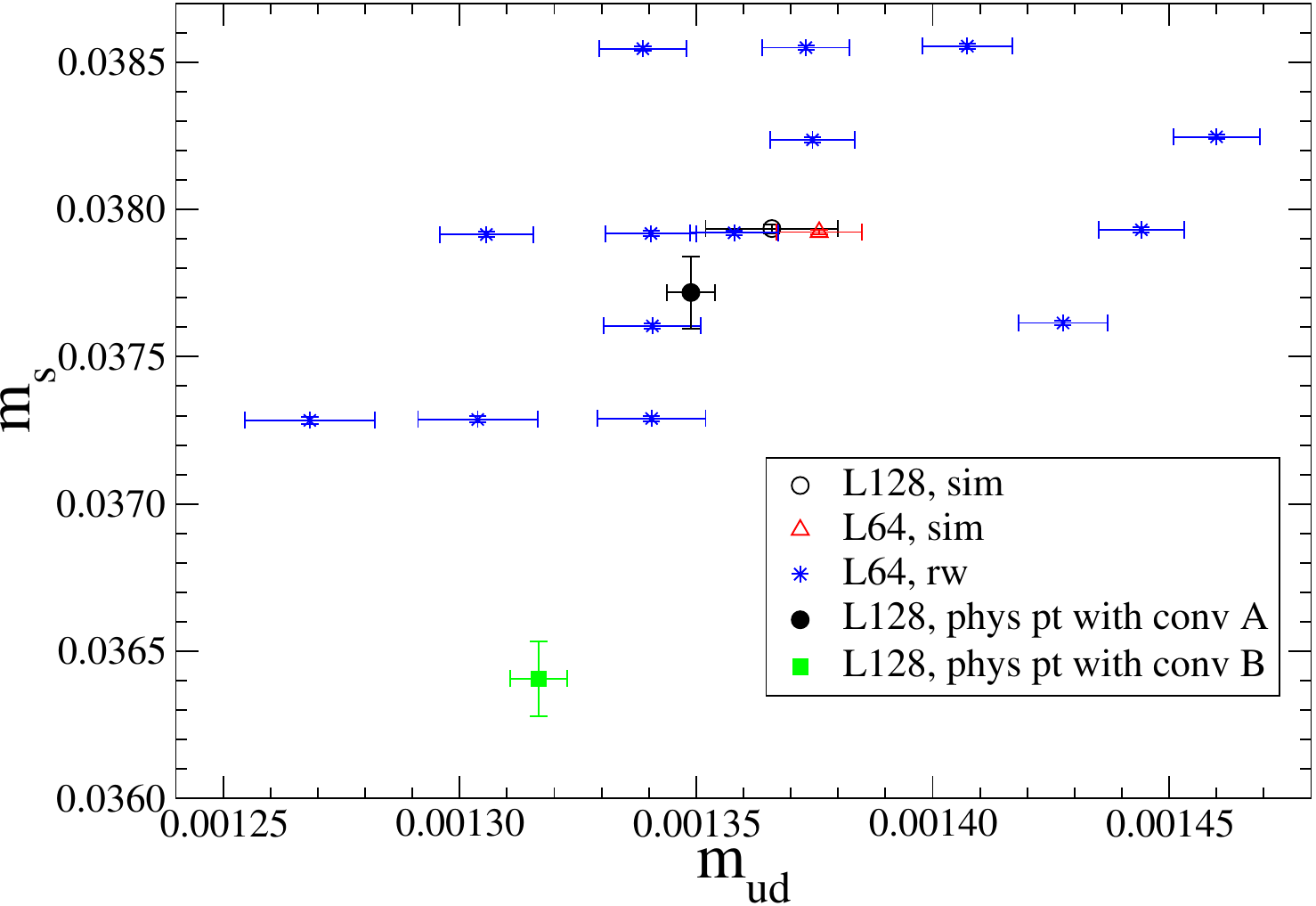}\\
\end{center}
\vspace{-5mm}
\caption{Simulation point, reweighted points, and physical points with conventions A and B in $m_{\rm ud}$-$m_{\rm s}$ plane. The errors of $m_{\rm s}$ in the blue symbols are smaller than their symbols.}
\label{fig:physicalpoint}
\end{figure}

In order to determine the physical point on the L128 lattice, we use the Taylor expansion of $m_\pi^2$, $m_K^2$ and $m_\Xi$ in terms of the AWI quark masses around the simulation point:
\ben
X&=&X\vert_{\rm sim}+\left.\frac{\partial X}{\partial m_{\rm ud}}\right\vert_{\rm sim}\left(m_{\rm ud}- m_{\rm ud}\vert_{\rm sim}\right)\nn\\
&&\hspace{15mm}+\left.\frac{\partial X}{\partial m_{\rm s}}\right\vert_{\rm sim}\left(m_{\rm s}- m_{\rm s}\vert_{\rm sim}\right)
\label{eq:hmass_taylor}
\een
with $X=m_\pi^2, m_K^2, m_\Xi$.
The coefficients $\partial X/\partial m_{\rm ud}$ and $\partial X/\partial m_{\rm s}$ are determined from the simulated and reweighted results on the L64 lattice. Figure~\ref{fig:fit_64} shows the AWI quark mass dependence of $m_\pi^2-m_\pi^2|_{\rm sim}$, $m_K^2-m_K^2|_{\rm sim}$ and $m_\Xi-m_\Xi|_{\rm sim}$ on the L64 lattice. We employ the uncorrelated fit with the linear function of  Eq.~(\ref{eq:hmass_taylor}). The fit results in Fig.~\ref{fig:fit_64} well describe the quark mass dependence of $m_\pi^2$, $m_K^2$, and $m_\Xi$. Using $\partial X/\partial m_{\rm ud}$ and $\partial X/\partial m_{\rm s}$ on the L64 lattice, which are listed in Table~\ref{tab:fit_64}, we determine the physical point on the L128 lattice to reproduce the pseudoexperimental values of $m_\pi=0.1350$ GeV, $m_K=0.4976$ GeV, and $m_\Xi=1.3148$ GeV (convention A).  In Table~\ref{tab:hmass_ph_A} we present the results for the lattice cutoff and the bare ud and s quark masses at the physical point. Figure~\ref{fig:physicalpoint} illustrates the locations of the simulation point, the reweighted points, and the physical point in the $m_{\rm ud}$-$m_{\rm s}$ plane, which confirms that the simulation point with convention A is well tuned to the physical one.

\begin{table}[t!]
\caption{Fit results for $\partial X/\partial m_{\rm ud}$ and $\partial X/\partial m_{\rm s}$ in Eq.~(\ref{eq:hmass_taylor}) on the L64 lattice.}
\begin{center}
\begin{tabular}{ccc}
\hline
X  & $\partial X/\partial m_{\rm ud}$ & $\partial X/\partial m_{\rm s}$  \\\hline
$m_{\pi}^2$ & 2.457(19) & 1.9(3.2)\\
$m_{K}^2$   & 1.321(33) & 1.220(08)\\
$m_{\Xi}$   & 4.25(35)  & 3.998(72)\\\hline
$f_\pi$       & 2.11(28) & 0.095(54)\\
$f_K$ & 1.132(77) & 0.319(15)\\\hline
$m_N$       & 10.4(1.3) & 0.48(19)\\
$m_\Lambda$ & 10.4(1.1) & 2.34(27)\\
$m_\Sigma$  &  6.77(86) & 2.60(22)\\\hline
\end{tabular}
\end{center}
\label{tab:fit_64}
\end{table}%

  \begin{table*}
    \begin{center}
      \caption{Quark masses, PS meson decay constants, and octet baryon masses at the physical point with convention A. $m_{\rm ud}$ and $m_{\rm s}$ are given in lattice units.
      \label{tab:hmass_ph_A}}
      \begin{tabular}{c|ccc|r}
        \hline\hline
        Data set & L128 & L160 & L256 & Continuum limit\\\hline
      $a^{-1}$ [GeV] & 2.325(04) & 3.125(05) & 4.790(21) & \multicolumn{1}{c}{$\infty$}\\
  $m_{\rm ud}$ & 0.0013489(50)  & 0.0010012(40)  & 0.0006170(60)  & \multicolumn{1}{c}{$-$} \\
  $m_{\rm s}$  & 0.037718(123)  & 0.027831(091)  & 0.017128(154) & \multicolumn{1}{c}{$-$}\\
  $m_{\rm s}/m_{\rm ud}$ & 27.963(60)  & 27.799(59)  & 27.762(92)  & 27.659(98)\\
  $f_{\pi}$ & 132.20(23)(1.55)  & 130.78(25)(1.10) & 129.30(54)(39) & 128.41(72) [MeV] \\
  $f_K$     & 157.45(19)(1.84)  & 156.13(21)(1.31) & 155.04(41)(46) & 154.29(56) [MeV]\\
  $f_{K}/f_{\pi}$  & 1.1910(16)(02) & 1.1938(21)(01) & 1.1991(36)(00) & 1.1997(36) \\
  $f_{K^\pm}$ & $-$ & $-$ & $-$ & 153.98(56)(31) [MeV]\\
  $f_{K^\pm}/f_{\pi^\pm}$  & $-$ & $-$ & $-$ & 1.1972(36)(25) \\
  $m_N$       &  950.4(6.1)  &  943.3(3.9) &  951.4(19.8) & 939.2(10.8) [MeV]\\
  $m_\Lambda$ & 1116.3(1.4)  & 1113.5(1.8) & 1112.6(4.9) & 1110.5(3.7) [MeV] \\
  $m_\Sigma$  & 1195.9(1.5)  & 1194.6(1.7) & 1194.3(4.6) & 1193.3(3.6) [MeV] \\
    \hline\hline
      \end{tabular}
    \end{center}
  \end{table*}

Having determined the physical quark masses we now obtain the PS decay constants and the octet baryon masses at the physical point employing the formula of Eq.~(\ref{eq:hmass_taylor}). The coefficients for the PS decay constants and the octet baryon masses are determined by using the reweighted results on the L64 lattice. The fit results of the coefficients are given in Table~\ref{tab:fit_64}. Combined with the quark mass difference between the physical point in Table~\ref{tab:hmass_ph_A} and the simulation one in Table~\ref{tab:hmass}, the formula of Eq.~(\ref{eq:hmass_taylor}) yields the values for the PS decay constants and the octet baryon masses at the physical point, which are presented in Table~\ref{tab:hmass_ph_A}. 

\section{Results at $\beta=2.00$ and $2.20$}
\label{sec:results_b200b220}

\begin{figure}[h]
\begin{center}
\includegraphics[width=75mm,angle=0]{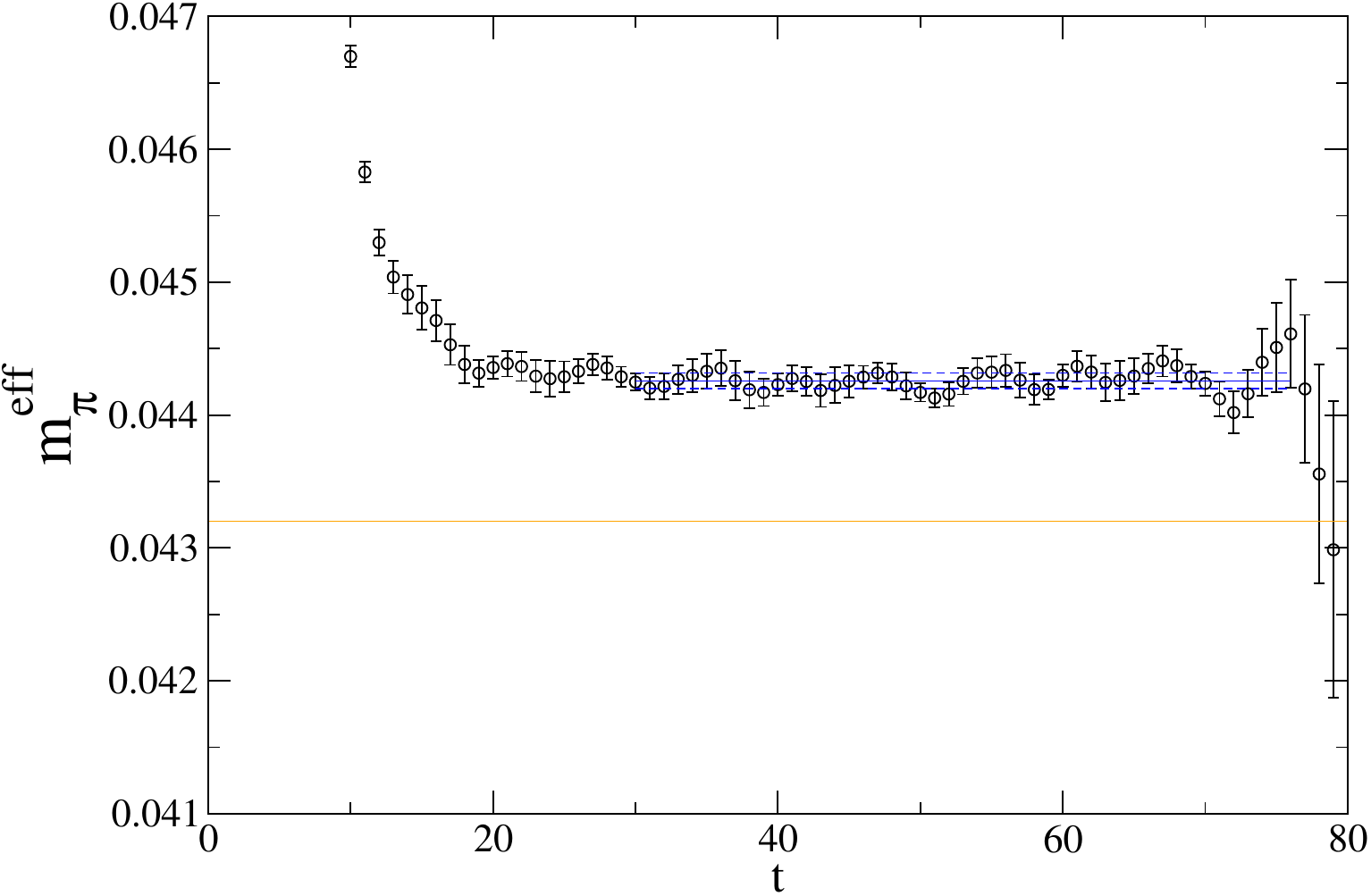}\\
\includegraphics[width=75mm,angle=0]{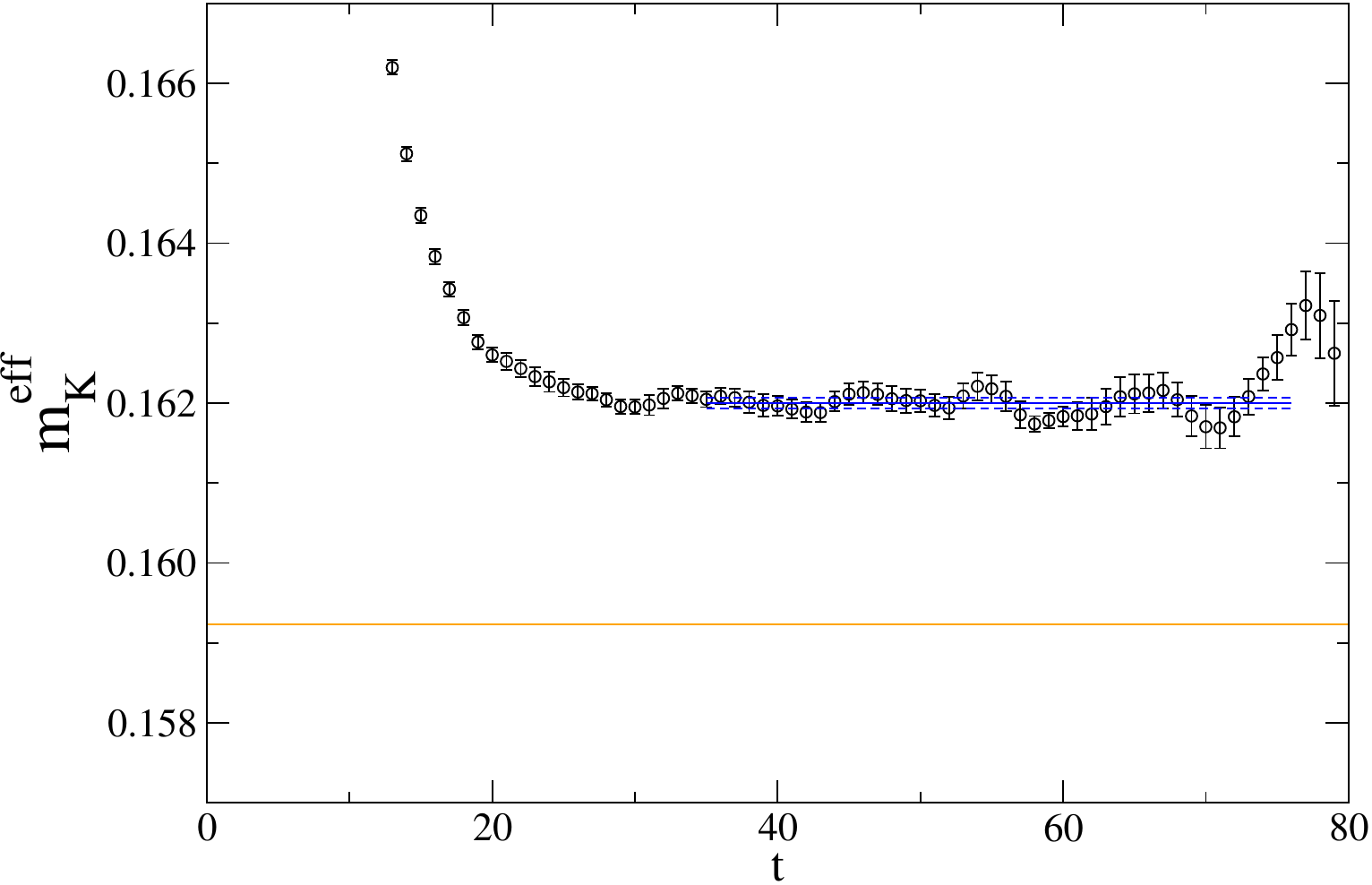}
\end{center}
\vspace{-5mm}
\caption{Effective masses for $\pi$ (top) and $K$ (bottom) mesons on the L160 lattice at $\beta=2.00$ with $(\kappa_{\rm ud},\kappa_{\rm s})=(0.125814,0.124925)$. The horizontal line denotes the pseudoexperimental value.}
\label{fig:effmass_ps_b200}
\end{figure}

\begin{figure}[h]
\begin{center}
\includegraphics[width=70mm,angle=0]{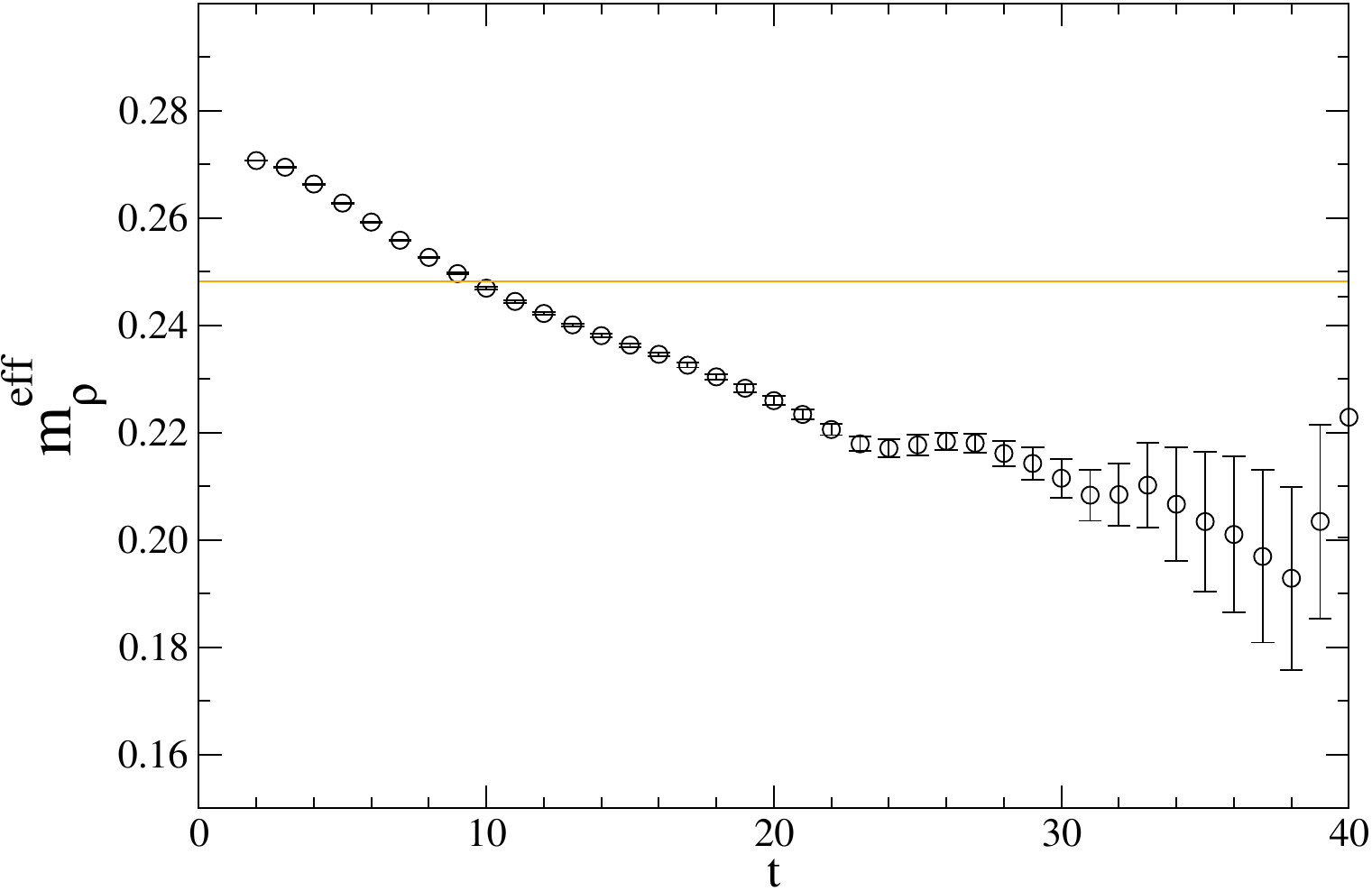}\\
\includegraphics[width=70mm,angle=0]{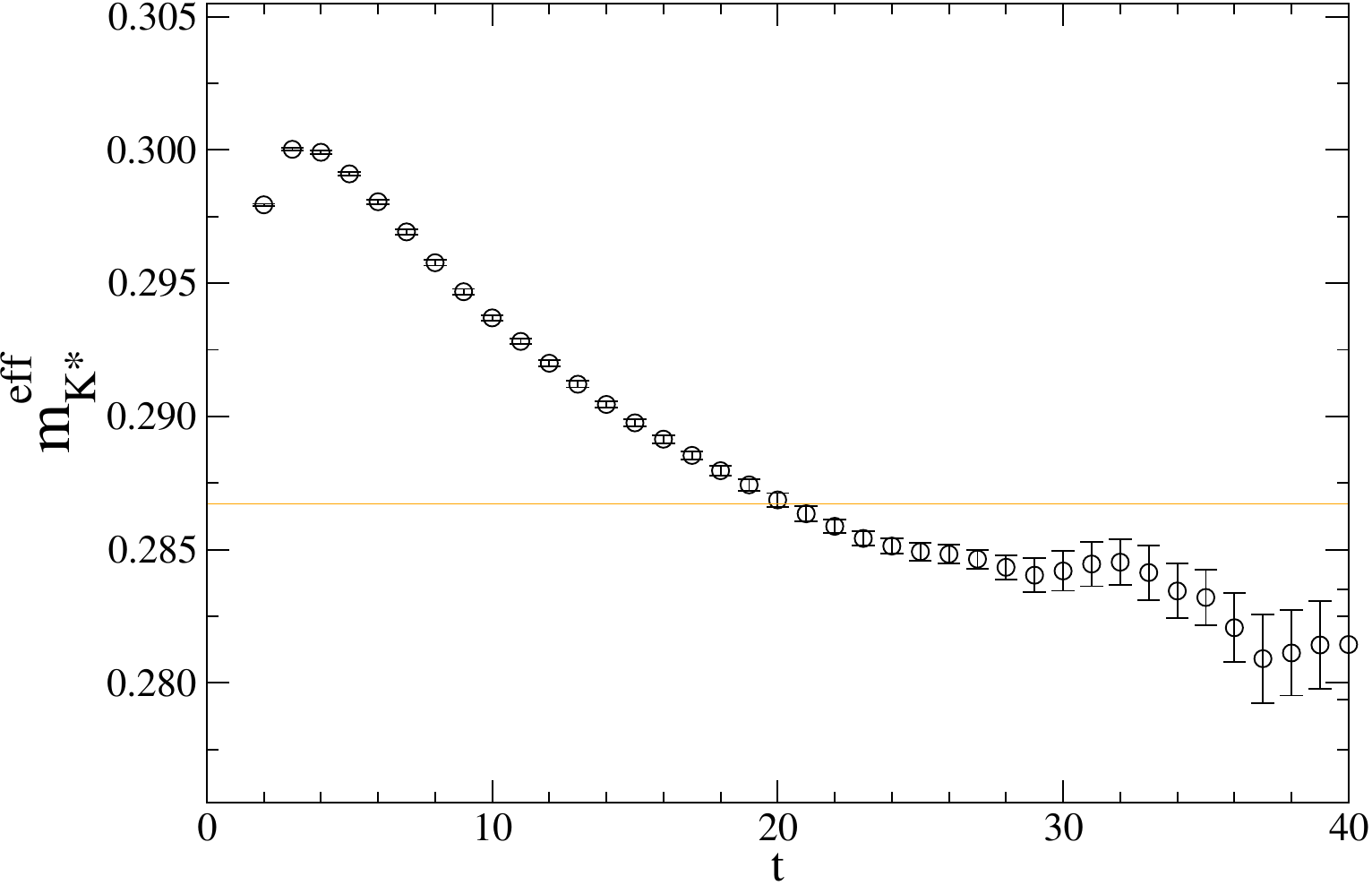}\\
\includegraphics[width=70mm,angle=0]{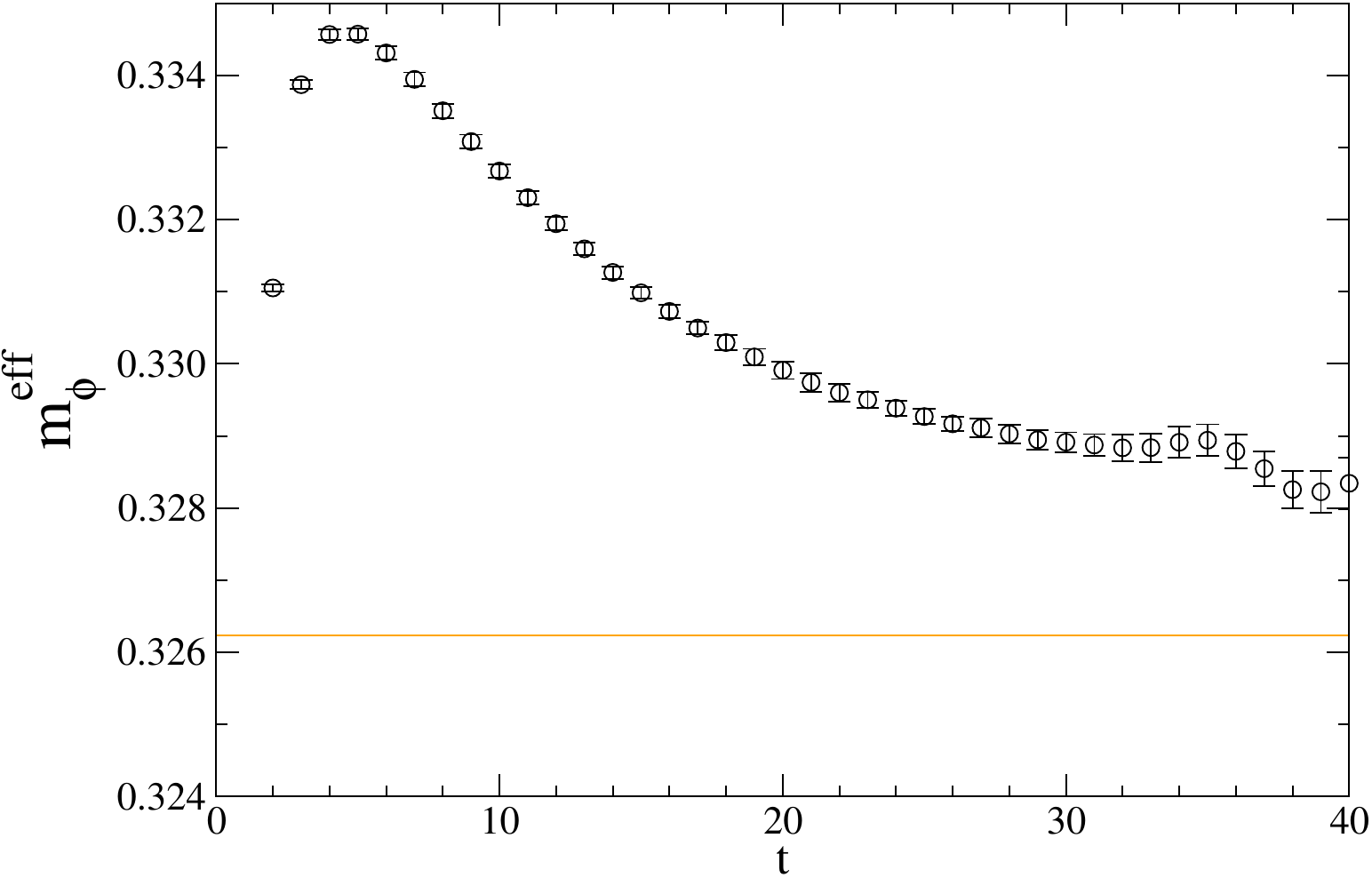}\\
\end{center}
\vspace{-5mm}
\caption{Effective masses for $\rho$, $K^*$ and $\phi$ (from top to bottom) mesons on the L160 lattice at $\beta=2.00$ with $(\kappa_{\rm ud},\kappa_{\rm s})=(0.125814,0.124925)$. The horizontal line denotes the pseudoexperimental resonance level.}
\label{fig:effmass_v_b200}
\end{figure}

\begin{figure}[t!]
\begin{center}
\includegraphics[width=70mm,angle=0]{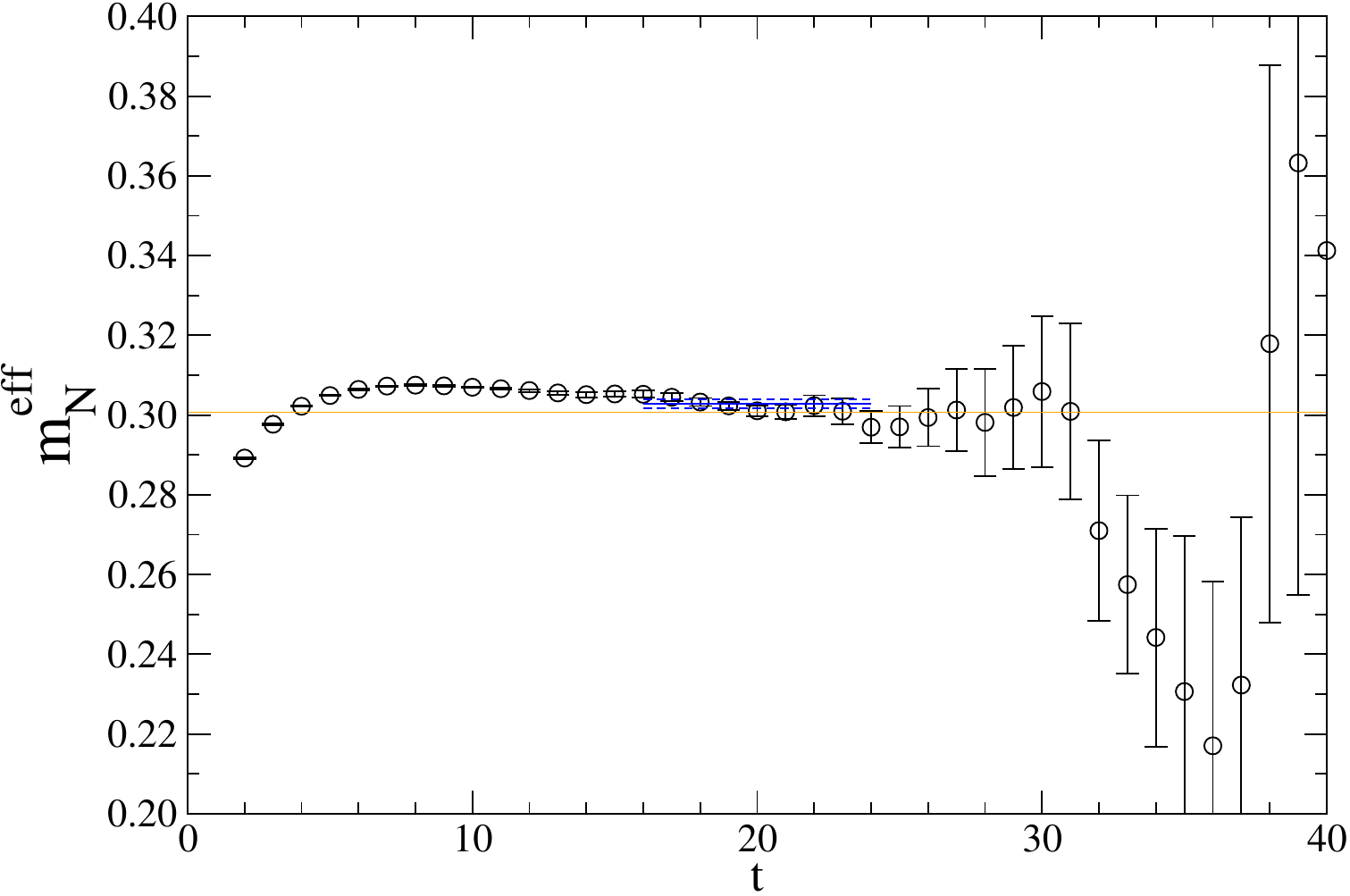}\\
\includegraphics[width=70mm,angle=0]{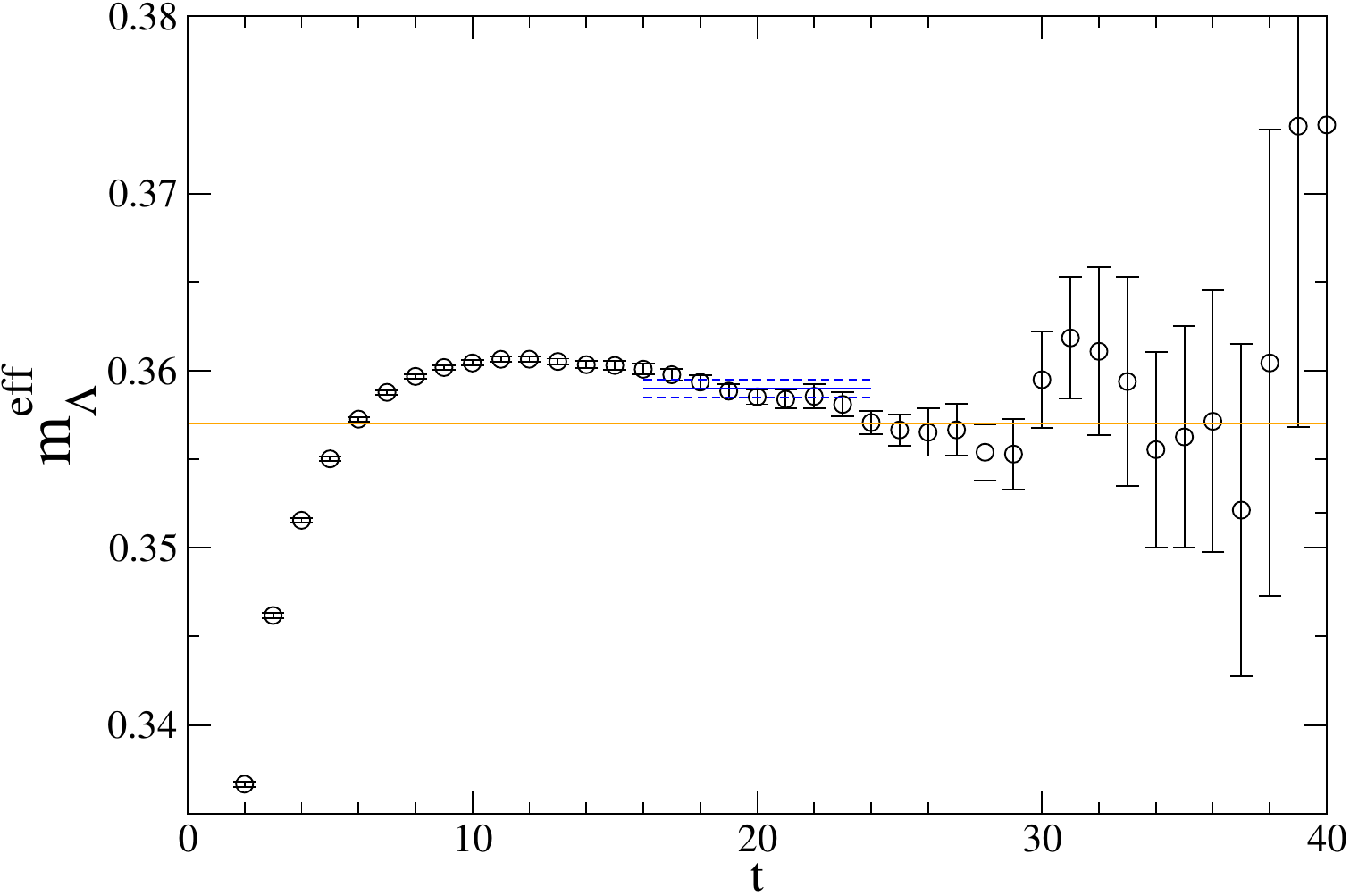}\\
\includegraphics[width=70mm,angle=0]{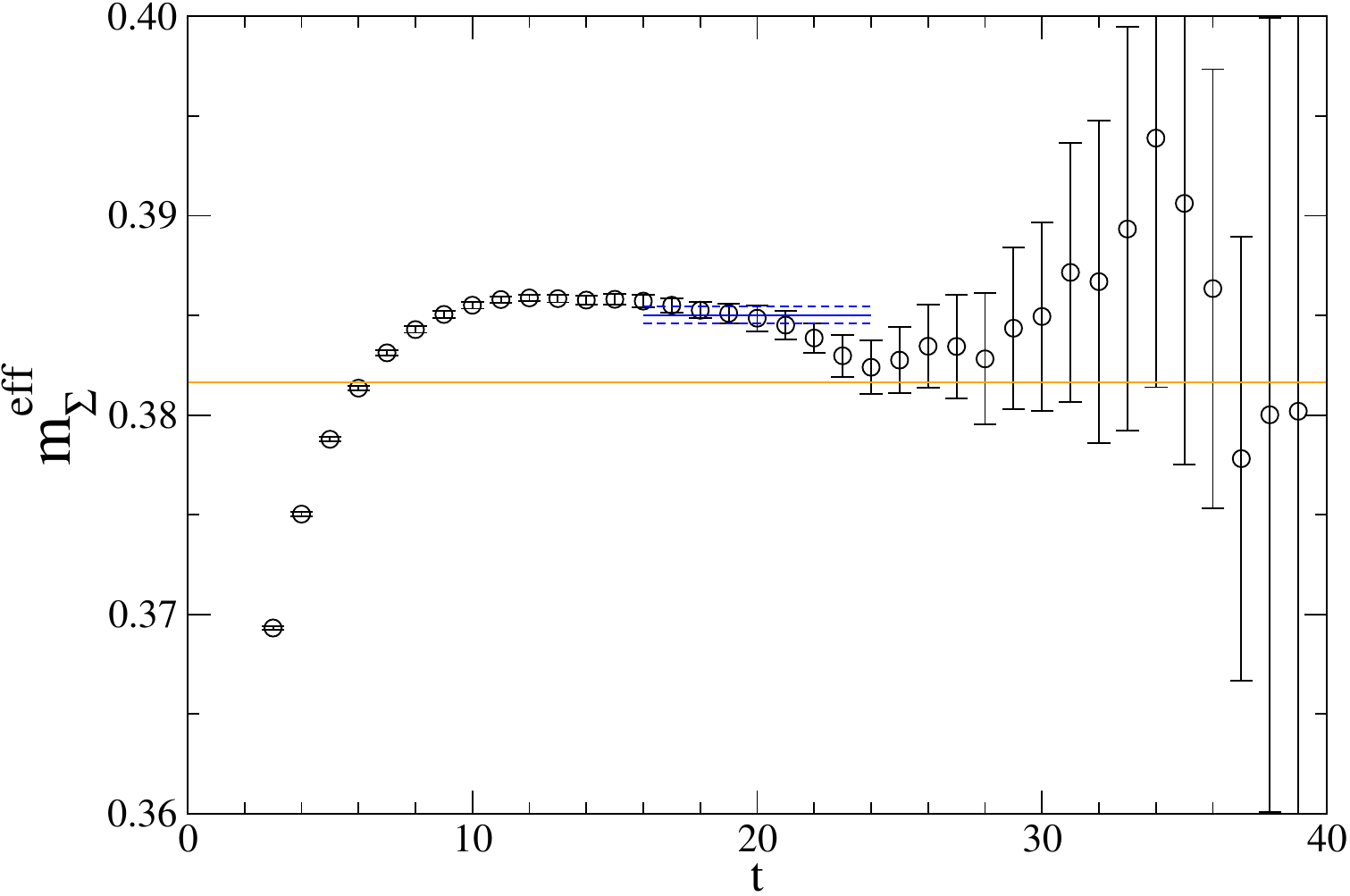}\\
\includegraphics[width=70mm,angle=0]{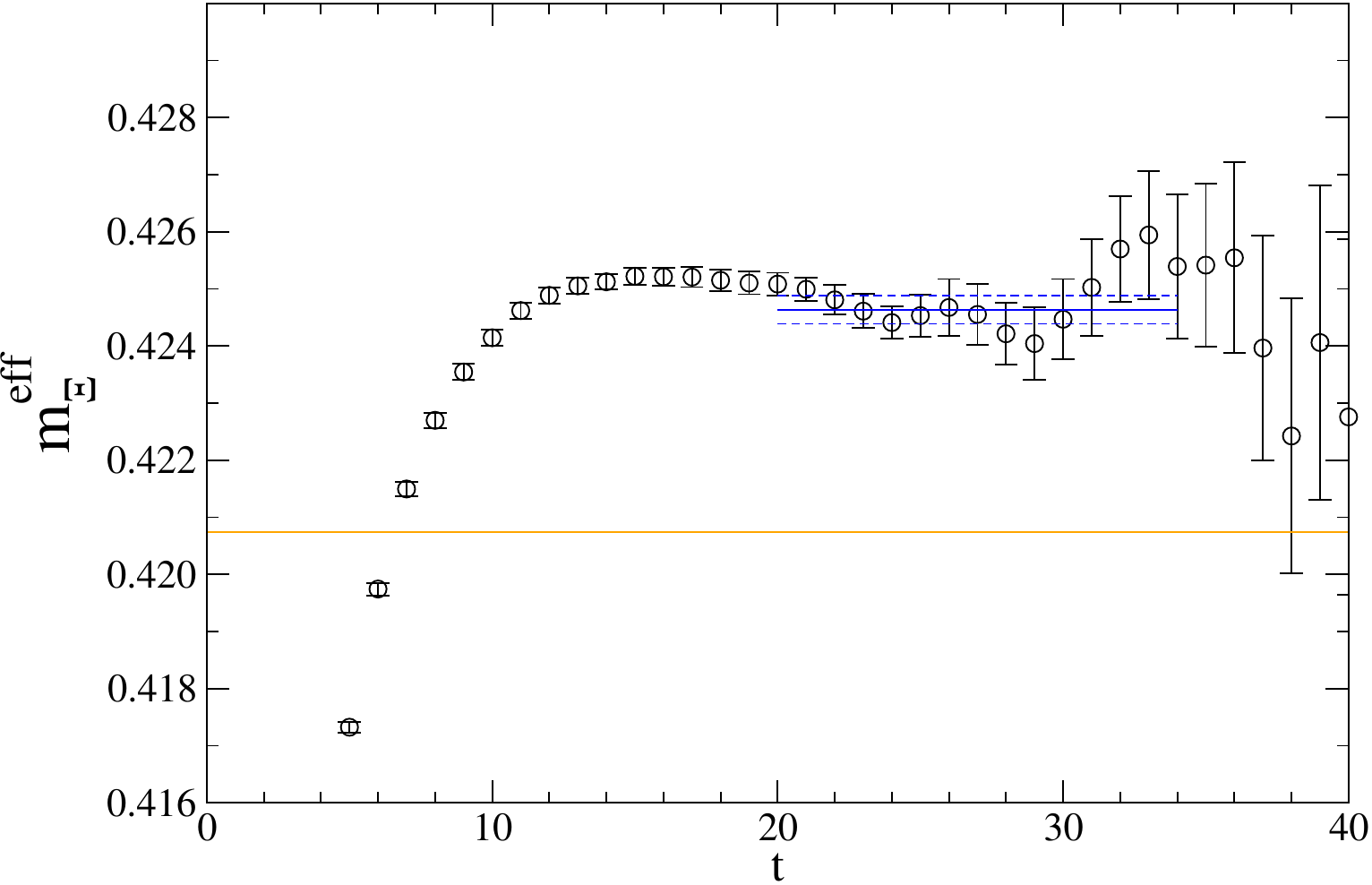}\\
\end{center}
\vspace{-5mm}
\caption{Effective masses for $N$, $\Lambda$, $\Sigma$ and $\Xi$ (from top to bottom) baryons on the L160 lattice at $\beta=2.00$ with $(\kappa_{\rm ud},\kappa_{\rm s})=(0.125814,0.124925)$. The horizontal line denotes the pseudoexperimental mass.}
\label{fig:effmass_8b_b200}
\end{figure}

\begin{figure}[t!]
\begin{center}
\includegraphics[width=70mm,angle=0]{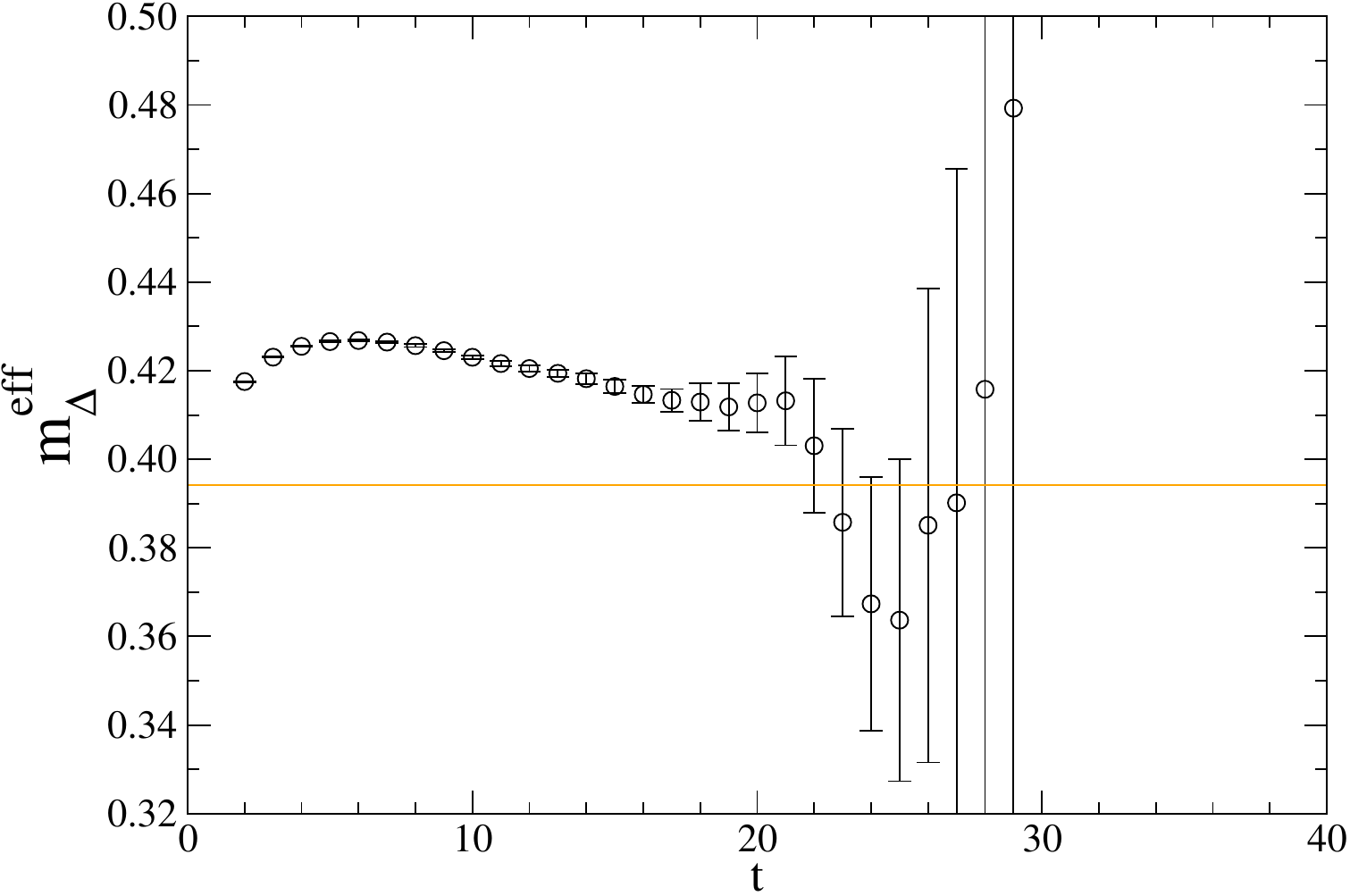}\\
\includegraphics[width=70mm,angle=0]{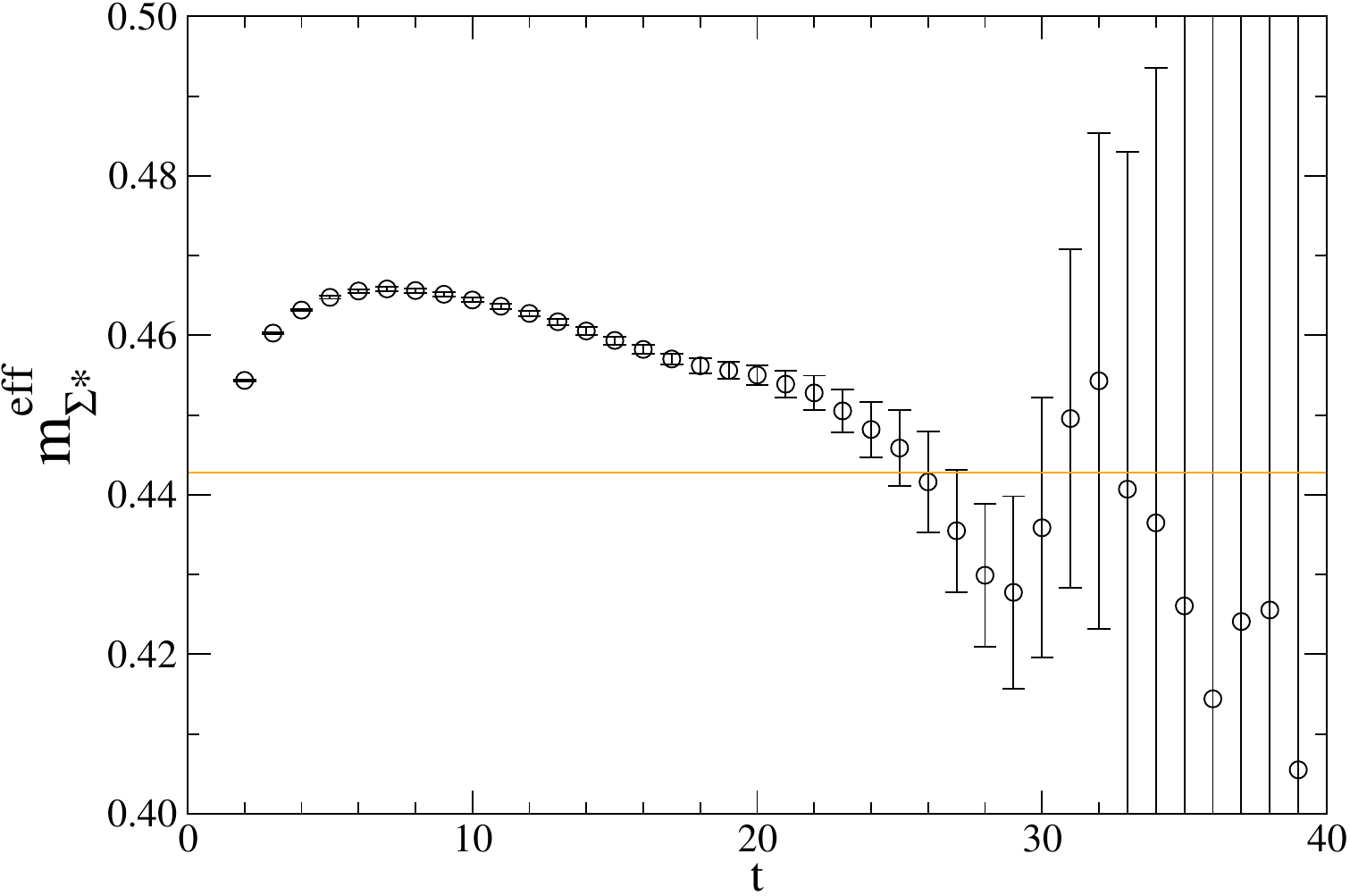}\\
\includegraphics[width=70mm,angle=0]{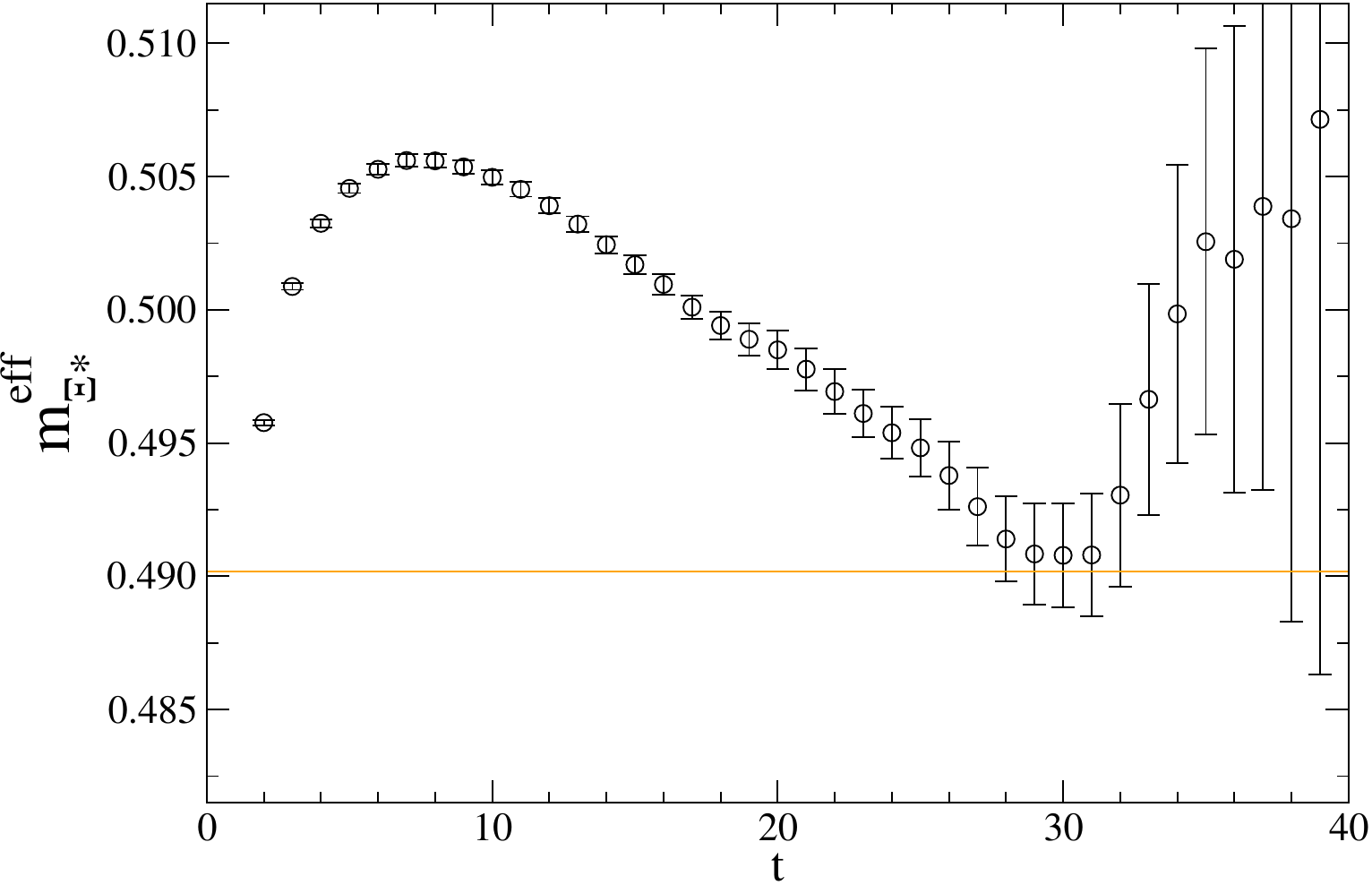}\\
\includegraphics[width=70mm,angle=0]{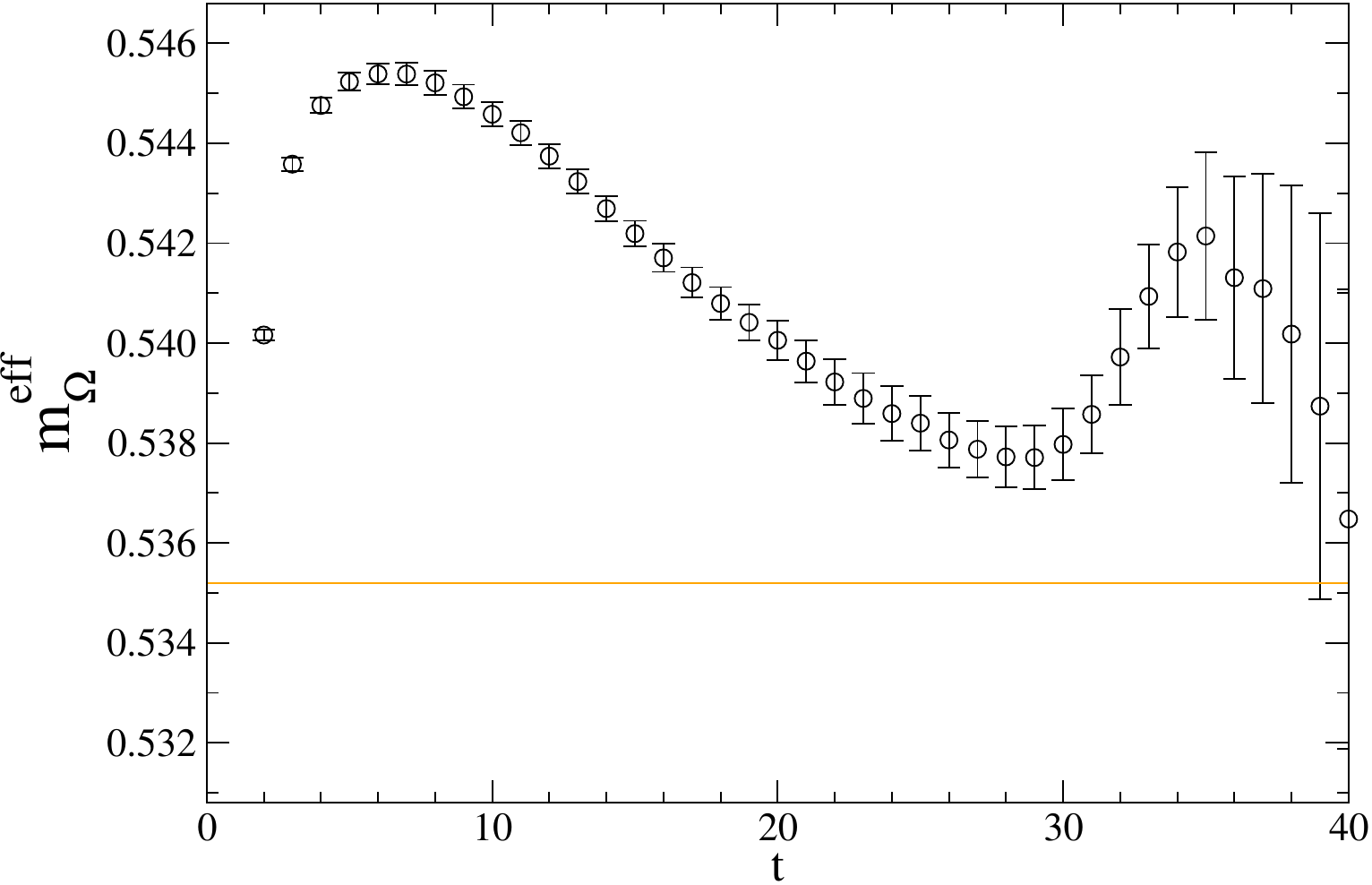}\\
\end{center}
\vspace{-5mm}
\caption{Effective masses for $\Delta$, $\Sigma^*$, $\Xi^*$ and $\Omega$ (from top to bottom) baryon channels on the L160 lattice at $\beta=2.00$ with $(\kappa_{\rm ud},\kappa_{\rm s})=(0.125814,0.124925)$. The horizontal line denotes the pseudoexperimental resonance level for $\Delta$, $\Sigma^*$ and $\Xi^*$ and the pseudoexperimental mass for $\Omega$.}
\label{fig:effmass_10b_b200}
\end{figure}

We show the effective masses for the PS mesons, the vector mesons, the octet baryons, and the decuplet baryons on the L160 lattice at $\beta=2.00$ with $(\kappa_{\rm ud},\kappa_{\rm s})=(0.125814,0.124925)$ in Figs.~\ref{fig:effmass_ps_b200}, \ref{fig:effmass_v_b200}, \ref{fig:effmass_8b_b200}, and \ref{fig:effmass_10b_b200}, respectively.
We find that qualitative features are similar to the $\beta=1.82$ case: good plateaus for the PS mesons and gradual decreasing behavior for the vector mesons and the decuplet baryons including the $\Omega$ baryon. We extract the masses for the PS meson and the octet baryon by applying a single exponential fit to their propagators. The results are presented in Table~\ref{tab:hmass} together with the AWI quark masses $m_{\rm ud,s}$ and the decay constants. 

The physical ud and s quark masses and the lattice cutoff are determined with the Taylor expansion of $m_\pi^2$, $m_K^2$, and $m_\Xi$ given in Eq.~(\ref{eq:hmass_taylor}), where we use the values of $\partial X/\partial m_{\rm ud}$ and $\partial X/\partial m_{\rm s}$ in Table~\ref{tab:fit_64} determined on the L64 lattice at $\beta=1.82$. Note that $\partial X/\partial m_{\rm ud}$ and $\partial X/\partial m_{\rm s}$ are dimensionful for $X=m_\pi^2$ and  $m_K^2$ so that the lattice spacing ratio of $\beta=2.00$ to 1.82 is additionally required. 
We present the results for the physical quark masses and the lattice cutoff in Table~\ref{tab:hmass_ph_A}. We also  obtain the PS meson decay constants and the octet baryon masses at the physical point using the formula of Eq.~(\ref{eq:hmass_taylor}) with the coefficients in Table~\ref{tab:fit_64}. Their results are given in Table~\ref{tab:hmass_ph_A}.

\begin{figure}[h]
\begin{center}
\includegraphics[width=75mm,angle=0]{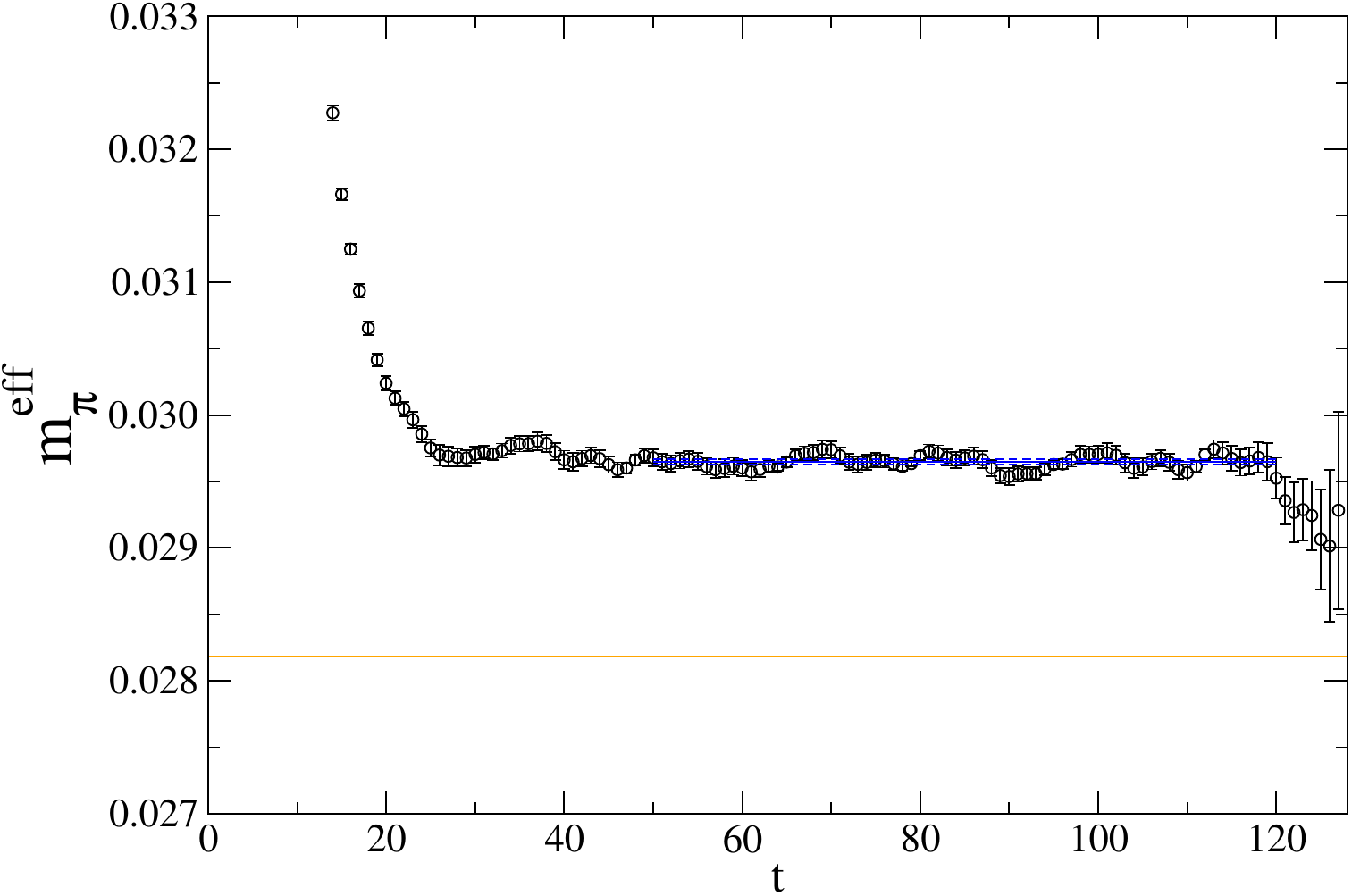}\\
\includegraphics[width=75mm,angle=0]{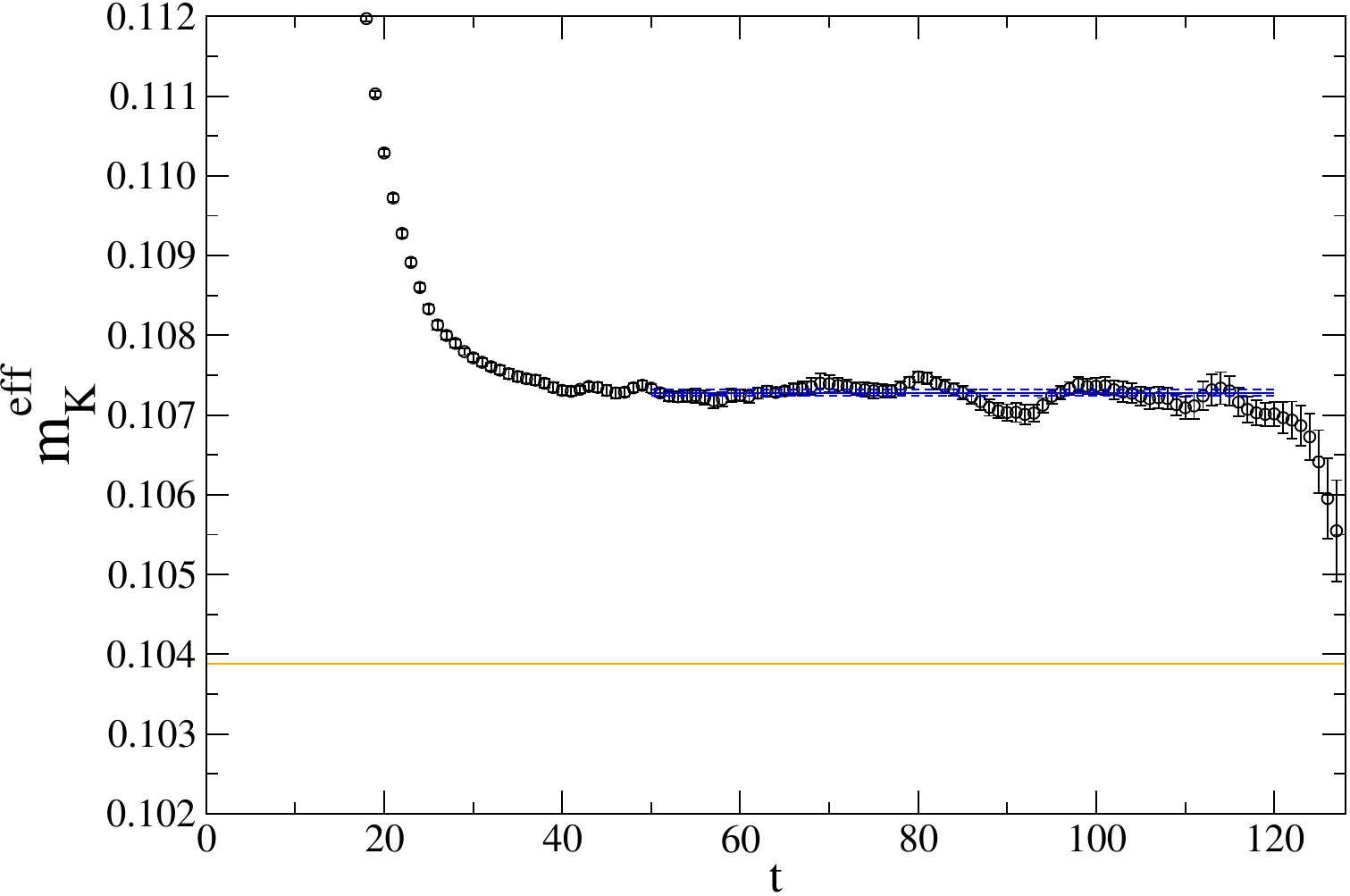}
\end{center}
\vspace{-5mm}
\caption{Same as Fig.~\ref{fig:effmass_ps_b200} but on the L256 lattice at $\beta=2.20$ with $(\kappa_{\rm ud},\kappa_{\rm s})=(0.1254872,0.1249349)$.}
\label{fig:effmass_ps_b220}
\end{figure}

\begin{figure}[h]
\begin{center}
\includegraphics[width=70mm,angle=0]{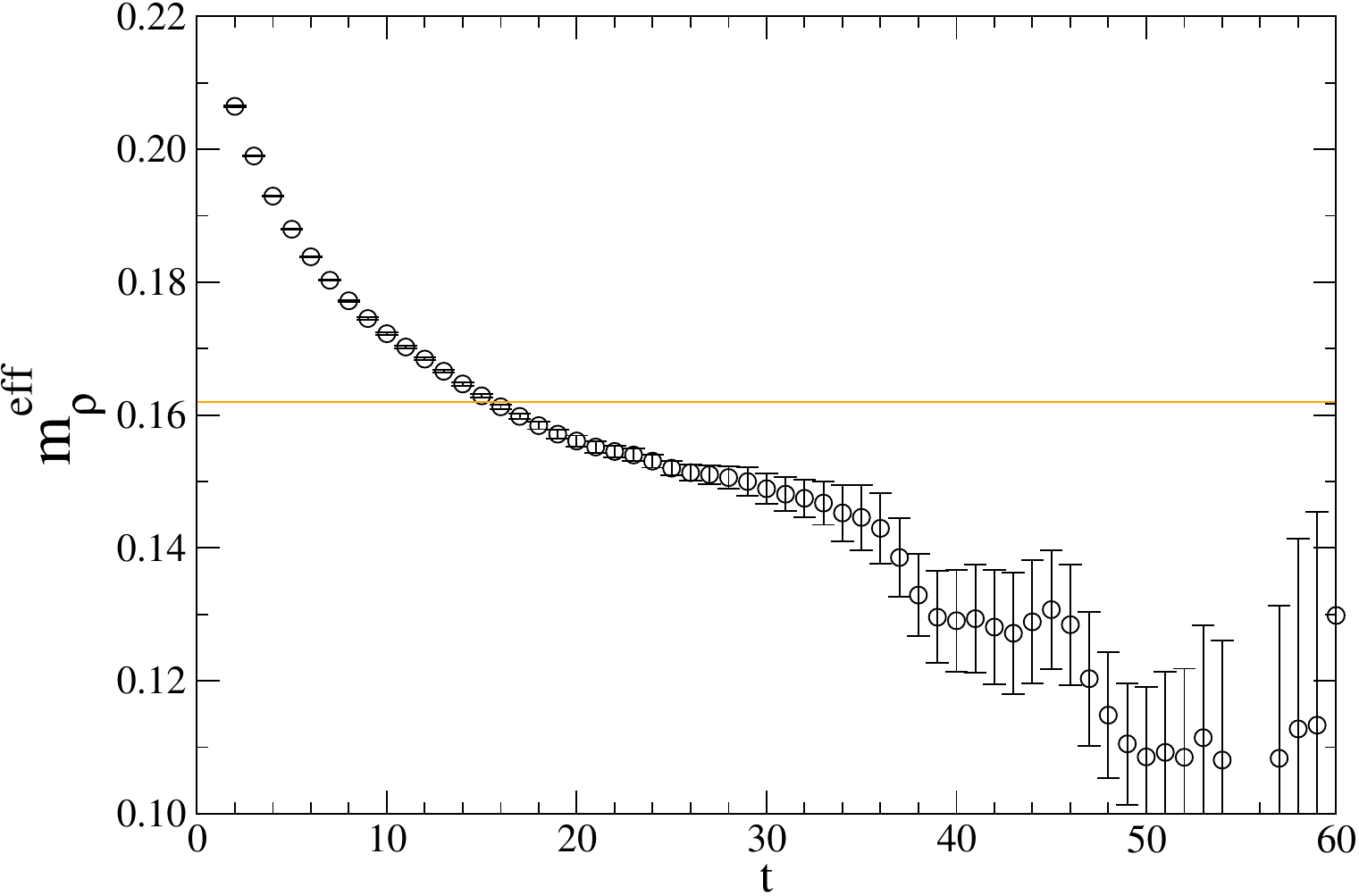}\\
\includegraphics[width=70mm,angle=0]{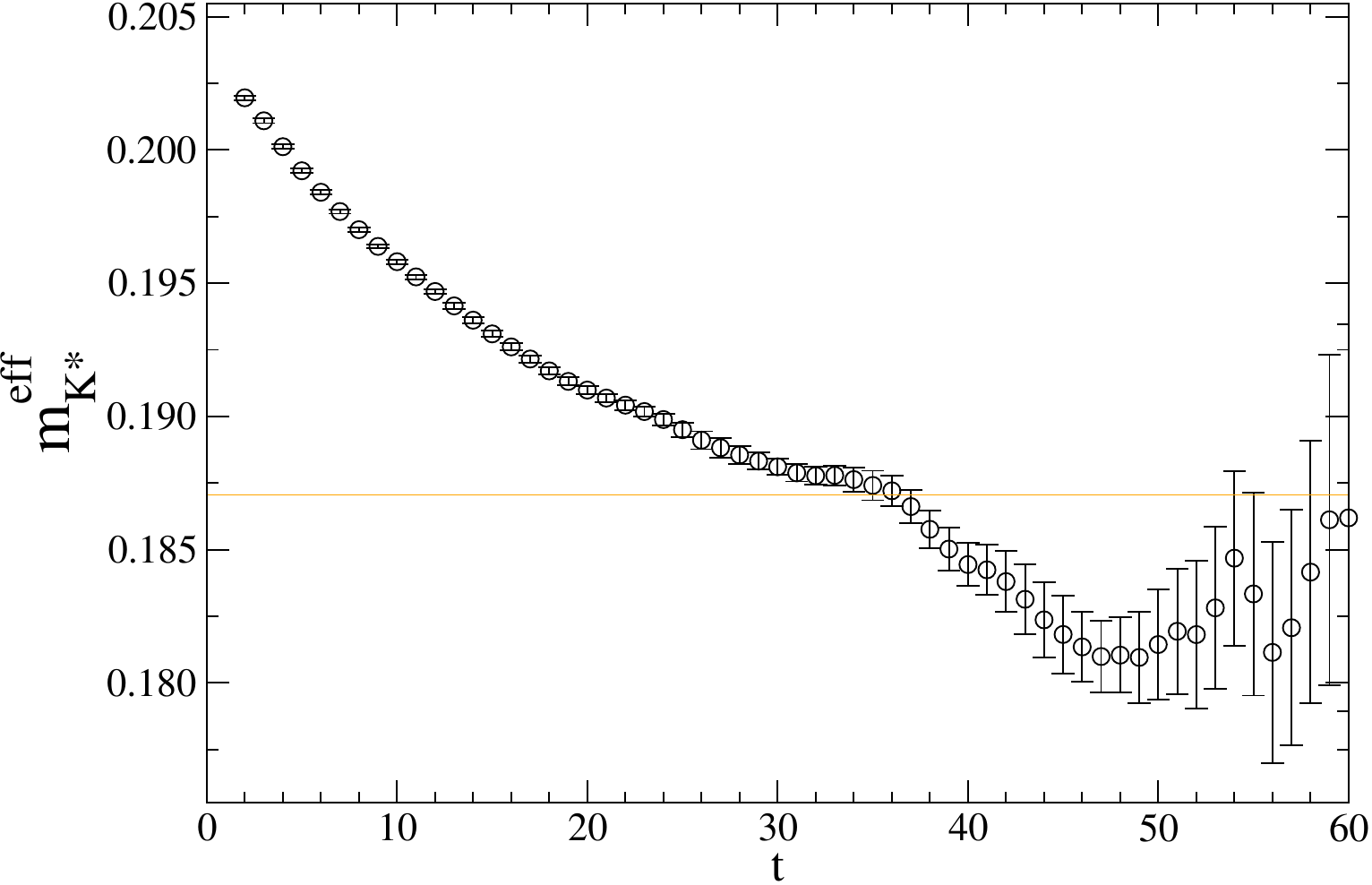}\\
\includegraphics[width=70mm,angle=0]{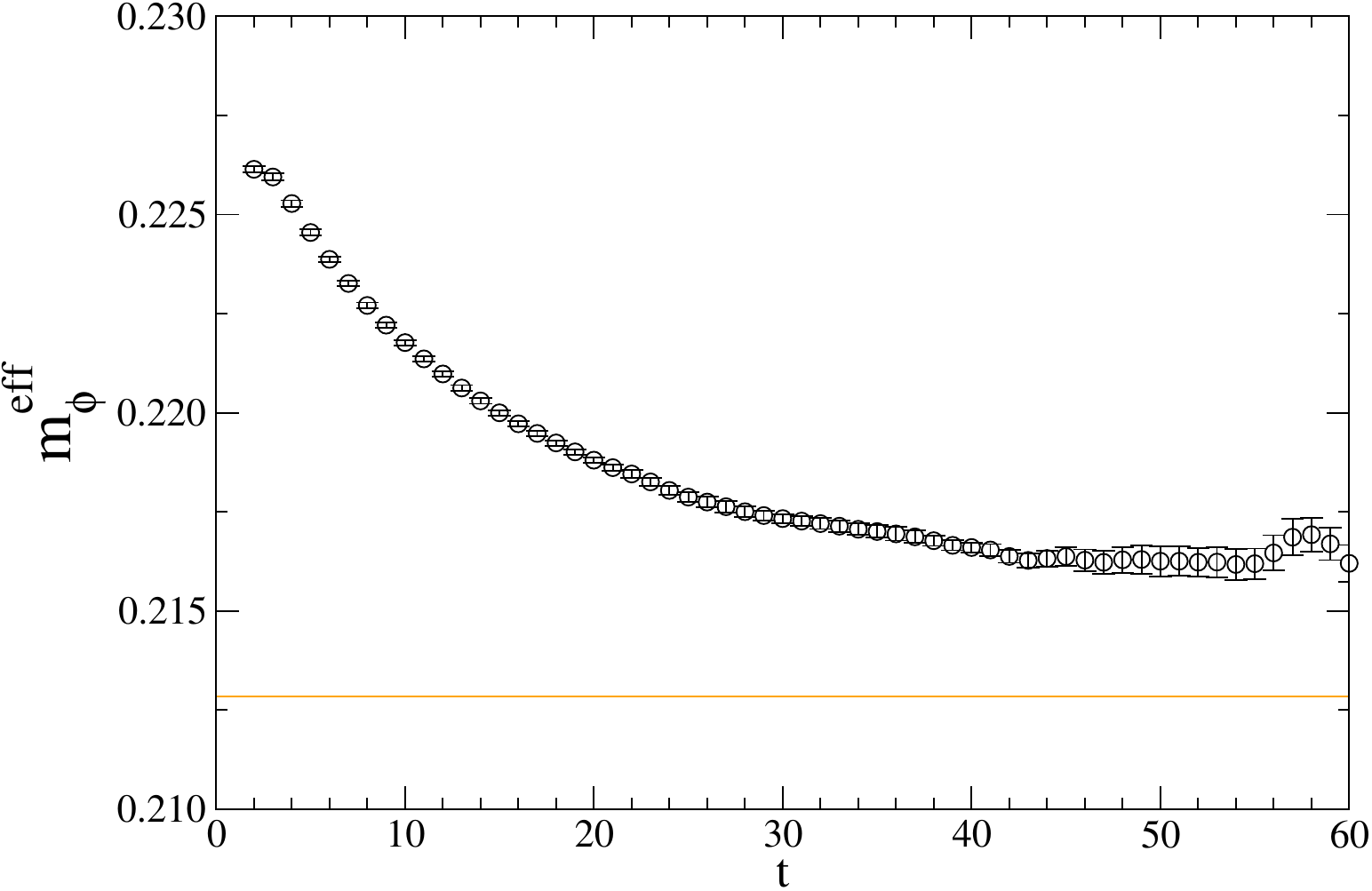}\\
\end{center}
\vspace{-5mm}
\caption{Same as Fig.~\ref{fig:effmass_v_b200} but on the L256 lattice at $\beta=2.20$ with $(\kappa_{\rm ud},\kappa_{\rm s})=(0.1254872,0.1249349)$.}
\label{fig:effmass_v_b220}
\end{figure}

\begin{figure}[t!]
\begin{center}
\includegraphics[width=70mm,angle=0]{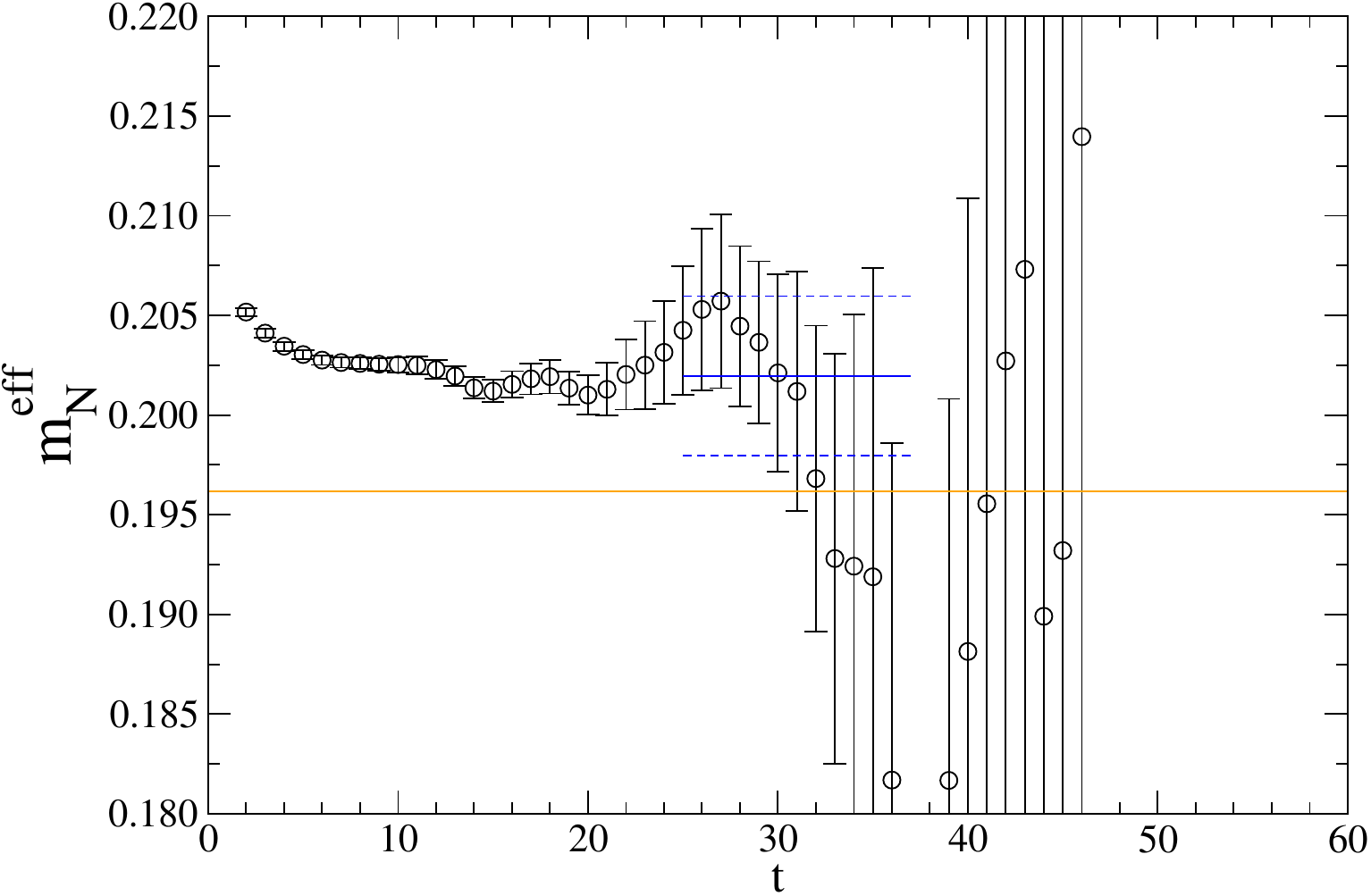}\\
\includegraphics[width=70mm,angle=0]{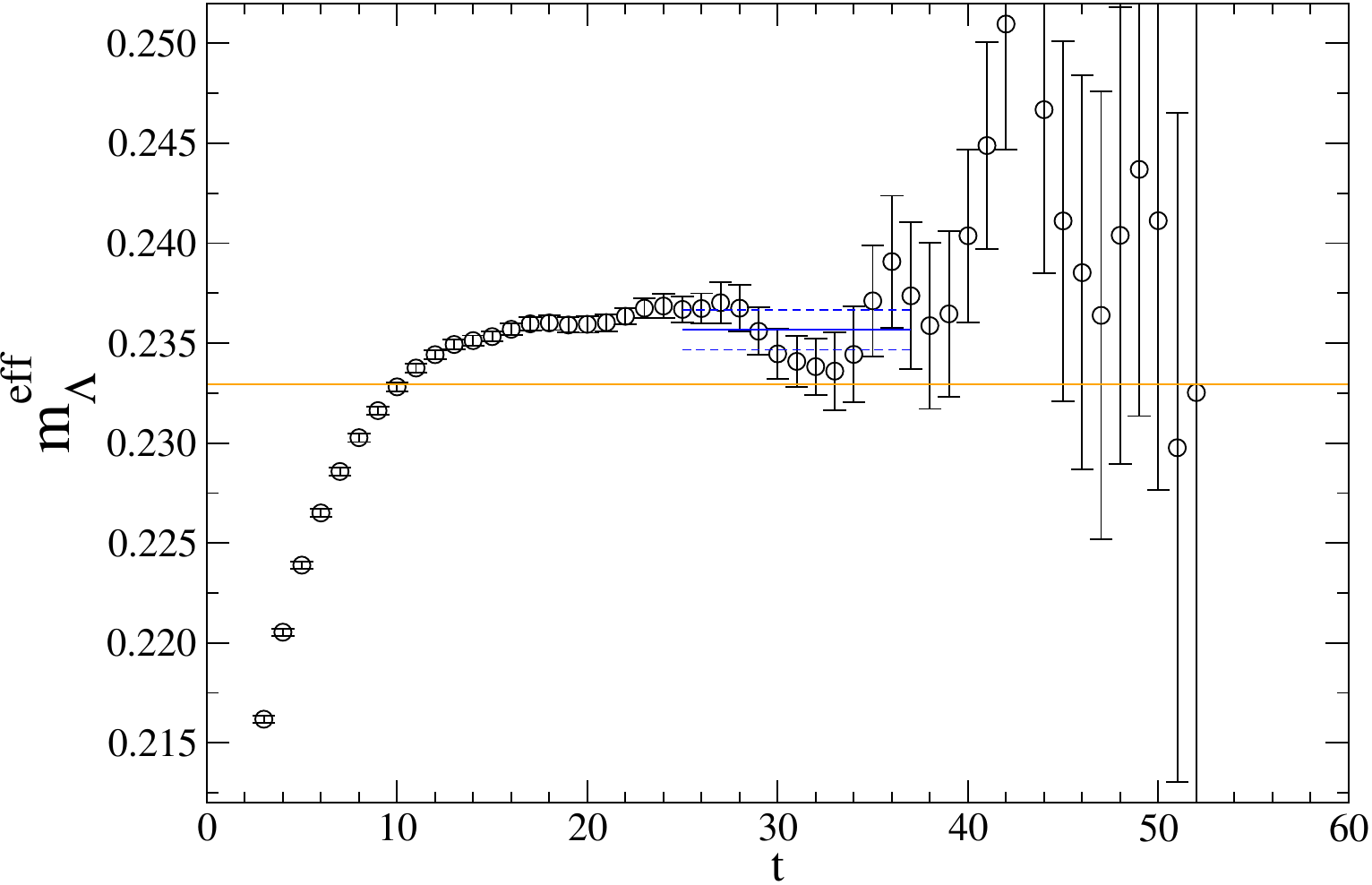}\\
\includegraphics[width=70mm,angle=0]{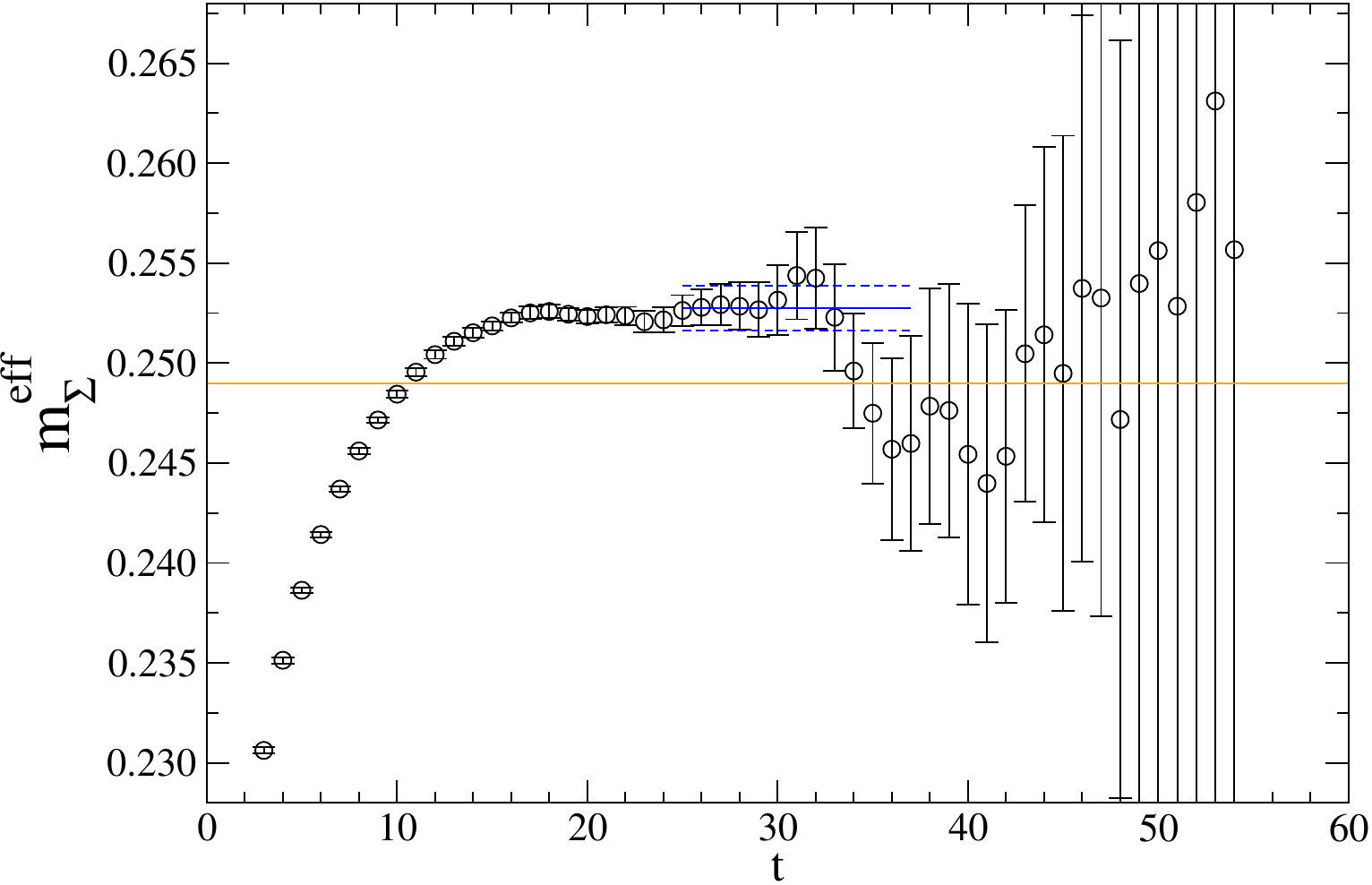}\\
\includegraphics[width=70mm,angle=0]{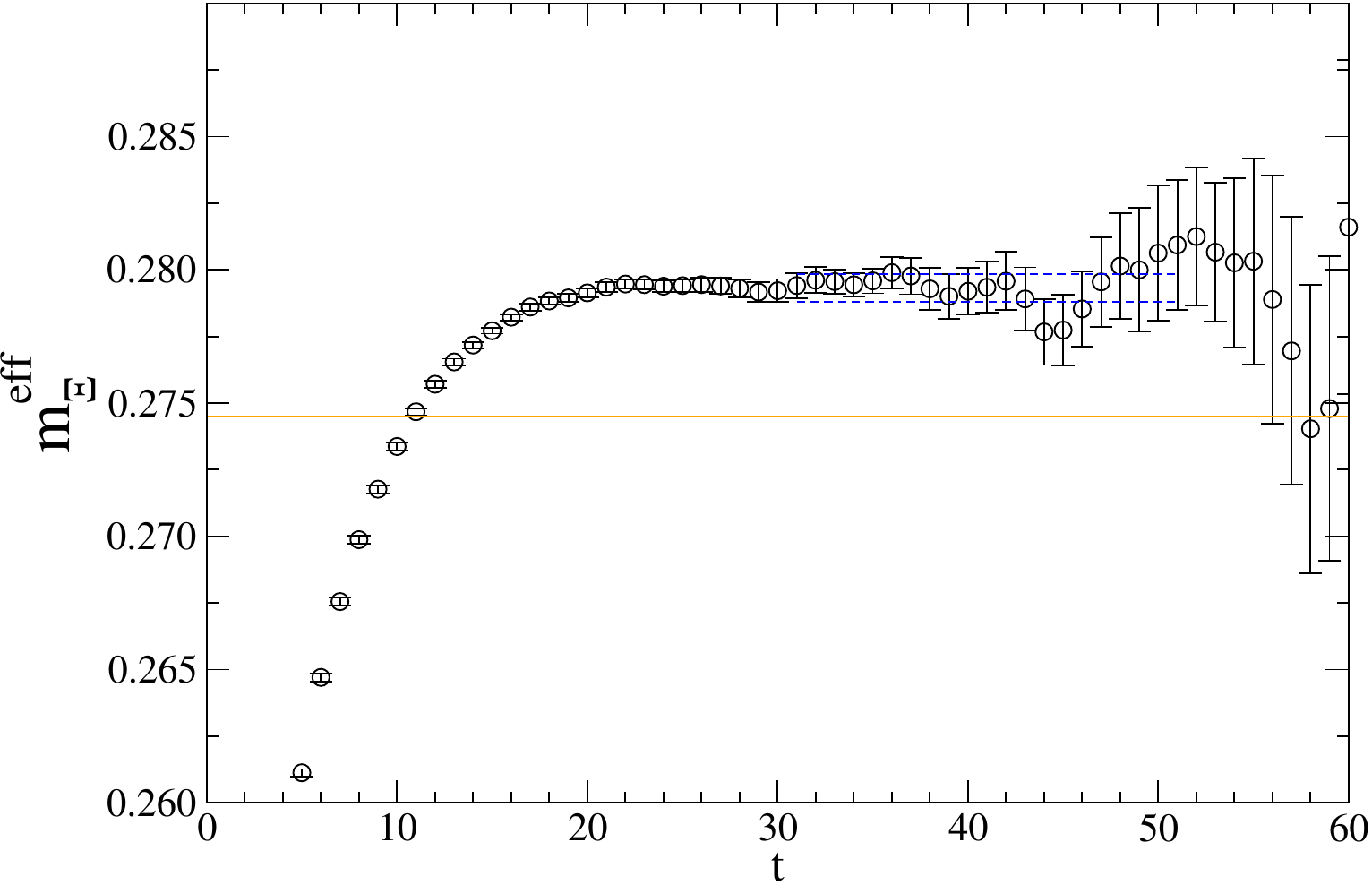}\\
\end{center}
\vspace{-5mm}
\caption{Same as Fig.~\ref{fig:effmass_8b_b200} but on the L256 lattice at $\beta=2.20$ with $(\kappa_{\rm ud},\kappa_{\rm s})=(0.1254872,0.1249349)$.}
\label{fig:effmass_8b_b220}
\end{figure}

\begin{figure}[t!]
\begin{center}
\includegraphics[width=70mm,angle=0]{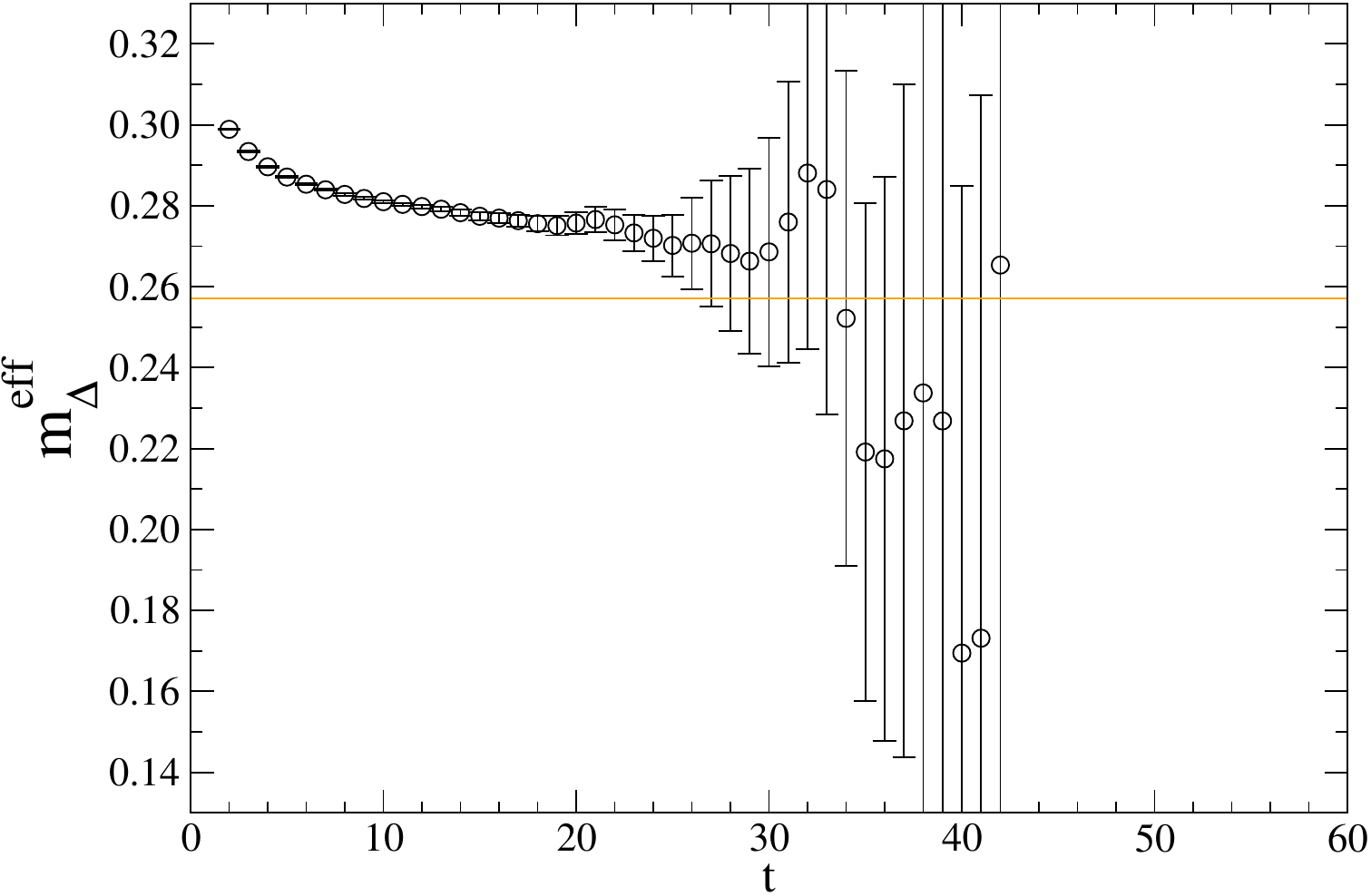}\\
\includegraphics[width=70mm,angle=0]{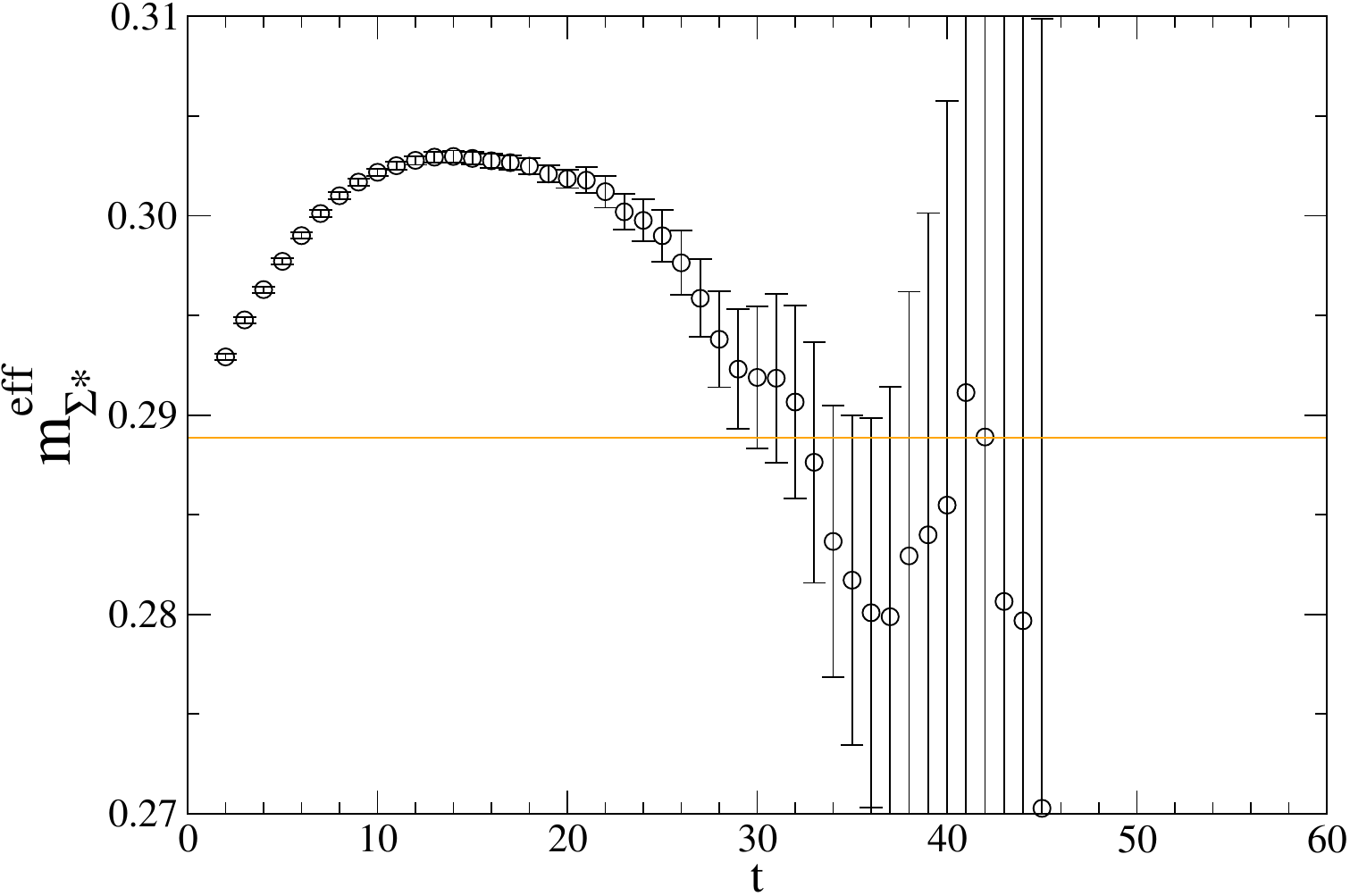}\\
\includegraphics[width=70mm,angle=0]{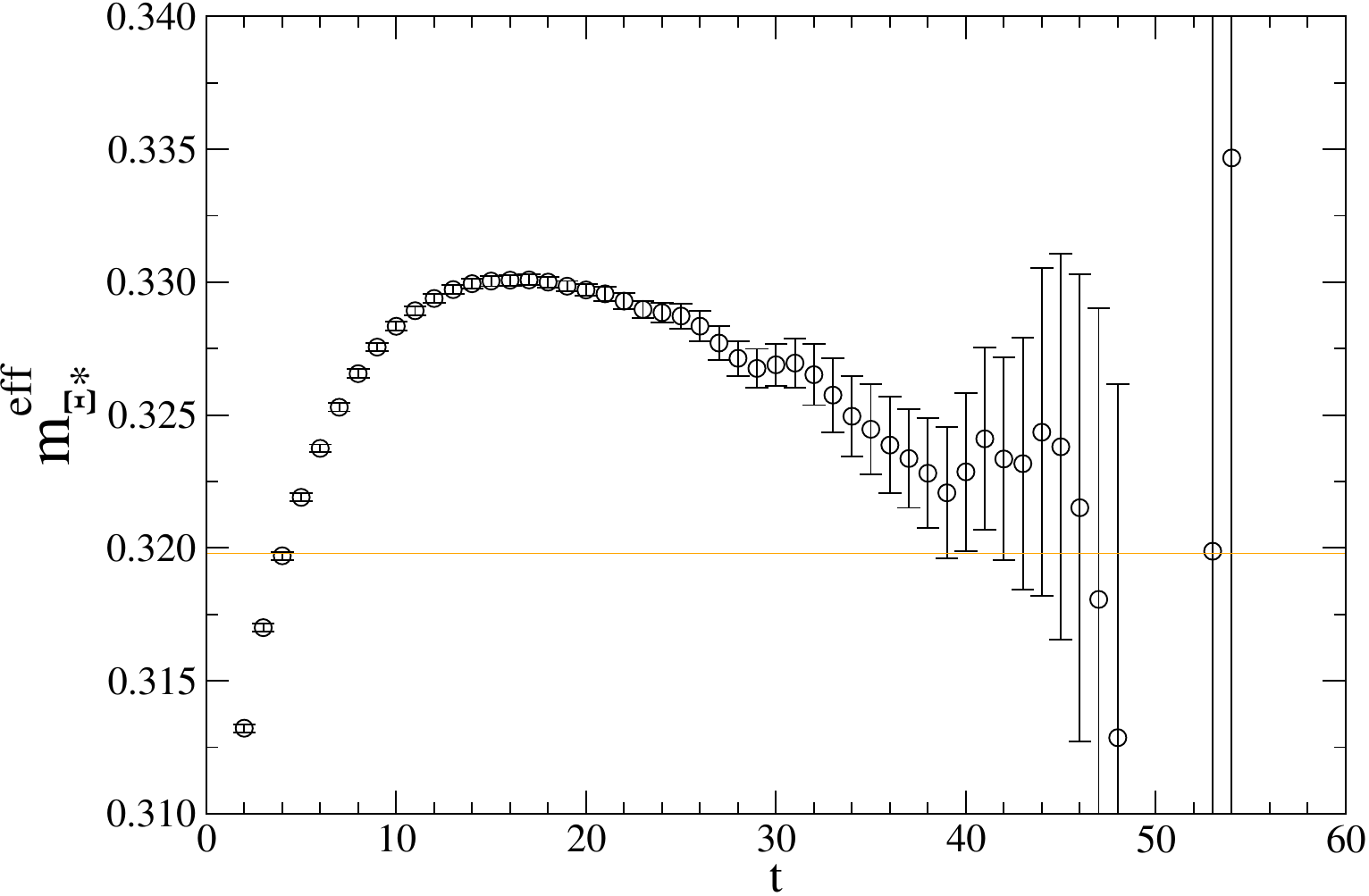}\\
\includegraphics[width=70mm,angle=0]{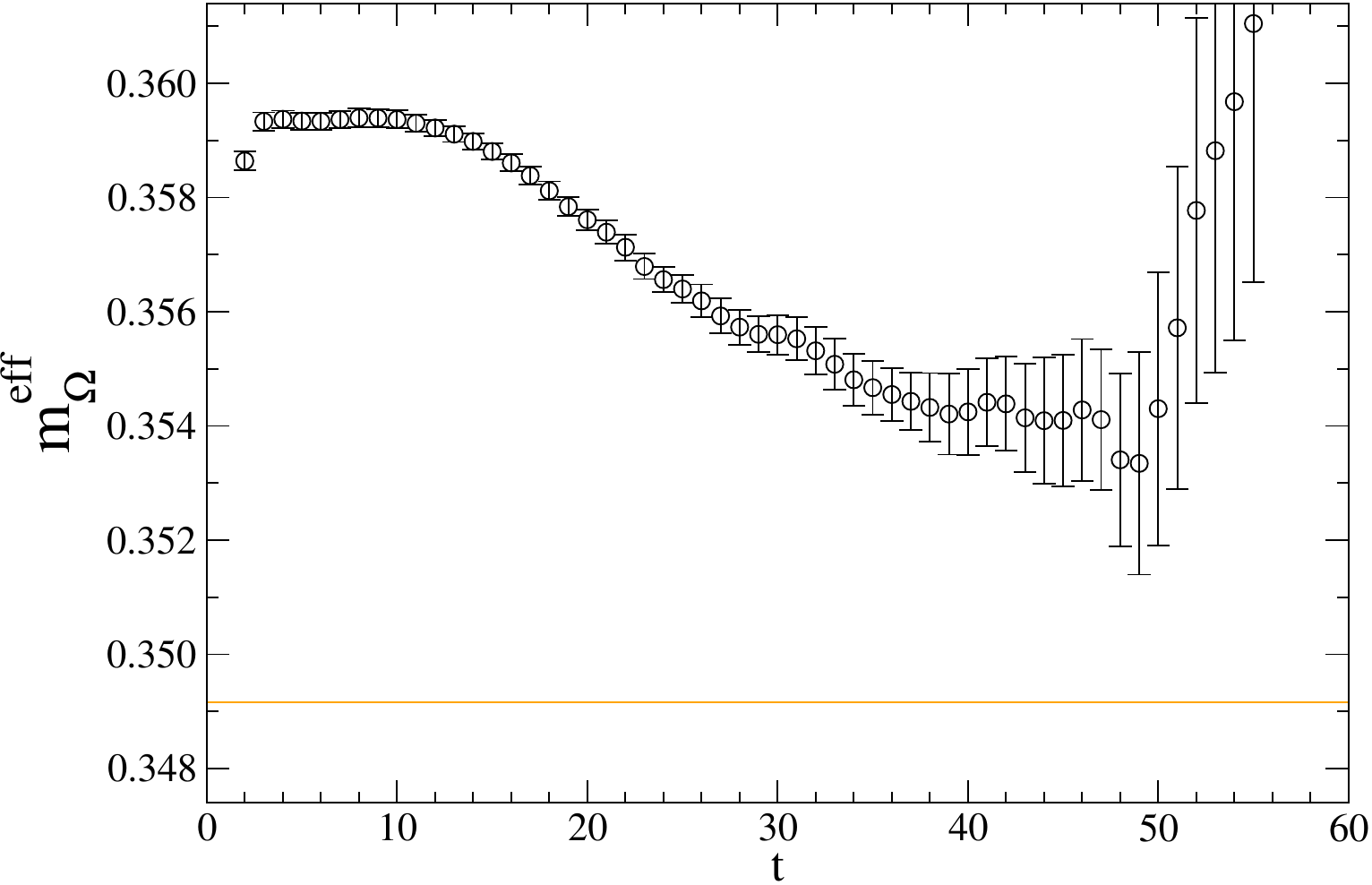}\\
\end{center}
\vspace{-5mm}
\caption{Same as Fig.~\ref{fig:effmass_10b_b200} but on the L256 lattice at $\beta=2.20$ with $(\kappa_{\rm ud},\kappa_{\rm s})=(0.1254872,0.1249349)$.}
\label{fig:effmass_10b_b220}
\end{figure}

The effective masses on the L256 lattice at $\beta=2.20$ with $(\kappa_{\rm ud},\kappa_{\rm s})=(0.1254872,0.1249349)$ are shown in Figs.~\ref{fig:effmass_ps_b220}, \ref{fig:effmass_v_b220}, \ref{fig:effmass_8b_b220}, and \ref{fig:effmass_10b_b220}. We again observe similar features to the $\beta=1.82$ and 2.00 cases, though the magnitude of the statistical error becomes somewhat larger. The fit results for the hadron masses including the AWI quark masses and the PS meson decay constants are given in Table~\ref{tab:hmass}.
Employing the same strategy as in the $\beta=2.00$ case we determine the physical quark masses, the lattice cutoff, and the PS meson decay constants and the octet baryon masses at the physical point, whose values are listed in Table~\ref{tab:hmass_ph_A}.

\section{Finite size effect due to fixed topological sectors}
\label{sec:topologicalcharge}

\begin{figure}[h]
\begin{center}
\includegraphics[width=70mm,angle=0]{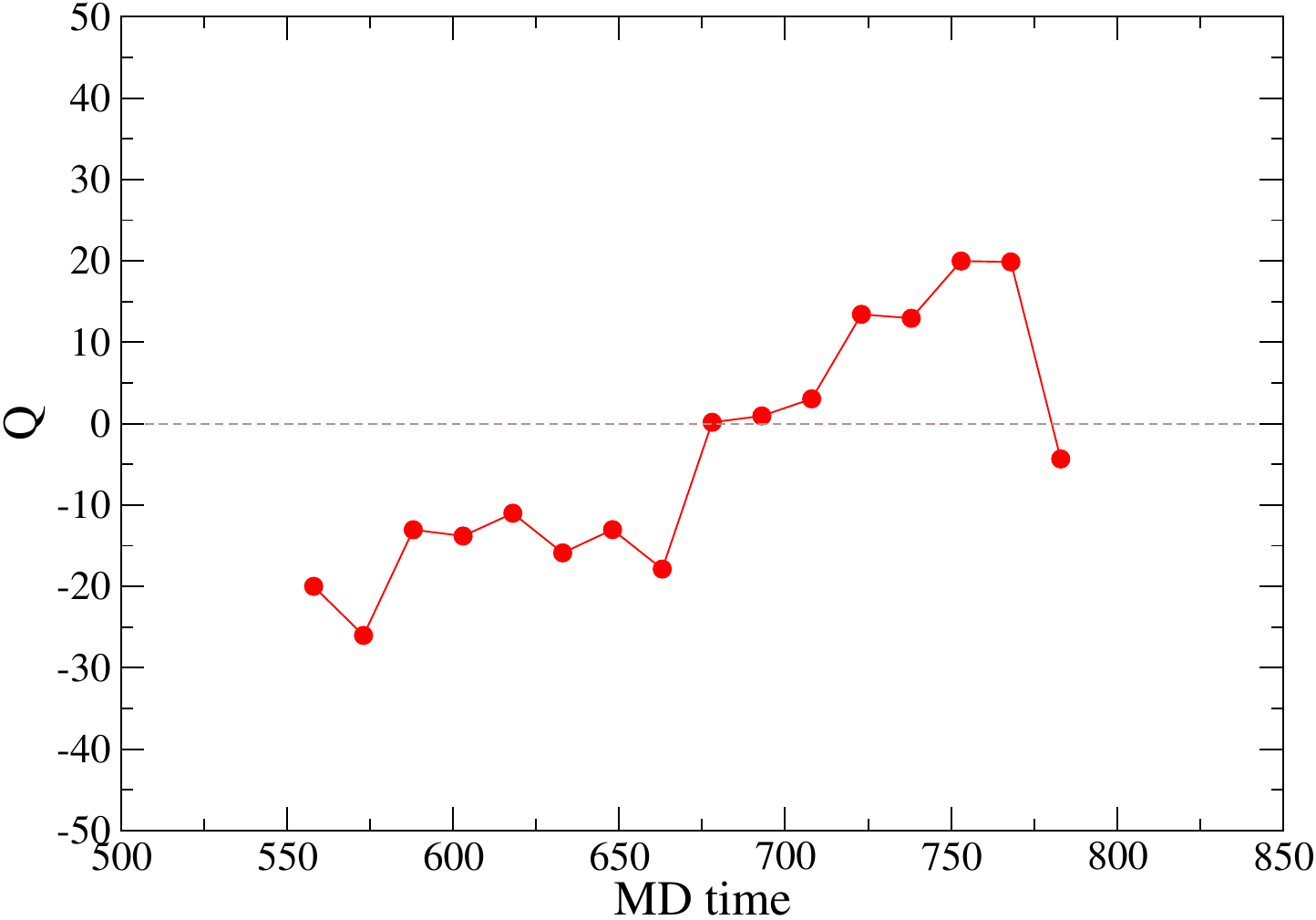}\\
\end{center}
\vspace{-5mm}
\caption{Monte Carlo history of topological charge on the L256 lattice as a function of MD time.}
\label{fig:tcharge_L256}
\end{figure}

It is widely known that the topological fluctuations tend to be suppressed toward the continuum limit in the HMC algorithm. Figure~\ref{fig:tcharge_L256} shows the Monte Carlo history of the topological charge $Q$ with a geometrical definition on the L256 lattice at the finest lattice spacing as a function of the MD time which corresponds to the trajectory multiplied by $\tau$. The topological charge is measured with the use of the APE smearing ($\alpha_{\rm APE}=0.1$ and $n_{\rm APE}=500$) method~\cite{APE:1987ehd}. This is the least frequent fluctuation of the topological charge among the three sets of PACS10 configurations, which may cause possible finite size effects. However, in principle the systematic effect due to a fixed topological sector vanishes in the infinite volume limit so that the large volume of the PACS10 configurations should reduce the finite size effects. 

We may give a rough estimate for the magnitude of the finite size effects on the pion mass on the L256 lattice following the formula given by Ref.~\cite{Brower:2003yx}:  
\ben
m_{\pi}^Q&=&m_\pi(0)+\frac{1}{2}m_\pi^{\prime\prime}(0)\left\{\frac{1}{\chi_t V}-\left(\frac{Q}{\chi_t V}\right)^2\right\}\nn\\
&&+O\left(\frac{1}{(\chi_t V)^2}\right),
\label{eq:m_q}
\een
where the $Q$-dependent effective mass $m_{\pi}^Q$ is related to the true mass $m_\pi(\theta=0)$ and its second derivative $m^{\prime\prime}_\pi(\theta=0)$ with respect to the $\theta$ parameter. $\chi_t=\langle Q^2 \rangle/V$ is the topological susceptibility. With the use of the leading order result of $m^{\prime\prime}_\pi(\theta=0)=-m_\pi(\theta=0)/8$ in the chiral perturbation theory~\cite{Brower:2003yx}, Eq.~(\ref{eq:m_q}) is rewritten as
\ben
\delta m_\pi&=&\frac{m_{\pi}^Q-m_\pi(0)}{m_\pi(0)}\\
&=&-\frac{1}{16}\left\{\frac{1}{\chi_t V}-\left(\frac{Q}{\chi_t V}\right)^2\right\}+O\left(\frac{1}{(\chi_t V)^2}\right)\nn
\een
with $\chi_t=m_\pi^2 f_\pi^2/(4N_f)$.  Using the numerical results for $m_\pi$, $f_\pi$ and $a^{-1}$ on the L256 lattice in Tables~\ref{tab:hmass} and \ref{tab:hmass_ph_A}, we obtain $\delta m_\pi=-0.000025$ in case that $|Q|=20$, which roughly amounts to the fluctuation amplitude in Fig.~\ref{fig:tcharge_L256}. The magnitude of $\delta m_\pi$ in our simulation is less than 0.00067 found in Table~\ref{tab:hmass} so that the topological charge effects on $m_\pi$ could be negligible. Note that other hadron masses have larger relative errors than the pion mass.

\section{Continuum limit for physical observables with convention A}
\label{sec:continuum}

\begin{figure}[h]
\begin{center}
\includegraphics[width=70mm,angle=0]{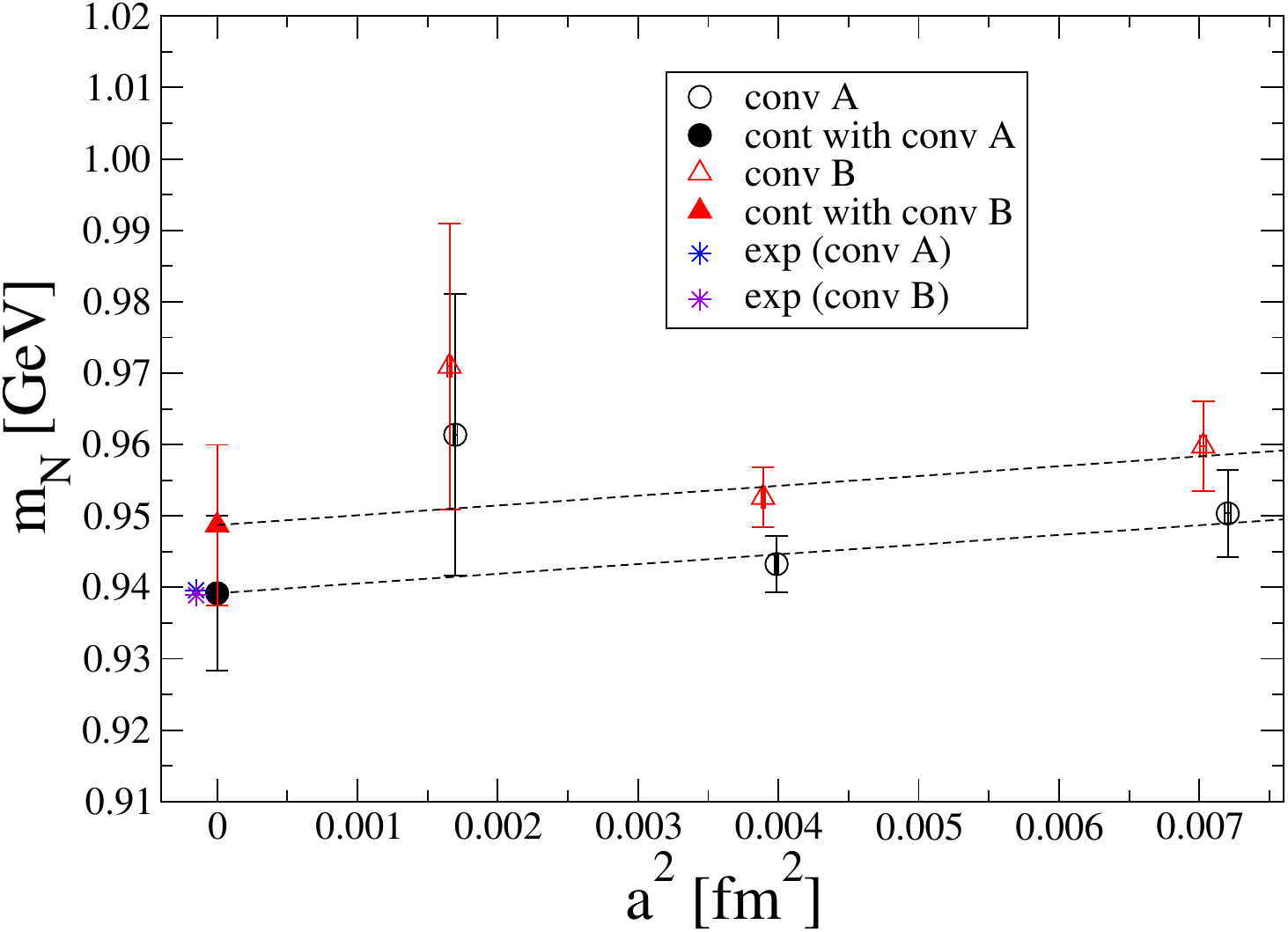}\\
\includegraphics[width=70mm,angle=0]{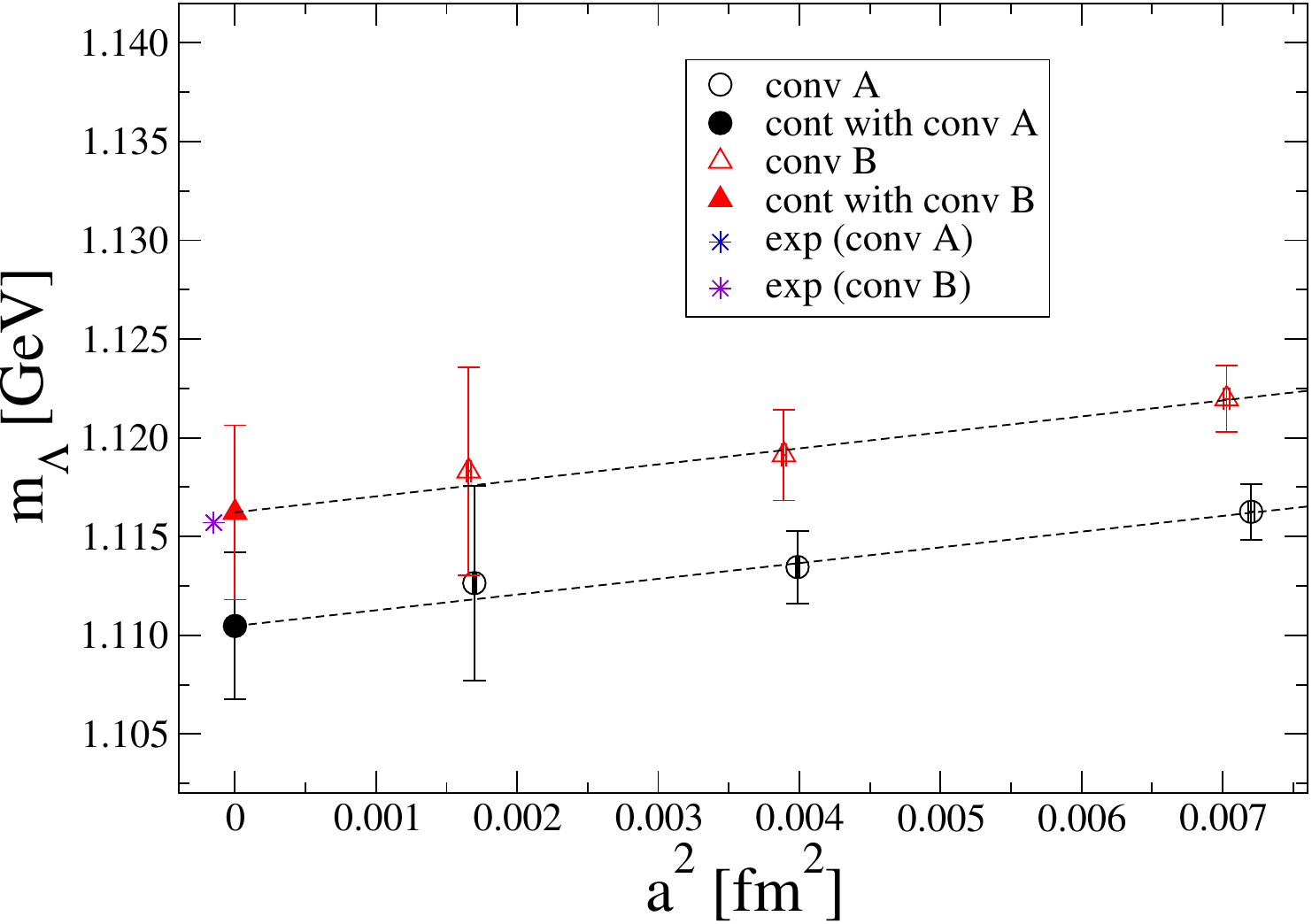}\\
\includegraphics[width=70mm,angle=0]{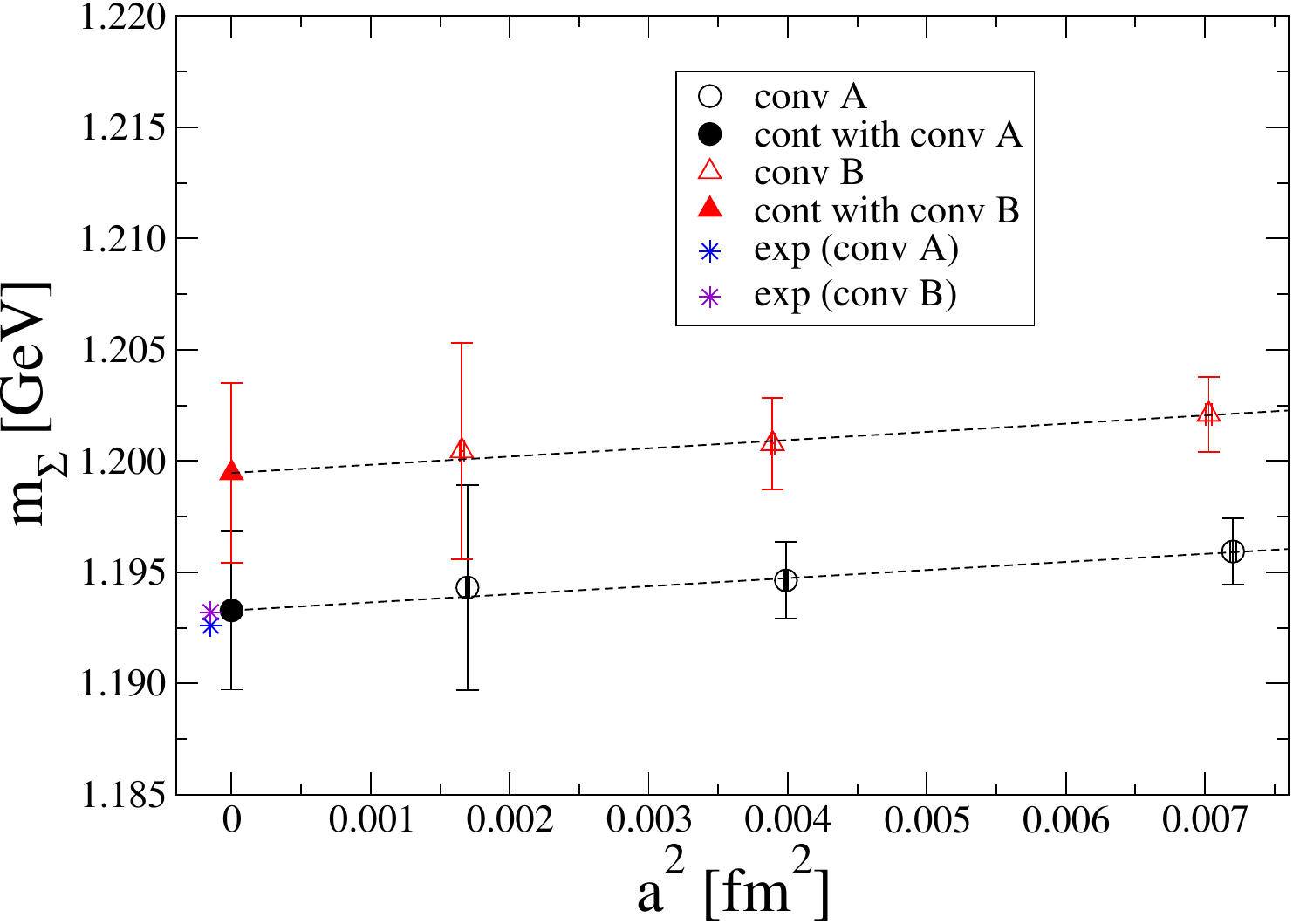}\\
\end{center}
\vspace{-5mm}
\caption{Continuum extrapolation of the masses for $N$, $\Lambda$, and $\Sigma$ (from top to bottom) baryons with conventions A (circle) and B (triangle). The asterisk denotes the pseudoexperimental values in Table~\ref{tab:pexp_isoqcd}.}
\label{fig:hmass_cont}
\end{figure}

 We plot the octet baryon masses in the physical unit as a function of $a^2$ in Fig.~\ref{fig:hmass_cont}, where the asterisk represents the pseudoexperimental values in Table~\ref{tab:pexp_isoqcd}. Since we assume the leading cutoff effect is $O(a^2)$ with the use of the nonperturbatively determined $c_{\rm SW}$, the continuum extrapolation of the octet baryon masses is performed with a linear function in terms of $a^2$.  The results in the continuum limit, whose numerical values are listed in Table~\ref{tab:hmass_ph_A}, show consistency with the pseudoexperimental values. 

\begin{figure}[h]
\begin{center}
\includegraphics[width=70mm,angle=0]{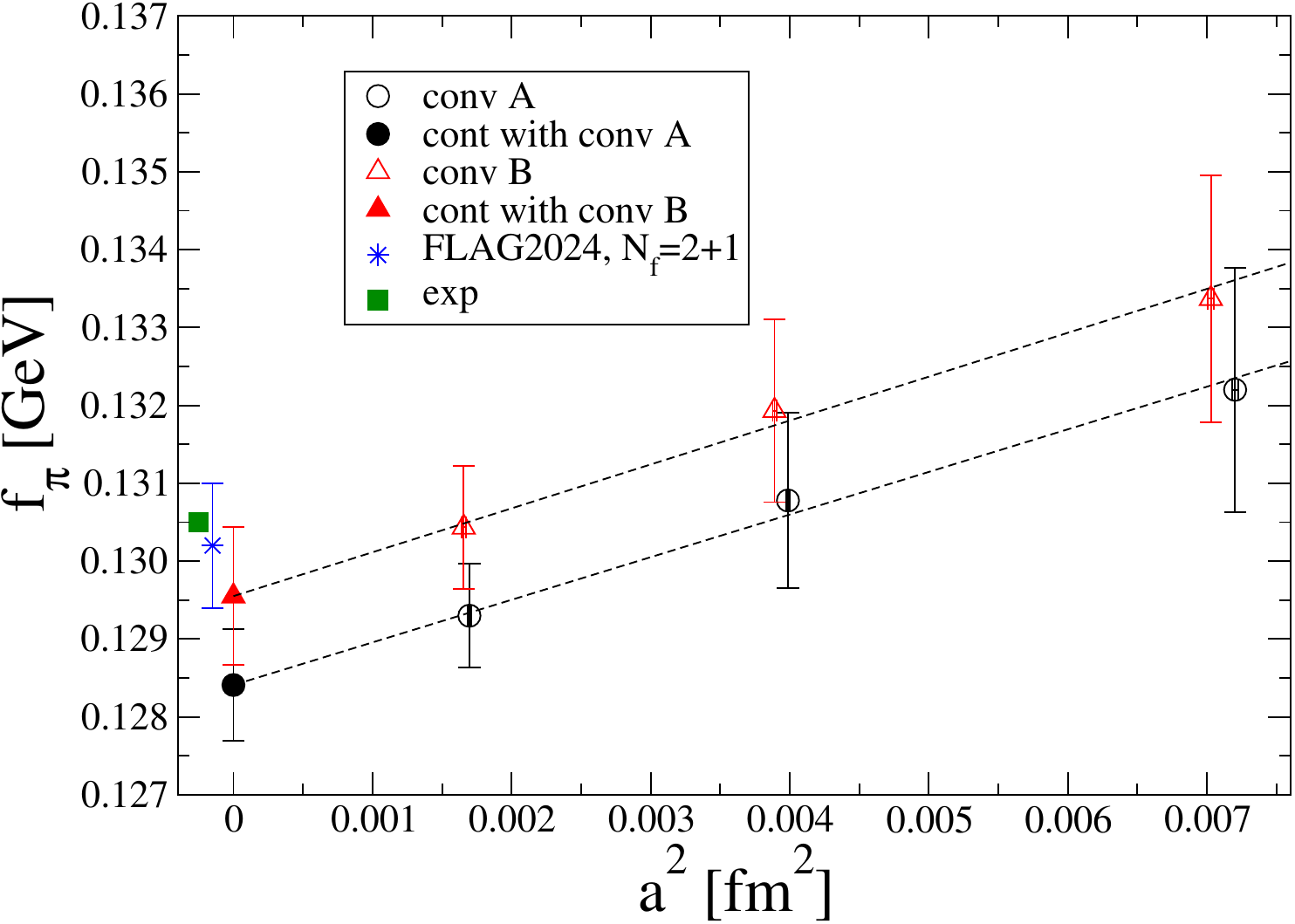}\\
\includegraphics[width=70mm,angle=0]{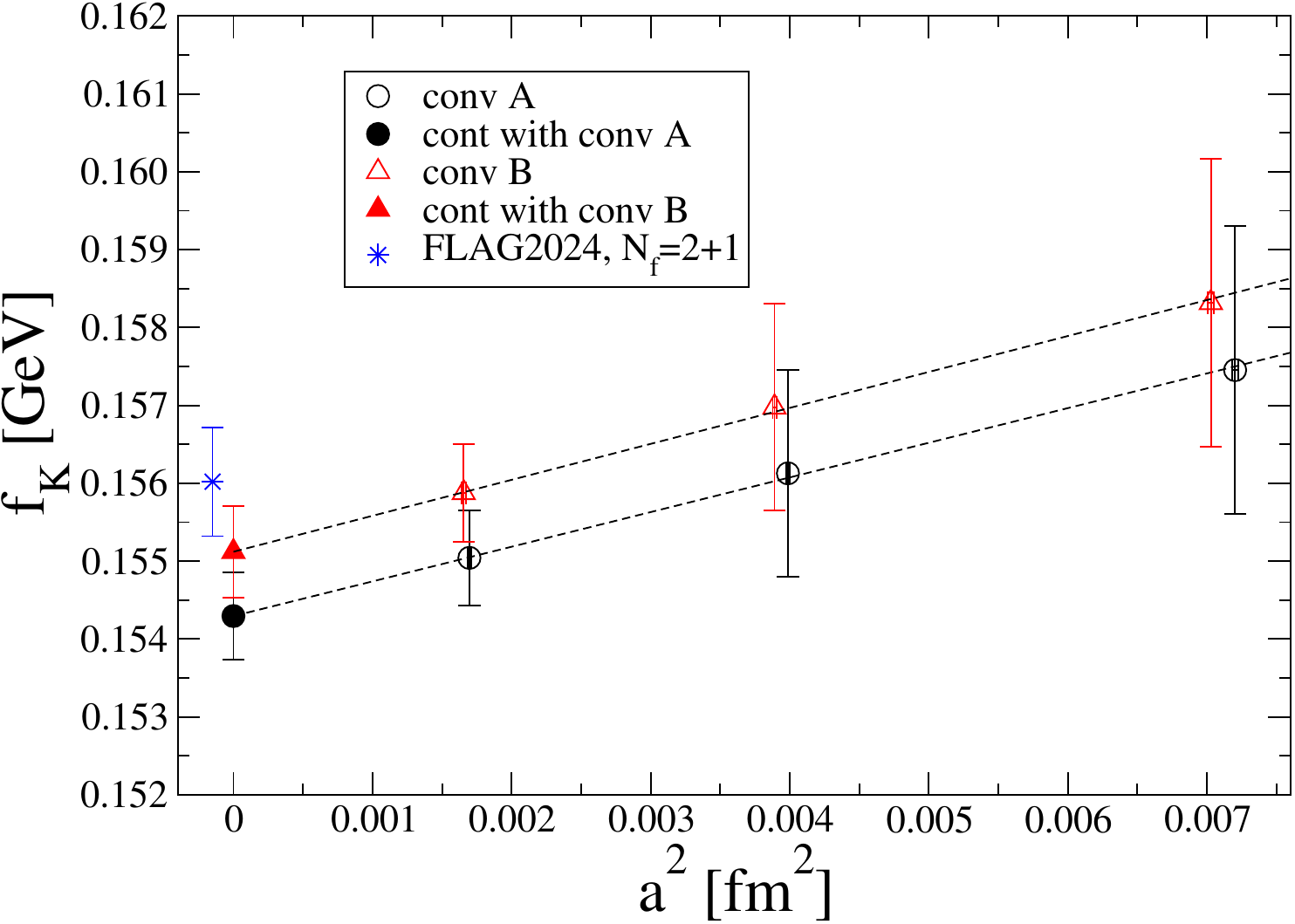}\\
\includegraphics[width=70mm,angle=0]{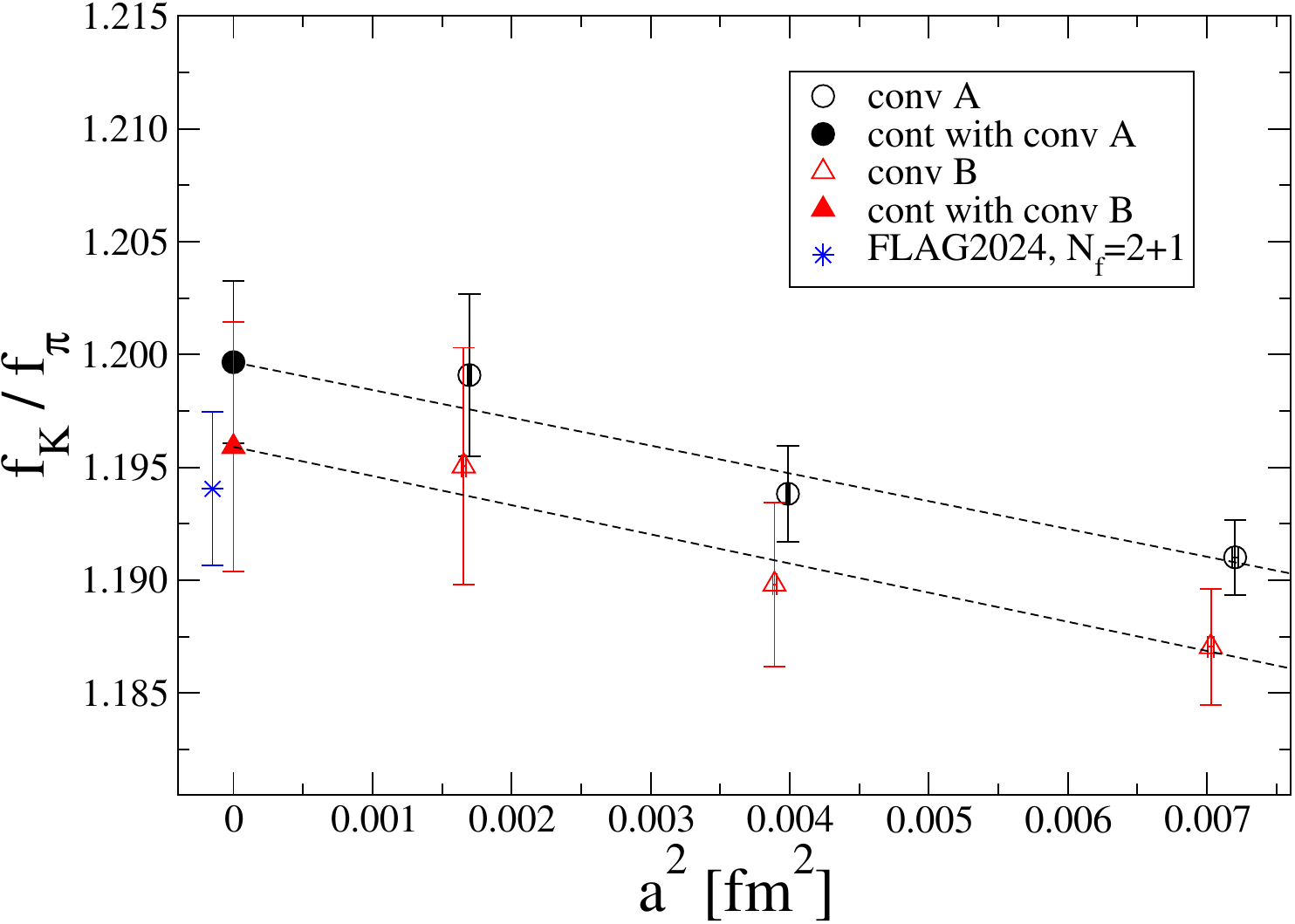}\\
\end{center}
\vspace{-5mm}
\caption{Continuum extrapolation of $f_\pi$, $f_K$, and $f_K/f_\pi$ (from top to bottom) with conventions A (circle) and B (triangle). The asterisk denotes the FLAG summary~\cite{nfzp-p5dn}.  We also plot the pseudoexperimental value of $f_\pi=0.1305$ GeV in the top panel.}
\label{fig:decayconst_cont}
\end{figure}

Figure~\ref{fig:decayconst_cont} shows the scaling property of the PS meson decay constants. Since the value of $c_A$, which is the $O(a)$ improvement factor for the local axial vector current, is consistent with zero~\cite{Taniguchi:2012gew}, we remove the leading cutoff effect of $O(a^2)$ with a linear extrapolation of the PS meson decay constants in term of $a^2$.
We find that both results for $f_\pi$ and $f_K$ in the continuum limit are about 1\% below the latest FLAG summary~\cite{nfzp-p5dn} beyond the error bars. On the other hand, the ratio of $f_K/f_\pi$ in the continuum limit is about 0.7\% larger than the latest FLAG summary~\cite{nfzp-p5dn}. Numerical values for $f_\pi$, $f_K$, and $f_K/f_\pi$ in the continuum limit are given in Table~\ref{tab:hmass_ph_A}.
The strong-isospin-breaking corrections to $f_\pi$, $f_K$ based on NLO SU(3) ChPT can be parametrized as $f_{\pi^\pm}=f_{\pi}$, $f_{K^\pm}=f_{K}\sqrt{1+\delta_{\rm SU(2)}}$, where $\delta_{\rm SU(2)}=-0.0041(6)$ estimated from Eq. (75) in \cite{nfzp-p5dn}. 
We add a 100\% error to the correction for $f_{K^\pm}$ and $f_{K^\pm}/f_{\pi^\pm}$ as the last error in Table~\ref{tab:hmass_ph_A}.

\begin{figure}[h]
\begin{center}
\includegraphics[width=70mm,angle=0]{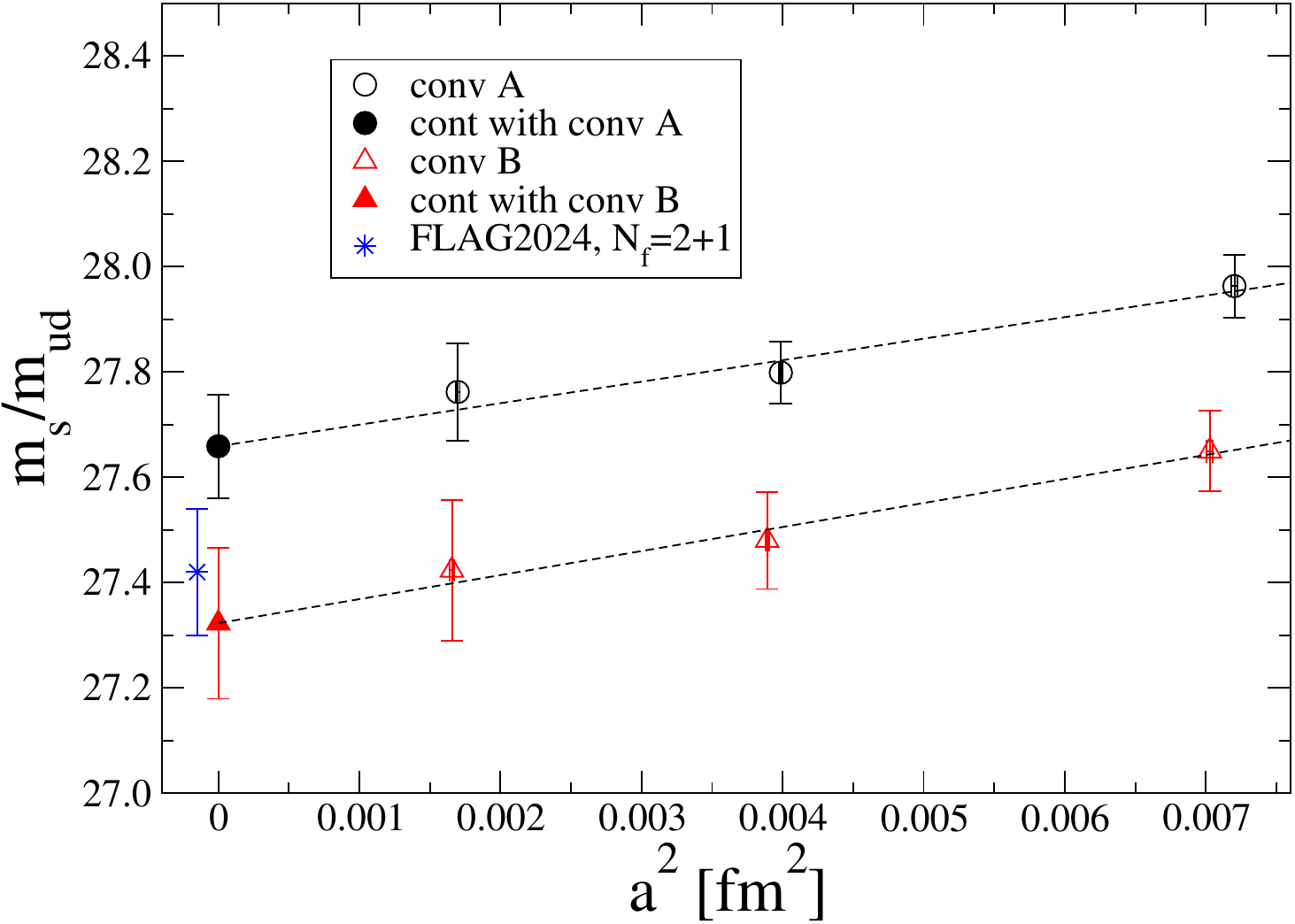}\\
\end{center}
\vspace{-5mm}
\caption{Continuum extrapolation of $m_{\rm s}/m_{\rm ud}$ with conventions A (circle) and B (triangle). The asterisk denotes the FLAG summary~\cite{nfzp-p5dn}.}
\label{fig:qmass_cont}
\end{figure}

As for the AWI quark masses, the nonperturbative renormalization factor for the PS operator in Eq.~(\ref{eq:zm}) is not available yet. We plot only the  quark mass ratio $m_{\rm s}/m_{\rm ud}$, where the renormalization factors for the axial vector current and the pseudoscalar operator are canceled  between the numerator and denominator, as a function of $a^2$ in Fig.~\ref{fig:qmass_cont}. The result in the continuum limit is slightly larger than the latest FLAG summary~\cite{nfzp-p5dn} beyond the error bar.

\section{Results with convention B}
\label{sec:isoqcd_ed}

We repeat the analysis of data employing the pseudoexperimental values of $m_\pi=0.1350$ GeV, $m_K=0.4946$ GeV, and $m_\Xi=1.3183$ GeV as the physical inputs (convention B).  
Table~\ref{tab:hmass_ph_B} summarizes the results, which should be compared with those in Table~\ref{tab:hmass_ph_A} obtained with the convention A. There are a couple of noteworthy points. Firstly, the octet baryon masses in lattice units with the convention B is slightly smaller than those with the convention A. In physical units, however, the results with the convention B become larger than those with the convention A. This is caused by the cutoff difference between two conventions. Figure~\ref{fig:hmass_cont} plots both results for conventions A and B at all $\beta$, whose difference is retained even in the continuum limit. 
Secondly, the results for $f_\pi$ and $f_K$ are in a similar situation to those for the octet baryon masses. As for the ratio $f_K/f_\pi$, which is dimensionless, the result with the convention B is smaller than those with the convention A. The comparison between two conventions at all $\beta$ is made in Fig.~\ref{fig:decayconst_cont}. We observe that the results for  $f_\pi$, $f_K$ and $f_K/f_\pi$ show consistency with the latest FLAG summary~\cite{nfzp-p5dn}.
Thirdly, the s quark mass with the convention B is reduced by about 3.5\% at all $\beta$ compared to that with the convention A. This is mainly due to the change of the $m_K$ input from 497.6 MeV to 494.6 MeV. The ud quark mass in lattice units is also reduced by about 2.4\% at all $\beta$, even though the $m_\pi$ input remains unchanged. Accordingly, the ratio $m_{\rm s}/m_{\rm ud}$ is systematically reduced by about 1\% at all $\beta$ and this deviation remains even in the continuum limit, which is observed in Fig.~\ref{fig:qmass_cont}. The continuum result  with the convention B becomes consistent with the latest FLAG summary~\cite{nfzp-p5dn} within the error bar.

  \begin{table*}
    \begin{center}
      \caption{Quark masses, PS meson decay constants, and octet baryon masses at the physical point with convention B. $m_{\rm ud}$ and $m_{\rm s}$ are given in lattice units. 
      \label{tab:hmass_ph_B}}
      \begin{tabular}{c|ccc|r}
        \hline\hline
        Data set & L128 & L160 & L256 & Continuum limit\\\hline
      $a^{-1}$ [GeV] & 2.354(04) & 3.164(05) & 4.849(21) & \multicolumn{1}{c}{$\infty$}\\
  $m_{\rm ud}$ & 0.0013167(61)  & 0.0009773(50)  & 0.0006014(70)  & \multicolumn{1}{c}{$-$}\\
  $m_{\rm s}$  & 0.036407(127)  & 0.026856(098)  & 0.016491(156) & \multicolumn{1}{c}{$-$}\\
  $m_{\rm s}/m_{\rm ud}$ & 27.650(77)  & 27.480(92)  & 27.423(134)  & 27.323(143)\\
  $f_{\pi}$ & 133.37(32)(1.56)  & 131.93(41)(1.11) & 130.43(69)(39) & 129.55(88) [MeV] \\
  $f_K$     & 158.32(21)(1.85)  & 156.98(23)(1.32) & 155.88(43)(47) & 155.12(59) [MeV]\\
  $f_{K}/f_{\pi}$  & 1.1871(25)(02) & 1.1898(36)(01) & 1.1951(52)(00) & 1.1959(55) \\
  $f_{K^\pm}$ & $-$ & $-$ & $-$ & 154.80(59)(32) [MeV]\\
  $f_{K^\pm}/f_{\pi^\pm}$  & $-$ & $-$ & $-$ & 1.1935(55)(24) \\
  $m_N$       &  959.8(6.2)  &  952.6(4.2) &  970.9(20.1) & 948.7(11.2) [MeV]\\
  $m_\Lambda$ & 1122.0(1.7)  & 1119.1(2.3) & 1118.3(5.3) & 1116.2(4.4) [MeV] \\
  $m_\Sigma$  & 1202.1(1.7)  & 1200.8(2.1) & 1200.5(4.9) & 1199.5(4.0) [MeV] \\
    \hline\hline
      \end{tabular}
    \end{center}
  \end{table*}

\section{Summary and outlook}
\label{sec:summary}

The PACS10 gauge configurations at three lattice spacings are generated with the $O(a)$-improved Wilson quark action and the Iwasaki gauge action in 2+1 flavor QCD keeping the space-time volumes over (10 fm$)^4$ at the physical point. We have investigated the finite size effect on the hadron masses and the PS meson decay constants using the additionally generated (5.4 fm$)^4$ lattice at the coarsest lattice spacing of $a=0.08487(14)$~fm. We hardly observe the finite size effect on the hadron effective masses. On the other hand, the results for the PS meson decay constants on the (10.9 fm$)^4$ deviates from those on the (5.4 fm$)^4$ lattices beyond the error bars. We have also investigated the continuum limit of the  hadron masses, the PS meson decay constants and the quark mass ratio employing two conventions for the pseudoexperimental values in the isosymmetric QCD. 
In both conventions the octet baryon masses agree  with the pseudoexperimental values within 2$\sigma$ error bands. While the PS meson decay constants and the quark mass ratio with the convention A show slight deviation from the latest FLAG summary beyond the error bars, those with the convention B are consistent with it. 

The PACS10 configurations are already used to calculate various  physical quantities directly at the physical point. Especially large spatial extent of the PACS10 configurations allows us investigate the hadron structures with finer momentum resolution~\cite{Shintani:2018ozy,Shintani:2019wai,PACS:2019hxd,Ishikawa:2022ulx,Tsuji:2022ric,Tsuji:2023llh,Aoki:2025taf}. Currently we are generating the ``PACS10$_c$'' configurations in the 2+1+1 flavor QCD, which incorporate the dynamical charm quark effects  with volumes exceeding (10 fm$)^4$ at the physical point. The new configurations should contribute to remove the uncertainties due to the missing dynamical charm quark.

\begin{acknowledgments}
Numerical calculations in this work were performed on Oakforest-PACS in Joint Center for Advanced High Performance Computing (JCAHPC) and Cygnus and Pegasus in Center for Computational Sciences at University of Tsukuba under Multidisciplinary Cooperative Research Program of Center for Computational Sciences, University of Tsukuba, and Wisteria/BDEC-01 in the Information Technology Center, the University of Tokyo. This research also used computational resources of  the Supercomputer Fugaku (Project ID: hp200167, hp210112, hp220079, hp230199, hp240207, hp240089, hp250218) provided by RIKEN Center for Computational Science (R-CCS), as well as Oakforest-PACS (Project ID: hp170022, hp170093, hp180051, hp190081, hp200062) provided by the Information Technology Center of the University of Tokyo / JCAHPC. We employ Bridge++ code  (http://bridge.kek.jp/Lattice-code/)~\cite{Ueda:2014rya,Akahoshi:2021gvk} for calculation of the topological charge. This work is supported by the JLDG constructed over the SINET6 of NII. This work was also supported in part by Grants-in-Aid for Scientific Research from the Ministry of Education, Culture, Sports, Science and Technology (Nos.~16K13798,  16H06002, 18K03605, 19H01892, 20K03924, 22K03612, 23H01195, 24K07037, 25KJ0404) and MEXT as “Program for Promoting Researches on the Supercomputer Fugaku” (Search for physics beyond the standard model using large-scale lattice QCD simulation and development of AI technology toward next-generation lattice QCD; Grant Number JPMXP1020230409).
\end{acknowledgments}

\bibliography{bib/for_this_paper}

\end{document}